\documentclass[sts]{imsart}

\newif\ifpreprint
\preprinttrue 

\makeatletter
\ifpreprint
    \def\journal@name{Preprint Version}%
    \AtBeginDocument{\setpkgattr{copyright}{text}{\itshape\journal@name}}%
\fi
\makeatother

\RequirePackage{amsthm,amsmath,amsfonts,amssymb}
\RequirePackage[dvipsnames]{xcolor}
\RequirePackage[numbers]{natbib}
\RequirePackage[colorlinks,citecolor=blue,urlcolor=blue,linkcolor=blue]{hyperref}
\RequirePackage{graphicx}
\RequirePackage{booktabs}
\RequirePackage{multirow}
\RequirePackage{algorithm}
\RequirePackage{algpseudocode}
\RequirePackage{tikz}
\usetikzlibrary{arrows.meta,positioning,calc,fit,backgrounds,shapes.geometric,
                decorations.pathreplacing,decorations.markings,patterns,shadows.blur}

\startlocaldefs
\theoremstyle{plain}

\newtheorem{theorem}{Theorem}[section]

\theoremstyle{definition}

\newtheorem{remark}[theorem]{Remark}

\DeclareMathOperator*{\argmin}{arg\,min}
\DeclareMathOperator*{\argmax}{arg\,max}
\DeclareMathOperator{\AR}{AR}
\DeclareMathOperator{\ESS}{ESS}
\DeclareMathOperator{\IACT}{IACT}
\DeclareMathOperator{\MCSE}{MCSE}
\DeclareMathOperator{\GRADe}{GRAD\text{-}e}
\DeclareMathOperator{\Var}{Var}

\DeclareMathOperator{\CoLSI}{CoLSI}

\newcommand{\vth}{\boldsymbol{\theta}}          
\newcommand{\vp}{\mathbf{p}}                    
\newcommand{\vmu}{\boldsymbol{\mu}}             
\newcommand{\vx}{\mathbf{x}}                    
\newcommand{\vz}{\mathbf{z}}                    
\newcommand{\vc}{\mathbf{c}}                    
\newcommand{\vzero}{\mathbf{0}}
\newcommand{\Hmod}{\widetilde{H}}               
\newcommand{\Hext}{\widehat{H}}                 
\newcommand{\pimod}{\widetilde{\pi}}            
\newcommand{\Rhat}{\widehat{R}}
\newcommand{\code}[1]{\textsf{#1}}

\definecolor{cSampler}{RGB}{31,84,140}
\definecolor{cInteg}{RGB}{191,84,32}
\definecolor{cTune}{RGB}{27,120,90}
\definecolor{cDiag}{RGB}{124,60,140}
\definecolor{cSoft}{RGB}{120,120,128}
\definecolor{cAccent}{RGB}{214,158,46}
\definecolor{cLight}{RGB}{242,244,247}

\endlocaldefs

\begin{document}

\begin{frontmatter}
\title{\textit{Sampling the Funnel:} A Comprehensive Survey of Hamiltonian Monte Carlo for Hierarchical Bayesian Inference}
\runtitle{Survey of Hamiltonian Monte Carlo for Hierarchical Bayesian Inference}

\begin{aug}
\author[A]{\fnms{Miguel}~\snm{Fernandez-de-Retana}\thanks{\textbf{Corresponding author.}}\ead[label=e1]{m.fernandezderetana@deusto.es}\orcid{0009-0002-0883-1303}}
\author[B]{\fnms{Aitor}~\snm{Almeida}\ead[label=e2]{aitor.almeida@deusto.es}\,\orcid{0000-0002-1585-4717}}
\address[A]{Miguel Fernandez-de-Retana is a Researcher at the Faculty of Engineering, University of Deusto (Bilbao, Spain)\printead[presep={\ }]{e1}.}

\address[B]{Aitor Almeida is a Lecturer and Researcher at the Faculty of Engineering, University of Deusto (Bilbao, Spain)\printead[presep={\ }]{e2}.}
\end{aug}

\begin{abstract}
Hierarchical models are central to applied Bayesian statistics, yet the variance structures they depend on can produce posterior geometries that remain difficult to explore even with Hamiltonian Monte Carlo (HMC), a designation now covering a large and increasingly heterogeneous family of algorithms. This survey uses Generalized HMC as an organizing framework to describe the many variants of HMC, their methodological and computational properties, and their empirical consequences. Varying the trajectory length, the degree of momentum refreshment, and the presence of a Metropolis correction recovers several familiar Hamiltonian and Langevin samplers as special cases. Three further, largely independent directions account for much of the remaining literature: modifying the metric, as in Riemannian HMC; modifying the Hamiltonian used by the acceptance test, as in modified or shadow HMC; and modifying the numerical integrator, as in multi-stage splitting schemes. We review these variants alongside the procedures used to tune and diagnose them, with particular attention to the effective sample size, whose standard estimators assume a reversibility that generalized and modified samplers lack, and ignore the importance weights that modified Hamiltonians introduce. We close with the open problems these methods face for hierarchical models, taking joint longitudinal and time-to-event models as the motivating example.

\end{abstract}

\begin{keyword}
\kwd{Bayesian Inference}
\kwd{Geometric Integration}
\kwd{Hamiltonian Monte Carlo}
\kwd{Hierarchical Models}
\kwd{Markov Chain Monte Carlo}
\end{keyword}

\end{frontmatter}

\section{Introduction}\label{sec:introduction}

Hierarchical models have become a standard tool for the analysis of data possessing \textit{grouped} or \textit{nested} structures such as repeated measurements on individuals, clustered observations (e.g., patients within hospitals in a clinical trial), or multilevel experimental designs. In all cases, whenever data arrive in \textit{groups}, a hierarchical model provides a natural framework for their analysis, allowing for the \textit{joint} estimation of global parameters common to the whole population, group-level parameters that capture systematic differences between groups, or individual-level parameters describing within-group variations, with each level informing the others through their shared structural dependences. In the Bayesian setting, hierarchical models are constructed by specifying a population distribution for the group-specific parameters, whose hyperparameters are themselves assigned a prior, so that the population level acts as a prior for the groups below it. The resulting structure is a modeling language that is both flexible and interpretable, and expressive enough to describe partial pooling of information across groups and repeated measurements, heterogeneity in group effects, uncertainty propagation, and the sharing of information across multiple levels \cite{gelman2013bayesian}. This hierarchical structure underlies random-effects models, multilevel regression, latent variable models, and the joint models for longitudinal and survival data that motivate the present review, and it accounts for a substantial part of the practical appeal of Bayesian methods in the applied sciences.

\begin{figure*}[t]
\centering
\resizebox{\textwidth}{!}{%
\begin{tikzpicture}[>=Latex, line join=round]

\begin{scope}[x=0.40cm, y=0.31cm]
  \begin{scope}
    \clip (-6.4,-9.6) rectangle (6.4,4.6);
    \fill[cSampler!14]
      plot[domain=-9.6:4.6,samples=140,smooth] ({2*exp(\x/2)},{\x})
      -- plot[domain=4.6:-9.6,samples=140,smooth] ({-2*exp(\x/2)},{\x}) -- cycle;
    \fill[cSampler!30]
      plot[domain=-9.6:4.6,samples=140,smooth] ({1*exp(\x/2)},{\x})
      -- plot[domain=4.6:-9.6,samples=140,smooth] ({-1*exp(\x/2)},{\x}) -- cycle;
    \foreach \c in {1,2,3}{
      \draw[cSampler!85!black, line width=0.5pt, domain=-9.6:4.6, samples=140, smooth]
        plot ({\c*exp(\x/2)},{\x});
      \draw[cSampler!85!black, line width=0.5pt, domain=-9.6:4.6, samples=140, smooth]
        plot ({-\c*exp(\x/2)},{\x});
    }
  \end{scope}
  \draw[->, gray!70] (-6.6,-9.6) -- (6.9,-9.6) node[below, black, font=\footnotesize] {$\theta_j$};
  \draw[->, gray!70] (-6.6,-9.6) -- (-6.6,5.2) node[left, black, font=\footnotesize] {$\log\tau$};

  \draw[cInteg, line width=0.9pt, <->] (-0.95,-7.2) -- (0.95,-7.2);
  \node[cInteg, font=\scriptsize\bfseries, anchor=west] at (0.95,-7.2) {\textit{Neck:} Small $\varepsilon$};
  \draw[cInteg, line width=0.9pt, <->] (-5.6,2.6) -- (5.6,2.6);
  \node[cInteg, font=\scriptsize\bfseries, anchor=south] at (0,2.8) {\textit{Mouth:} Large $\varepsilon$};

  \node[font=\footnotesize\bfseries, anchor=north] at (0,-11.5)
    {(a) Centered: $\theta_j \sim \mathcal{N}(0,\tau^2)$};
\end{scope}

\begin{scope}[xshift=8.2cm, x=0.40cm, y=0.31cm]
  \begin{scope}
    \clip (-6.4,-9.6) rectangle (6.4,4.6);
    \fill[cTune!14] (0,-2.5) ellipse [x radius=4.0, y radius=4.60];
    \fill[cTune!30] (0,-2.5) ellipse [x radius=2.0, y radius=2.30];
    \foreach \c in {1,2,3}{
      \draw[cTune!85!black, line width=0.5pt]
        (0,-2.5) ellipse [x radius={2.0*\c}, y radius={2.30*\c}];
    }
  \end{scope}
  \draw[->, gray!70] (-6.6,-9.6) -- (6.9,-9.6) node[below, black, font=\footnotesize] {$\tilde{\theta}_j$};
  \draw[->, gray!70] (-6.6,-9.6) -- (-6.6,5.2) node[left, black, font=\footnotesize] {$\log\tau$};

  \draw[cInteg, line width=0.9pt, <->] (-2.0,-2.5) -- (2.0,-2.5);
  \node[cInteg, font=\scriptsize\bfseries, anchor=north] at (0,-0.75) {\textit{Global $\varepsilon$}};

  \node[font=\footnotesize\bfseries, anchor=north] at (0,-11.5)
    {(b) Non-Centered: $\tilde{\theta}_j \sim \mathcal{N}(0,1),\; \theta_j = \tau\tilde{\theta}_j$};
\end{scope}

\end{tikzpicture}}
\caption{Geometry of a hierarchical posterior under two standard parameterizations, displayed in the plane spanned by a single group-level parameter and the logarithm of the population scale, with shading and contours marking the envelopes at one, two and three conditional standard deviations. In panel~(a) the group parameters are \emph{centered} upon the population distribution, so that their conditional dispersion is proportional to $\tau$ and the envelopes $\theta_j = \pm k\tau$ are exponential in $\log\tau$; the posterior thus contracts into a narrow neck as $\tau$ approaches zero and widens into a broad mouth as $\tau$ grows. Since no single integration step size $\varepsilon$ is at once stable within the neck and efficient within the mouth, a sampler governed by one global scale will either produce divergent trajectories or fail to descend into the neck at all, which is the difficulty isolated by Neal \cite{neal2003slice} and known since as the \textit{``funnel''}. In panel~(b) the \emph{non-centered} parameterization $\theta_j = \mu + \tau\tilde{\theta}_j$ renders $\tilde{\theta}_j$ independent of $\varphi$ \textit{a priori} and restores an approximately isotropic geometry, although it does so only for those groups whose posterior is dominated by the prior. Where the data are informative, the likelihood couples the two coordinates once more, and the funnel reappears in the non-centered coordinates, so that neither parameterization is uniformly preferable \cite{papaspiliopoulos2007general,betancourt2015hamiltonian}. The general case is treated in Section~\ref{sec:hierarchical}.}
\label{fig:funnel}
\end{figure*}
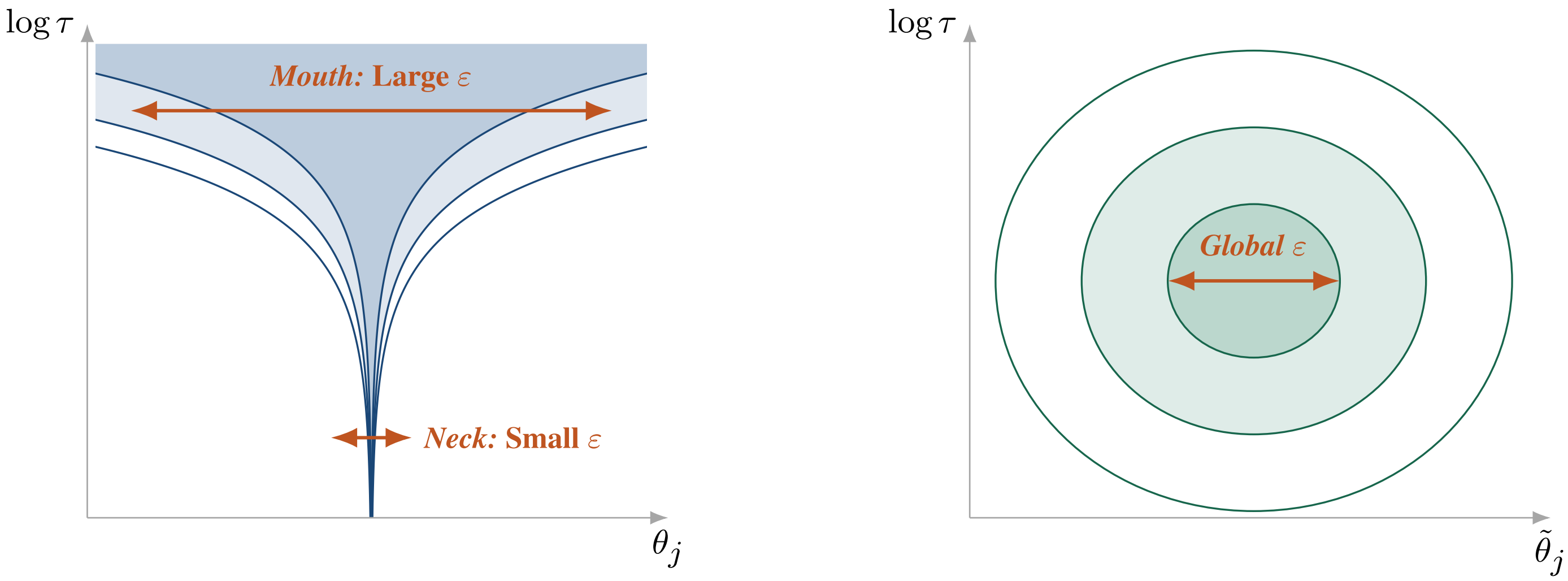

The very flexibility and expressiveness that make hierarchical models attractive can, however, make their posterior distributions remarkably difficult to explore computationally, since a variance component linking group parameters to their population distribution may produce regions with profoundly different local scales. The familiar illustration is Neal's \textit{funnel} \cite{neal2003slice}, reproduced in Figure~\ref{fig:funnel}, in which the conditional spread of the group parameters is proportional to a population scale $\tau$ and therefore contracts sharply near $\tau=0$ while broadening as $\tau$ increases. A sampler governed by a single proposal scale, integration step size, or preconditioning matrix must then trade numerical stability in the narrow region against meaningful progress in the broader one. More troublingly, a chain confined to the latter may nevertheless appear well mixed in ordinary diagnostics such as \textit{trace plots}, concealing the bias that results from its failure to traverse the funnel \cite{betancourt2016diagnosing}.

Hamiltonian Monte Carlo (HMC) \cite{duane1987hybrid,neal2011mcmc} addresses random-walk behavior by augmenting the \textit{parameters} $\vth$ with \textit{auxiliary momenta} $\vp$ and generating proposals by numerically integrating the Hamiltonian dynamics from a fictitious physical system whose potential energy is the negative log target density. Proposed moves are thus guided by gradients of the log target density rather than by purely random-walk increments, subject to a Metropolis accept/reject correction. The benefit is most evident in high dimensions, where for a $D$-dimensional target with \textit{independent and identically distributed} (i.i.d.) components, Beskos et al.\ \cite{beskos2013optimal} show that the leapfrog step size $\varepsilon$ must only shrink as $D^{-1/4}$ to keep the acceptance rate bounded away from zero, compared with $D^{-1/2}$ for the proposal standard deviation of Random-Walk Metropolis. Betancourt and Girolami \cite{betancourt2015hamiltonian} give a further detailed account of the benefits and limitations of HMC for hierarchical models.

The designation \textit{``Hamiltonian Monte Carlo''} has, nevertheless, long ceased to identify a single algorithm. Over the course of three decades, the method has been developed within several research communities whose concerns, motivations, and vocabularies differ considerably: \emph{computational statistics} has been occupied principally with adaptivity and with the diagnosis of convergence, \emph{molecular simulation} with energy conservation and the design of integrators, \emph{machine learning} (ML) with scalability and with learned proposal mechanisms, and \emph{numerical analysis} with the geometric structure of the underlying flows. A practitioner approaching a hierarchical model is accordingly confronted with generalized, Riemannian, modified, magnetic, relativistic or microcanonical variants of the sampler, among others, with several schemes for the automatic selection of its hyperparameters, with families of splitting integrators of varying order and stability properties, and with a number of mutually inconsistent definitions of well-known diagnostic metrics. The relations among these developments are seldom stated explicitly, and a choice among them is correspondingly difficult to make on principled grounds.

\subsection{Aims \& Organizing Principles}

The present article is intended as a \textit{narrative} rather than a systematic review and, accordingly, makes no attempt to enumerate the literature exhaustively. Our purpose is instead to offer a connected account of the developments that bear most directly upon hierarchical Bayesian inference, arranging them so that their logical relations become apparent and, \textit{hopefully}, in a form that is both useful and accessible to practitioners and researchers alike. The exposition is organized around three related ideas, introduced here to orient the discussion that follows.

At the center of this account lies the observation that most Hamiltonian samplers in practical use may be regarded as particular instances of a single generalization of the original method. Generalized Hamiltonian Monte Carlo (GHMC) \cite{horowitz1991generalized,kennedy2001cost} replaces the complete refreshment of the momentum performed at every iteration of standard HMC with a partial refreshment governed by a parameter $\phi \in (0,1]$, together with a reversal, or \textit{flip}, of the momentum whenever a proposal is rejected to ensure reversibility. As $\phi$ ranges from one to values approaching zero, the update interpolates between complete refreshment and near-total retention of the previous momentum. Once this choice is considered jointly with the number of integration steps $L$ and the presence, or absence, of a Metropolis correction, standard HMC, the Metropolis-Adjusted Langevin algorithm (MALA), second-order Langevin Monte Carlo (L2MC), deterministic molecular dynamics (MDMC), or stochastic Langevin dynamics (SLDMC) all emerge as particular cases. Section~\ref{sec:ghmc} and Table~\ref{tab:ghmc-cases} revisit these relationships in detail, while Figure~\ref{fig:taxonomy} uses them to organize a wider taxonomy.

Additionally, beyond the $(L, \phi, \text{MH-test})$ axes of this generalized sampler, much of the remaining literature can be understood in terms of three further directions that are largely independent of one another. Riemannian manifold Hamiltonian Monte Carlo (RMHMC) \cite{girolami2011riemann}, considered in Section~\ref{sec:geometry}, replaces the constant mass matrix $M$ with a position-dependent \emph{metric} $G(\vth)$, thereby rendering the Hamiltonian \textit{non-separable} but allowing it to adapt to local variations in curvature such as those encountered in the funnel of Figure~\ref{fig:funnel}. \textit{Modified}, or shadow, Hamiltonian methods \cite{izaguirre2004shadow,akhmatskaya2008gshmc,radivojevic2020modified} begin instead from the fact that a numerical symplectic integrator conserves not the exact Hamiltonian but a nearby \emph{modified} (or ``shadow'') Hamiltonian $\Hmod$ to much higher order \cite{hairer2006geometric}. By sampling from a modified distribution $\pimod \propto \exp(-\Hmod)$ and \textit{reweighting} the resulting draws back to the original target, they exchange the additional work associated with a truncation of the formal series in $\varepsilon$ for improved energy conservation and higher acceptance rates, as discussed in Section~\ref{sec:mhmc}. Multi-stage splitting integrators (MSSIs) \cite{blanes2014numerical,radivojevic2018multi,nagar2024adaptive}, presented in Section~\ref{sec:integrators}, generalize the standard Verlet-Leapfrog scheme to palindromic compositions of \textit{momentum kicks} and \textit{position drifts} with tunable coefficients, again trading additional gradient evaluations for more accurate energy conservation. Since these three modifications are largely orthogonal and may be combined, one may, for example, construct a modified-Hamiltonian sampler with partial momentum refreshment and an adaptively tuned three-stage integrator, with each choice carrying its own consequences for the geometry of the target, the stability of the integration, and the efficiency of the resulting sampler.

The final organizing concern is \textit{tuning} and \textit{diagnosis}, a subject often relegated to a brief aside in the algorithmic literature despite its central importance in practice. The performance of a Hamiltonian sampler is governed by the step size $\varepsilon$, the number of integration steps $L$, the momentum noise parameter $\phi$, and the mass matrix $M$ (or, more generally, the metric $G(\vth)$), yet the optimal values best suited to these quantities depend upon a target geometry that is unknown \textit{a priori}. Section~\ref{sec:tuning} thus contrasts two particularly successful families of schemes that have developed around this problem of parameter selection and tuning: methods descended from the No-U-Turn sampler (NUTS) \cite{hoffman2014no,hoffman2021adaptive,bourabee2024gist}, which adapt online from the observed behavior of the chain, and those descended from the adaptive integration approach (AIA) \cite{nagar2024adaptive,akhmatskaya2026adaptive}, which fit a tractable surrogate during a \textit{burn-in} phase and derive their parameters from a closed-form analysis. Section~\ref{sec:diagnostics} then turns to the assessment of convergence, exploration, and computational efficiency, paying particular attention to the \textit{effective sample size} (ESS) metric, whose meaning and relation to sampler efficiency are considerably more subtle and context-dependent than its routine use might suggest. We consequently distinguish among the assumptions, limitations, and appropriate uses of the available estimators: Geyer's initial sequence estimators \cite{geyer1992practical}, for example, provide the default ESS in most widely used implementations but rely upon reversibility, a condition not satisfied by generalized and modified Hamiltonian methods, whereas estimators developed for \textit{importance sampling} \cite{kong1992note,elvira2022rethinking,martino2016effective} account for weight degeneracy while disregarding autocorrelation entirely. Modified HMC combines simultaneously autocorrelation, irreversibility, and importance reweighting, and no standard estimator presently addresses all three at once.

\subsection{Scope}

Our motivating application, treated in Section~\ref{sec:hierarchical}, is the joint modeling of longitudinal and time-to-event data \cite{rizopoulos2012joint}. Such models include subject-specific random effects beneath a variance hierarchy and can therefore exhibit funnel-like geometries. Moreover, their survival likelihoods commonly require numerical quadrature for the cumulative hazard, making gradients substantially more expensive to compute than in a comparable generalized linear mixed models. In this setting, posterior geometry and the number of gradients per effectively independent draw are both important. Higher-order integrators, modified Hamiltonians, and partial momentum refreshment are relevant insofar as they improve that trade-off.

We restrict attention throughout to continuous parameter spaces and to samplers that are, at least \textit{in intent}, asymptotically exact whether directly or after reweighting. Variational methods, together with the extensive literature on unadjusted stochastic gradient dynamics, appear only where they bear directly upon the samplers under discussion. Extensions to discrete parameters are surveyed briefly in Section~\ref{sec:geometry} and revisited among the open problems collected in Section~\ref{sec:challenges}.

\subsection{Notation \& Outline}

We write $\vth \in \mathbb{R}^D$ for the model parameters, which play the role of \textit{``position''} coordinates in the simulated system, and $\vp \in \mathbb{R}^D$ for the \textit{auxiliary momenta}; $\vmu$ denotes the Gaussian noise vector employed in partial momentum updates, and $\vz = (\vth, \vp)$ a point of the $2D$-dimensional phase space. The target density is $\pi(\vth)$, the potential energy $U(\vth) = -\log \pi(\vth)$ up to an additive constant, the kinetic energy $K(\vp) = \tfrac{1}{2}\vp^\top M^{-1}\vp$ with mass matrix $M$, and the Hamiltonian $H = U + K$. Numerical integrators are denoted by $\Psi_{\varepsilon, L}$, where $\varepsilon$ is the step size and $L$ the number of integration steps, and are constructed from position drifts $\varphi_t^A$ and momentum kicks $\varphi_t^B$. Finally, $\Hmod^{[k]}$ denotes a truncated \textit{modified} Hamiltonian, $\Hext$ the \textit{extended} Hamiltonian associated with a partial momentum update (PMU), $\phi$ the momentum noise parameter, $N$ the number of Monte Carlo samples, and $D$ the dimension of the problem, with the notation as a whole summarized in Table~\ref{tab:notation} for reference. The notation is chosen to be consistent with the literature, but also to be as uniform as possible across the several communities that have contributed to the development of Hamiltonian Monte Carlo.

The remainder of the article is organized as follows. Section~\ref{sec:background} reviews Bayesian inference, Markov Chain Monte Carlo, and the pathologies specific to hierarchical problems, after which Section~\ref{sec:hmc} develops HMC and its geometric foundations, and Section~\ref{sec:ghmc} introduces GHMC as the unifying framework. Modified Hamiltonians and importance-reweighted sampling are then treated in Section~\ref{sec:mhmc}, with Riemannian, non-separable, and alternative dynamics following in Section~\ref{sec:geometry}. Numerical integration, hyperparameter tuning, and sampling diagnostics, including the effective sample size, occupy Sections~\ref{sec:integrators}, \ref{sec:tuning}, and \ref{sec:diagnostics}, respectively. Section~\ref{sec:parallelism} asks how these cost accountings change once the samplers are placed upon parallel hardware and distinguishes the several meanings attached to the term \textit{``parallelism''}. We return to hierarchical models, and to joint models in particular, in Section~\ref{sec:hierarchical}. The available software is surveyed in Section~\ref{sec:software}, and Section~\ref{sec:challenges} collects open problems and directions for future work.

\begin{table}[t]
\centering
\caption{Notation used throughout the survey.}
\label{tab:notation}
\renewcommand{\arraystretch}{1.15}
\begin{tabular}{ll}
\toprule
\textbf{Symbol} & \textbf{Meaning} \\
\midrule
$\vth \in \mathbb{R}^D$ & Model parameters \textit{(position coordinates)} \\
$\vp \in \mathbb{R}^D$ & Auxiliary momenta \\
$\vmu \in \mathbb{R}^D$ & Gaussian noise vector (partial momentum update) \\
$\vz = (\vth, \vp)$ & Phase-space point \\
$\pi(\vth)$ & Target density (posterior) \\
$U(\vth)$ & Potential energy of the fictitious system \\
$K(\vp)$ & Kinetic energy of the fictitious system \\
$H(\vth,\vp)$ & Hamiltonian, $U(\vth) + K(\vp)$ \\
$\Hmod^{[k]}$ & $k$-th order truncated \textit{modified} Hamiltonian \\
$\Hext$ & Extended Hamiltonian for the PMU \\
$M$ & Mass matrix; $G(\vth)$ for a position-dependent metric \\
$\varepsilon$, $\bar{\varepsilon}$ & Integration step size; Dimensionless step size \\
$L$ & Number of integration steps (i.e., trajectory length) \\
$\phi \in (0,1]$ & Momentum noise or refreshment parameter \\
$k$ & Number of stages of a splitting integrator \\
$\Psi_{\varepsilon,L}$ & Numerical integrator \\
$\varphi_t^A$, $\varphi_t^B$ & Position drift and momentum kick flows \\
$\omega_i$, $\bar\omega_i$ & Importance weights; Normalized importance weights \\
$N$ & Number of Monte Carlo samples \\
$D$ & Dimension of the problem \\
\bottomrule
\end{tabular}
\end{table}

\section{Background}\label{sec:background}

\subsection{Bayesian Inference \& Monte Carlo Integration}

In contrast to the \textit{frequentist} approach, in which the parameters $\vth$ of a model $\mathcal{M}$ are regarded as \textit{fixed}, but unknown constants and inferential procedures are justified by their behavior under hypothetical repetitions of the experiment, Bayesian inference regards $\vth$ as random variables and the observed data $\vx$ as fixed, so that uncertainty about the parameters is expressed directly in probabilistic terms and every inferential statement derives from the \emph{posterior} distribution obtained by conditioning the prior on the data through \textit{Bayes' theorem},
\begin{equation}\label{eq:posterior}
  \pi(\vth) \;\equiv\; p(\vth \mid \vx)
  \;=\; \frac{p(\vth)\, p(\vx \mid \vth)}{p(\vx)},
\end{equation}
where $p(\vth)$ is the \textit{prior}, $p(\vx \vert \vth)$ the \textit{likelihood}, and $p(\vx) = \int p(\vx \mid \vth) p(\vth)\, d\vth$ the \textit{marginal likelihood}, or \textit{evidence}, which acts as a normalizing constant and is typically not available in closed form \cite{gelman2013bayesian,robert1999monte}. Moreover, the posterior density is seldom the ultimate object of interest, with quantities of interest for inference being more often \textit{expectations} of functions $f(\vth)$ with respect to that posterior, of the form
\begin{equation}\label{eq:expectation}
  I \;=\; \mathbb{E}_\pi\!\left[f(\vth)\right] \;=\; \int f(\vth)\, \pi(\vth)\, d\vth,
\end{equation}
a class of integrals broad enough to encompass posterior means and covariances (taking $f(\vth) = \vth$ and $f(\vth) = \vth\vth^\top$ respectively), masses of credible regions (taking $f$ to be an indicator function), or posterior predictive distributions of the form $p(\tilde{\vx} \mid \vx) = \int p(\tilde{\vx} \mid \vth)\, \pi(\vth)\, d\vth$ for prediction and model checking. The integrals in \eqref{eq:expectation} are typically high-dimensional, with $D = \dim(\vth)$ ranging from a few to several thousand in realistic applications, and they are often sharply concentrated in small regions of the parameter space, so that the integrand is negligible over most of the domain of integration. 

Thus, the standard approach to this problem is through Monte Carlo (MC) integration, whereby the integral is replaced by an average of the integrand evaluated at draws from the posterior,
\begin{equation}\label{eq:mc-estimator}
  \widehat{I} \;=\; \frac{1}{N}\sum_{n=1}^{N} f(\vth_n),
  \qquad \vth_n \sim \pi(\vth) \ \ \text{\textit{i.i.d.}}
\end{equation}
where, provided that such draws are available, the estimator in \eqref{eq:mc-estimator} is unbiased and has variance $\sigma_f^2/N$, independently of $D$ once the variance of $f(\vth)$ has been fixed, which makes the procedure attractive even in high-dimensional settings \cite{owen2013monte}. The difficulty, of course, is that the posterior is generally known only up to a normalizing constant $p(\vx)$, and consequently cannot usually be sampled from directly.

\subsection{Markov Chain Monte Carlo (MCMC)}

Markov Chain Monte Carlo (MCMC) methods address this lack of direct sampling by constructing a Markov Chain $\vth_0, \vth_1, \dots$ whose \textit{stationary} distribution is the target $\pi$, so that after an initial \emph{burn-in} period the states of the chain may be treated as draws from the posterior, albeit correlated ones. The Metropolis--Hastings schema \cite{metropolis1953equation,hastings1970monte} then proposes a candidate state $\vth'$ from a \emph{proposal distribution} $q(\vth' \mid \vth)$ and accepts it with probability
\begin{equation}\label{eq:mh}
  \alpha \;=\; \min\left\{1,\;
  \frac{\pi(\vth')\, q(\vth \mid \vth')}{\pi(\vth)\, q(\vth' \mid \vth)}\right\},
\end{equation}
which leaves $\pi$ invariant and involves the target only through unnormalized densities. Under mild irreducibility and aperiodicity conditions, the resulting chain is \textit{ergodic} and the estimator \eqref{eq:mc-estimator} remains consistent when the $\vth_n$ are states of the chain rather than \textit{i.i.d.}\ draws \cite{tierney1994markov}. Nevertheless, the resulting consecutive states are \textit{correlated} and the variance of the estimator is consequently inflated so the effective number of independent draws is smaller than $N$, the subject of Section~\ref{sec:diagnostics}.

This loss of \textit{effective} sample size makes the efficiency of the sampler, and hence the choice of proposal distribution $q$, especially important, as is already apparent for Random-Walk Metropolis--Hastings (RWMH), which takes $q(\vth' \mid \vth) = \mathcal{N}(\vth, \sigma^2 I)$. In high dimension, probability mass becomes concentrated in a thin shell of the parameter space known as the \emph{typical set} \cite{betancourt2017conceptual}, so that an isotropic proposal large enough to move appreciably will generally leave this shell and be rejected, whereas one small enough to obtain a reasonable acceptance probability advances each coordinate by only a short distance. Indeed, under the product-target asymptotics of Roberts and Rosenthal \cite{roberts2001optimal}, the proposal standard deviation must scale as $\sigma\propto D^{-1/2}$, with an optimal acceptance rate of $0.234$, and the chain consequently requires $\mathcal{O}(D)$ iterations to travel an $\mathcal{O}(1)$ distance in each coordinate.

The poor high-dimensional scaling of random-walk proposals motivates the introduction of gradient information, as in the Metropolis-Adjusted Langevin algorithm (MALA) \cite{rossky1978brownian,roberts1998optimal}, whose proposal discretizes a Langevin diffusion that is $\pi$-invariant and, under suitable conditions, geometrically ergodic \cite{roberts1996exponential}. MALA biases each proposal along $\nabla \log \pi(\vth)$ and achieves an Euler step size $h \propto D^{-1/3}$, equivalently, a proposal noise standard deviation $\sigma \propto D^{-1/6}$, at an optimal acceptance rate of $0.574$. HMC goes further by composing many gradient evaluations into a single consistent trajectory and, at an optimal acceptance rate of approximately $0.651$, achieves $\varepsilon \propto D^{-1/4}$ \cite{beskos2013optimal}. Table~\ref{tab:scaling} collects these two regimes together with the random-walk baseline.

\begin{table}[t]
\centering
\caption{Optimal scaling of Metropolis-type samplers under \textit{i.i.d.} asymptotics. ``Step size'' is the proposal standard deviation for RWMH, the Euler step for MALA, and the leapfrog step for HMC. Work counts the cost of moving an $\mathcal{O}(1)$ distance in each coordinate, charging $\mathcal{O}(D)$ for each target or gradient evaluation.}
\label{tab:scaling}
\renewcommand{\arraystretch}{1.2}
\begin{tabular}{lccc}
\toprule
\textbf{Sampler} & \textbf{Step Size} & \textbf{Optimal AR} & \textbf{Work} \\
\midrule
RWMH \cite{roberts2001optimal}   & $D^{-1/2}$ & $0.234$ & $\mathcal{O}(D^{2})$ \\
MALA \cite{roberts1998optimal} & $D^{-1/3}$ & $0.574$ & $\mathcal{O}(D^{4/3})$ \\
HMC \cite{beskos2013optimal}     & $D^{-1/4}$ & $0.651$ & $\mathcal{O}(D^{5/4})$ \\
\bottomrule
\end{tabular}
\end{table}

\subsection{Hierarchical Models \& Associated Pathologies}\label{sec:hier-pathology}

In its simplest and most widely studied form, a \textit{hierarchical model} assigns a parameter $\theta_j$ to each of $J$ groups and ties the resulting vector $\vth = (\theta_1, \dots, \theta_J)$ together through a common population distribution, whose parameters are themselves treated as unknown \textit{hyperparameters} $\varphi$ and assigned a prior,
\begin{equation}\label{eq:hier}
\begin{aligned}
  \varphi &\sim p(\varphi), &
  \theta_j \mid \varphi &\sim p(\theta_j \mid \varphi), \\
  y_j \mid \theta_j &\sim p(y_j \mid \theta_j), &
  j &= 1,\dots,J,
\end{aligned}
\end{equation}
in which $y_j$ denotes the observations associated with the $j$-th group, with extensions to multivariate settings and to further levels being straightforward. In the most common case, the population distribution is a location-scale family, typically $\theta_j \mid \varphi \sim \mathcal{N}(\mu, \tau^2)$ with $\varphi = (\mu, \tau)$, so that the scale $\tau$ controls the degree to which the groups are pooled toward the common mean $\mu$, we use this Gaussian case as a running example below. The posterior over the full vector $(\varphi, \vth)$, of dimension $J+2$ in the Gaussian case, exhibits three distinct features that stress samplers and render the scaling results of Table~\ref{tab:scaling} relevant even for moderate values of the number of groups $J$ and the dimension of the hyperparameters $D$.

\medskip\noindent\textbf{Varying curvature.} Conditionally on $\tau$, the group parameters are dispersed on a scale proportional to $\tau$. When the posterior supports a wide range of values for this hyperparameter, as can happen with few groups or weakly informative ones, their conditional widths vary accordingly. The joint density can then assume the \textit{funnel} shape of Figure~\ref{fig:funnel}, with a narrow, high-curvature region at small $\tau$, and a broader region at large $\tau$. In terms of the potential energy $U(\vth)=-\log\pi(\vth)$, the local Hessian $\nabla^2U$ can consequently be \textit{ill conditioned}, i.e., have a large condition number $\kappa$, and the resulting geometry can be difficult to explore. This feature motivates the Riemannian methods of Section~\ref{sec:geometry} and the reparameterizations of Section~\ref{sec:hierarchical}.

\medskip\noindent\textbf{Strong global correlation.} The second feature is a strong global dependence that persists well away from the neck of the funnel. The population scale $\tau$ and the group parameters are strongly \textit{dependent}, since the spread of the $\theta_j$ about $\mu$ determines the plausible values of $\tau$, and the $\theta_j$ are mutually dependent through their common parent, with the consequence that samplers updating the coordinates in separate blocks are severely handicapped. In the Gaussian case with group-level estimates $\bar{y}_j \mid \theta_j \sim \mathcal{N}(\theta_j, \sigma_j^2)$, a Gibbs sampler alternating between the full conditionals $\tau \mid \vth$ and $\vth \mid \tau$ can take very small steps when each group carries little information. Given $\vth$, the scale $\tau$ is determined to within a relative error of order $J^{-1/2}$, whereas its posterior spread is larger by a factor of roughly $1/\lambda$, where $\lambda = \tau^2/(\tau^2+\sigma_j^2)$, so that the number of iterations needed to traverse the posterior of $\tau$ is of order $\min(\lambda^{-2}, J)$ \cite{papaspiliopoulos2007general}. The slow mixing of the variance components persists in deeper hierarchies \cite{zanella2021multilevel}. Samplers that update the entire parameter vector simultaneously, among them HMC, avoid this particular difficulty, although they remain fully subject to the first.

\medskip\noindent\textbf{Growing dimensionality.} As more data are collected, the dimension grows with the data, i.e., the number of groups $J$ and the number of parameters per group $q$ can increase with the sample size, so that the dimension of the target distribution is not fixed. Each additional group contributes parameters, so a joint model for $n$ subjects with $q$ random effects per subject has $nq$ latent parameters in addition to its fixed effects. As the dimension increases, samplers with random-walk behavior in even a subset of the coordinates can degrade sharply. The scaling results in Table~\ref{tab:scaling} are therefore practically relevant to hierarchical models.

\medskip

In order to address these pathologies, two broad families of complementary strategies have been developed. The first modifies the algorithm, whether by exploiting gradient information, by adapting the metric to the local geometry of the target, or by improving the numerical integrator, and occupies Sections~\ref{sec:hmc} through~\ref{sec:tuning}. The second modifies the coordinates in which the model is expressed, replacing the centered specification of \eqref{eq:hier} by a \emph{non-centered} parameterization
\begin{equation}\label{eq:ncp}
  \tilde{\theta}_j \sim \mathcal{N}(0, 1), \qquad \theta_j = \mu + \tau\tilde{\theta}_j,
\end{equation}
in which $\tilde{\theta}_j$ is \textit{a priori} independent of the hyperparameters $\varphi$ and the funnel is consequently absent from the prior, as illustrated in the right panel of Figure~\ref{fig:funnel}. Neither parameterization dominates the other in general: as Betancourt and Girolami \cite{betancourt2015hamiltonian} established, the preferable choice depends upon whether the prior or the likelihood dominates the posterior of each individual group, and in models where that balance differs from one group to another neither specification is adequate on its own. Section~\ref{sec:hierarchical} returns to this question in detail.

\section{Hamiltonian Monte Carlo (HMC)}\label{sec:hmc}

\subsection{The Augmented Target Distribution}

Hamiltonian Monte Carlo \cite{duane1987hybrid,neal2011mcmc} circumvents the random-walk behavior described in Section~\ref{sec:background} by augmenting the parameters with auxiliary momenta and using Hamiltonian trajectories to construct distant proposals informed by the geometry of the target distribution. The construction borrows from statistical mechanics the notion of a \textit{canonical distribution}: given an energy function $E(\vx)$ defined over the states of a physical system, the associated canonical distribution at temperature $T$ has density $P(\vx) \propto \exp(-E(\vx)/T)$, so that, read in the opposite direction, any density of interest may be realized as a canonical distribution at $T = 1$ by taking $E(\vx) = -\log P(\vx)$ up to an additive constant \cite{neal2011mcmc}. Applying this correspondence to an augmented state consisting of the parameters $\vth$, which now play the role of a \textit{position} coordinate, together with auxiliary fictitious \emph{momenta} $\vp \in \mathbb{R}^D$ assigned a density independent of $\vth$, yields the joint target
\begin{equation}\label{eq:aug-target}
  \pi(\vth, \vp) \;=\; \pi(\vth)\, p(\vp) \;\propto\; \exp\!\left(-H(\vth, \vp)\right),
\end{equation}
whose energy function is the \emph{Hamiltonian}
\begin{equation}\label{eq:hamiltonian}
  \begin{aligned}
  H(\vth, \vp) &= U(\vth) + K(\vp)\\
  &= -\log \pi(\vth) \;+\; \tfrac{1}{2}\vp^\top M^{-1}\vp ,
  \end{aligned}
\end{equation}
where $U$ is the \textit{potential energy}, $K$ the \textit{kinetic energy}, and $M$ a symmetric positive definite \textit{mass matrix} of the system. Choosing $\vp \sim \mathcal{N}(\vzero, M)$ renders $K$ quadratic and the Hamiltonian \emph{separable}, in the sense that $U$ depends on $\vth$ alone and $K$ on $\vp$ alone. Separability is not theoretically demanded but allows the use of the computationally efficient explicit symplectic integrators of Section~\ref{sec:integrators}, and relinquishing it as the Riemannian methods of Section~\ref{sec:geometry}, carries a substantial computational cost.

Because the joint density in \eqref{eq:aug-target} factors into a product of the marginal $\pi(\vth)$ and the conditional $p(\vp)$, the position and momentum are independent under the augmented target and the marginal in $\vth$ is exactly the posterior $\pi(\vth)$: any chain that is invariant for $\pi(\vth, \vp)$ therefore yields, upon discarding the momenta, samples from the distribution of interest. Although the momenta are introduced purely as auxiliary variables, the whole advantage of HMC stems from exploiting the geometry of the resulting augmented target in order to propose moves in $\vth$ that remain coherent over comparatively long distances, thereby producing a Markov Chain far less diffusive than that obtained from conventional random-walk Metropolis--Hastings. To see how this advantage arises, we first describe the dynamics underlying those proposals and then turn to the manner in which they may be embedded within a valid MCMC algorithm.

\subsection{Hamiltonian Dynamics \& Its Properties}

In terms of the dynamics of the system, the augmented state $\vz = (\vth, \vp)$ evolves in a fictitious time $t$ according to \textit{Hamilton's equations},
\begin{equation}\label{eq:hamilton-eqs}
\begin{aligned}
  \dot{\vth} \;&=\; \frac{\partial H}{\partial \vp} \;=\; M^{-1}\vp, \\
  \dot{\vp} \;&=\; -\frac{\partial H}{\partial \vth} \;=\; -\nabla U(\vth),
\end{aligned}
\end{equation}
which may be written compactly as $\dot{\vz} = J\,\nabla H(\vz)$ with $J = \left(\begin{smallmatrix} 0 & I \\ -I & 0\end{smallmatrix}\right)$ the canonical symplectic matrix. Writing $\Phi_t$ for the exact flow map that advances the state over time, four properties make $\Phi_t$ an ideal proposal mechanism for MCMC \cite{neal2011mcmc,leimkuhler2004simulating,betancourt2017geometric}.

\medskip\noindent\textbf{Reversibility.} The map $\Phi_t$ is one-to-one and possesses an inverse $\Phi_{-t}$ obtained by negating the time derivatives in \eqref{eq:hamilton-eqs}. When the Hamiltonian is separable and the kinetic energy is even in the momentum, so that $K(\vp) = K(-\vp)$ as holds for the quadratic form adopted here, the inverse admits the more convenient representation
\begin{equation}\label{eq:reversibility}
  \Phi_{-t} \;=\; \mathcal{F} \circ \Phi_{t} \circ \mathcal{F},
  \qquad \mathcal{F}(\vth, \vp) = (\vth, -\vp),
\end{equation}
so that reversing the dynamics amounts to negating the momentum, evolving forward for the same interval, and negating the momentum once more. Reversibility allows a Metropolis proposal built from $\Phi_t$ to be symmetric, and hence the acceptance probability to reduce to a ratio of target densities alone.

\medskip\noindent\textbf{Conservation of the Hamiltonian.} Along any solution of \eqref{eq:hamilton-eqs} the energy is an invariant of the motion, since
\begin{equation}\label{eq:conservation}
\begin{aligned}
  \frac{dH}{dt}
  &\;=\; \sum_{i=1}^{D}\left(\frac{d\theta_i}{dt}\frac{\partial H}{\partial \theta_i}
     + \frac{dp_i}{dt}\frac{\partial H}{\partial p_i}\right) \\
  &\;=\; \sum_{i=1}^{D}\left(\frac{\partial H}{\partial p_i}\frac{\partial H}{\partial \theta_i}
     - \frac{\partial H}{\partial \theta_i}\frac{\partial H}{\partial p_i}\right)
  \;=\; 0 .
\end{aligned}
\end{equation}
As the augmented density in \eqref{eq:aug-target} is a monotone function of $H$ alone, an exactly simulated trajectory moves within a hypersurface of constant probability density, and a proposal generated in this manner would be accepted with probability one however far it travelled. In practice, the dynamics can only be integrated approximately, with the resulting energy error responsible for rejections.

\medskip\noindent\textbf{Volume Preservation.} The vector field defined by \eqref{eq:hamilton-eqs} is divergence free, i.e.,
\begin{equation}\label{eq:liouville}
  \nabla\cdot\dot{\vz}
  \;=\; \sum_{i=1}^{D}\left(\frac{\partial}{\partial\theta_i}\frac{\partial H}{\partial p_i}
     - \frac{\partial}{\partial p_i}\frac{\partial H}{\partial \theta_i}\right)
  \;=\; 0 ,
\end{equation}
from which it follows that $\Phi_t$ carries any region of phase space onto a region of equal volume, a result known as \textit{Liouville's theorem}, and equivalently, that the Jacobian determinant of $\Phi_t$ has unit absolute value. The significance for MCMC is that the Metropolis--Hastings ratio in \eqref{eq:mh} then requires no Jacobian correction, so that a deterministic and highly non-local proposal may be employed without evaluating a determinant that would in general be intractable.

\medskip\noindent\textbf{Symplecticness.} A stronger statement may be made by writing $B_t$ for the Jacobian matrix of $\Phi_t$ regarded as a map of $\vz$, in terms of which the flow satisfies the \textit{symplecticness} condition
\begin{equation}\label{eq:symplectic}
  B_t^\top J^{-1} B_t \;=\; J^{-1},
\end{equation}
from which volume preservation follows immediately, since \eqref{eq:symplectic} forces $\det(B_t)^2 = 1$. In practice, this symplecticness is strictly stronger than volume preservation, and HMC integrators are constructed to preserve it.

\medskip
The observation on which the whole method depends is that reversibility, volume preservation and symplecticness can be maintained \emph{exactly} by a suitable discretization of \eqref{eq:hamilton-eqs}, even though conservation of the Hamiltonian cannot \cite{neal2011mcmc,hairer2006geometric}. The two halves of this asymmetry are pursued separately in the sections that follow, the preservation of structure in Section~\ref{sec:integrators} and the failure of exact energy conservation in Section~\ref{sec:mhmc}.

\begin{remark}
In terms of the choice of momentum distribution, any density $p(\vp)$ yields a valid algorithm, but the Gaussian $\mathcal{N}(\vzero, M)$ is very nearly universal for two reasons. It renders the first of Hamilton's equations linear, $\dot{\vth} = M^{-1}\vp$, giving a drift that can be solved exactly, and it makes the mass matrix act as a linear preconditioner, since taking $M$ equal to the inverse posterior covariance transforms the target to approximate isotropy, which is why practical implementations estimate a diagonal or dense $M$ during warm-up. Other momentum distributions can be useful in specific settings, including Laplace momenta for discontinuous targets \cite{nishimura2020discontinuous} and relativistic kinetic energies that bound the particle speed and limit the effect of large gradients \cite{lu2017relativistic}. Both are discussed in Section~\ref{sec:geometry}.
\end{remark}

\subsection{The HMC Algorithm}

Since Hamilton's equations admit no closed-form solution for a general potential, the flow must be approximated by a numerical integrator $\Psi_{\varepsilon, L}$ applied for $L$ steps of size $\varepsilon$, after which a complete HMC iteration may be understood as the composition of two updates that each leave the augmented target \eqref{eq:aug-target} invariant \cite{neal2011mcmc}. In the first, the momentum is drawn afresh from $\vp \sim \mathcal{N}(\vzero, M)$ independently of the current position and, because $\vth$ remains unchanged while $\vp$ is drawn from its correct conditional distribution, which by independence coincides with its marginal, invariance is immediate. In the second, a Metropolis update is performed in which the proposal is obtained by integrating the dynamics from the current state and negating the resulting momentum, and is accepted with probability
\begin{equation}\label{eq:hmc-accept}
\begin{aligned}
  \alpha &\;=\; \min\left\{1,\, \exp\!\left(-\Delta H\right)\right\}, \\
  \Delta H &\;=\; H(\vth', \vp') - H(\vth, \vp).
\end{aligned}
\end{equation}
The negation is what makes the proposal \textit{symmetric} and hence \eqref{eq:hmc-accept} valid, although implementations routinely omit it: $K(\vp) = K(-\vp)$ leaves the acceptance probability unchanged, and the momentum is discarded at the start of the next iteration in any case. In Generalized HMC, treated in Section~\ref{sec:ghmc}, the momentum persists from one iteration to the next and this negation acquires a genuinely consequential role.

Moreover, provided that the integrator is volume preserving and time reversible, the Metropolis correction leaves the augmented target $\pi(\vth,\vp)$ invariant for any step size \cite{neal2011mcmc}, so numerical error affects the acceptance rate rather than changing in any way the \textit{stationary distribution} of the chain. Symplectic integrators are preferred because, in addition to satisfying these structural requirements, they control the energy error over long trajectories. The complete procedure is given in Algorithm~\ref{alg:hmc}.

\begin{algorithm}[t]
\caption{Hamiltonian Monte Carlo (HMC)}
\label{alg:hmc}
\begin{algorithmic}[1]
\State \textbf{Input:} \begin{tabbing}
        \hspace{0.4cm} \= $N$: \hspace{0.75cm} \= Number of Monte Carlo samples \\
        \> $\varepsilon$: \> Step size \\
        \> $L$: \> Number of integration steps (trajectory length) \\
        \> $M$: \> Mass matrix \\
        \> $\Psi_{\varepsilon, L}$: \> Symplectic, reversible numerical integrator
    \end{tabbing}
\State \textbf{Output:} \begin{tabbing}
        \hspace{0.4cm} \= $\{\vth_n\}_{n=1}^N$: \hspace{0.3cm} \= Samples from the target distribution
    \end{tabbing}
\State Initialize $\vth_0$
\For{$n = 1$ to $N$}
    \State $\vth = \vth_{n-1}$
    \State Sample momentum $\vp \sim \mathcal{N}(\vzero, M)$
    \State \textit{// Generate proposal by simulating Hamiltonian dynamics:}
    \State $(\vth', \vp') = \Psi_{\varepsilon, L}(\vth, \vp)$
    \State $\Delta H = H(\vth', \vp') - H(\vth, \vp)$
    \State \textit{// Metropolis--Hastings test:}
    \State $\alpha = \min\lbrace 1, \exp(-\Delta H)\rbrace$
    \State Sample $u \sim \mathrm{Uniform}(0, 1)$
    \If{$u < \alpha$}
        \State $\vth_n = \vth'$ \textit{// Proposal accepted}
    \Else
        \State $\vth_n = \vth$ \textit{// Proposal rejected}
    \EndIf
    \State Discard momentum $\vp'$
\EndFor
\end{algorithmic}
\end{algorithm}

The sampler has three tuning parameters. The \textit{step size} $\varepsilon$ governs the energy error, with excessively large values producing large $\Delta H$ and low acceptance rates, or, beyond the stability limit, divergent trajectories; while excessively small values spend costly gradient evaluations on negligible movement. The \textit{trajectory length} $L$, meanwhile, determines how far a proposal travels, so that short trajectories leave successive samples strongly correlated whereas overly long trajectories may double back towards their starting neighbourhood. For a near-Gaussian target, the total integration time $\varepsilon L$ is often chosen near half the period of the slowest mode, although certain integration times remain pathological because of the resonances discussed in Section~\ref{sec:integrators}. The \textit{mass matrix} $M$, finally, determines the relative time scales of the coordinates, bringing their characteristic frequencies closer together when it is well chosen and leaving them widely separated when it is not. Sections~\ref{sec:integrators} and~\ref{sec:tuning} are devoted to the selection of these quantities, whose optimal values depend upon the \textit{local} geometry and may therefore vary substantially across the posterior of a hierarchical model, an observation that motivates much of the discussion to follow.

\subsection{Mechanisms and Limitations}

Rather than attempting to locate the typical set by means of a random walk, HMC uses gradients of the log target to construct a trajectory that is likely to follow it \cite{betancourt2017conceptual}, with each trajectory remaining close to a level set of $H=U+K$ that will ordinarily intersect the typical set of $\pi$ as a newly drawn momentum contributes kinetic energy of order $D$. The numerical trajectory explores one such level set, while the subsequent refreshment of the momentum changes the total energy and thereby allows the chain to pass between different level sets. Without that refreshment, $H$ would remain nearly constant and, since $K$ is non-negative, $U(\vth)$ could never rise above the value allowed by the initial energy, leaving parts of the posterior inaccessible \cite{neal2011mcmc}. The energy Bayesian fraction of missing information introduced in Section~\ref{sec:diagnostics} is intended precisely to assess whether this mechanism produces sufficient movement between energy levels \cite{betancourt2016diagnosing}.

Furthermore, dimension enters the acceptance rate through the energy difference $\Delta=E(\vx')-E(\vx)$, for which a target proportional to $\exp(-E(\vx))$ satisfies $\mathbb{E}[\exp(-\Delta)]=1$ and, by Jensen's inequality, $\mathbb{E}[\Delta]\geq0$ \cite{neal2011mcmc}. For a product target, this difference is a sum of contributions from the individual components and the step size must consequently decrease with $D$ if the acceptance rate is to remain non-vanishing, leading to the rates summarized in Table~\ref{tab:scaling} and, for HMC in particular, to $\varepsilon\propto D^{-1/4}$.

The same construction also makes clear where standard HMC remains limited: because the momentum is refreshed completely after every iteration, the direction established by a successful trajectory is immediately discarded, motivating the partial refreshment examined in Section~\ref{sec:ghmc}. Moreover, because an order-$p$ integrator accumulates an error in $\Delta H$ of $\mathcal{O}(\varepsilon^{p})$ over a fixed integration time, reducing $\varepsilon$ improves acceptance only by trading that improvement with additional gradient evaluations, a trade-off reconsidered in Sections~\ref{sec:integrators} and~\ref{sec:mhmc}. Finally, because a single mass matrix and step size are used throughout the parameter space, the method cannot adjust directly to a funnel whose stable step size differs by orders of magnitude between its neck and mouth. The last of these limitations leads in turn to position-dependent metrics in Section~\ref{sec:geometry}, local step size adaptation in Section~\ref{sec:tuning}, and the reparameterization in Section~\ref{sec:hierarchical}.

\section{Generalized Hamiltonian Monte Carlo (GHMC)}\label{sec:ghmc}

\subsection{Partial Momentum Refreshment}

The complete refreshment of the momentum $\vp \sim \mathcal{N}(\vzero, M)$ performed in Algorithm~\ref{alg:hmc} is a simple and effective way to ensure that the chain does not become trapped in a single direction of travel, but it is also wasteful and discards useful trajectory information from one iteration to the next. A trajectory that has just travelled productively in some direction is followed by a momentum drawn without any reference to it, so that the direction is forgotten and the chain retains a residual random-walk. Yet, the momentum cannot simply be left unchanged, since its marginal must remain $\mathcal{N}(\vzero, M)$ for the chain to be reversible and have the correct invariant distribution, and it is the refreshment that renders the chain ergodic across the energy level sets of Section~\ref{sec:hmc}. Nothing, however, requires this refreshment to be complete.

\emph{Generalized Hamiltonian Monte Carlo} (GHMC), introduced by Horowitz \cite{horowitz1991generalized} and developed by Kennedy and Pendleton \cite{kennedy2001cost}, uses a \emph{partial momentum update} (PMU) to retain some of the previous momentum across iterations. A \textit{noise vector} $\vmu \sim \mathcal{N}(\vzero, M)$ is drawn and the pair $(\vp, \vmu)$ is rotated through an angle determined by a parameter $\phi \in (0,1]$,
\begin{equation}\label{eq:pmu}
  \begin{aligned}
    \vp^{*} &= \sqrt{1-\phi}\,\vp + \sqrt{\phi}\,\vmu, \\
    \vmu^{*} &= -\sqrt{\phi}\,\vp + \sqrt{1-\phi}\,\vmu .
  \end{aligned}
\end{equation}
Because the rotation in \eqref{eq:pmu} preserves volume and leaves the product $\mathcal{N}(\vzero, M) \otimes \mathcal{N}(\vzero, M)$ invariant, $\vp^{*}$ has the correct marginal distribution whenever $\vp$ does, while retaining a correlation of $\sqrt{1-\phi}$ with the previous momentum. Small values of $\phi$ therefore preserve the direction of travel over several iterations, with a complete refreshment at $\phi=1$, reducing GHMC to standard HMC, whereas as $\phi\to 0$ the previous momentum is retained almost entirely.

\begin{figure}[t]
\centering
\begin{tikzpicture}[
  >=Latex, line join=round,
  stage/.style = {draw=#1, fill=#1!10, rounded corners=3pt, align=center,
                  font=\scriptsize, inner sep=4pt,
                  minimum width=7.2cm, minimum height=1.00cm},
  half/.style  = {draw=#1, fill=#1!10, rounded corners=3pt, align=center,
                  font=\scriptsize, inner sep=4pt,
                  minimum width=3.40cm, minimum height=1.00cm},
  lbl/.style   = {font=\scriptsize\itshape, text=black!60, align=center,
                  inner sep=1.5pt},
  flow/.style  = {->, line width=0.8pt, draw=black!55},
  back/.style  = {->, line width=0.8pt, draw=cInteg!85,
                  dash pattern=on 2.4pt off 1.8pt}
]

\node[stage=cSampler] (pmu) at (0,0)
  {\textbf{(1)\ Partial Momentum Update (PMU)}\\[1.5pt]
   $\vp^{*}=\sqrt{1-\phi}\,\vp+\sqrt{\phi}\,\vmu$};

\node[stage=cTune, below=0.52cm of pmu] (mdyn)
  {\textbf{(2)\ Dynamics Integration}\\[1.5pt]
   $(\vth',\vp')=\Psi_{\varepsilon,L}(\vth,\vp^{*})$};

\node[stage=cAccent, below=0.52cm of mdyn] (test)
  {\textbf{(3)\ Metropolis--Hastings Test}\\[1.5pt]
   $\alpha=\min\{1,\exp(-\Delta H)\}$};

\node[half=cDiag,  anchor=north] (out)
  at ($(test.south)+(-1.90,-1.00)$)
  {\textbf{(4a)\ Accept}\\[1.5pt] $(\vth_n,\vp_n)=(\vth',\vp')$};
\node[half=cInteg, anchor=north] (flip)
  at ($(test.south)+( 1.90,-1.00)$)
  {\textbf{(4b)\ Reject \& Flip}\\[1.5pt] $(\vth_n,\vp_n)=(\vth,-\vp^{*})$};

\draw[flow] (pmu)  -- (mdyn);
\draw[flow] (mdyn) -- (test);
\draw[flow] (test.south) -- ++(0,-0.50) -| (out.north);
\draw[flow] (test.south) -- ++(0,-0.50) -| (flip.north);
\node[lbl, anchor=south] at ($(test.south)+(-1.00,-0.46)$) {$u<\alpha$};
\node[lbl, anchor=south] at ($(test.south)+( 1.00,-0.46)$) {$u\ge\alpha$};

\draw[back] (out.west)  -- ++(-0.60,0) |- (pmu.west);
\draw[back] (flip.east) -- ++( 0.60,0) |- (pmu.east);

\node[lbl, anchor=north] at ($(out.south)+(1.90,-0.42)$)
  {$\vp_n$ Persists Into the Next Iteration's Partial Update\\[1pt]
   ($\phi=1$ Collapses to a Full Momentum Refresh, Recovering HMC)};

\end{tikzpicture}
\caption{One GHMC iteration. The momentum is partially refreshed, the resulting state is propagated for $L$ integration steps, and the proposal is accepted or rejected by the Metropolis test. On rejection, the position is retained and the momentum is reversed. The dashed paths show that the resulting momentum is used in the next partial update. Setting $\phi=1$ gives the full momentum refresh of standard HMC.}
\label{fig:ghmc-anatomy}
\end{figure}
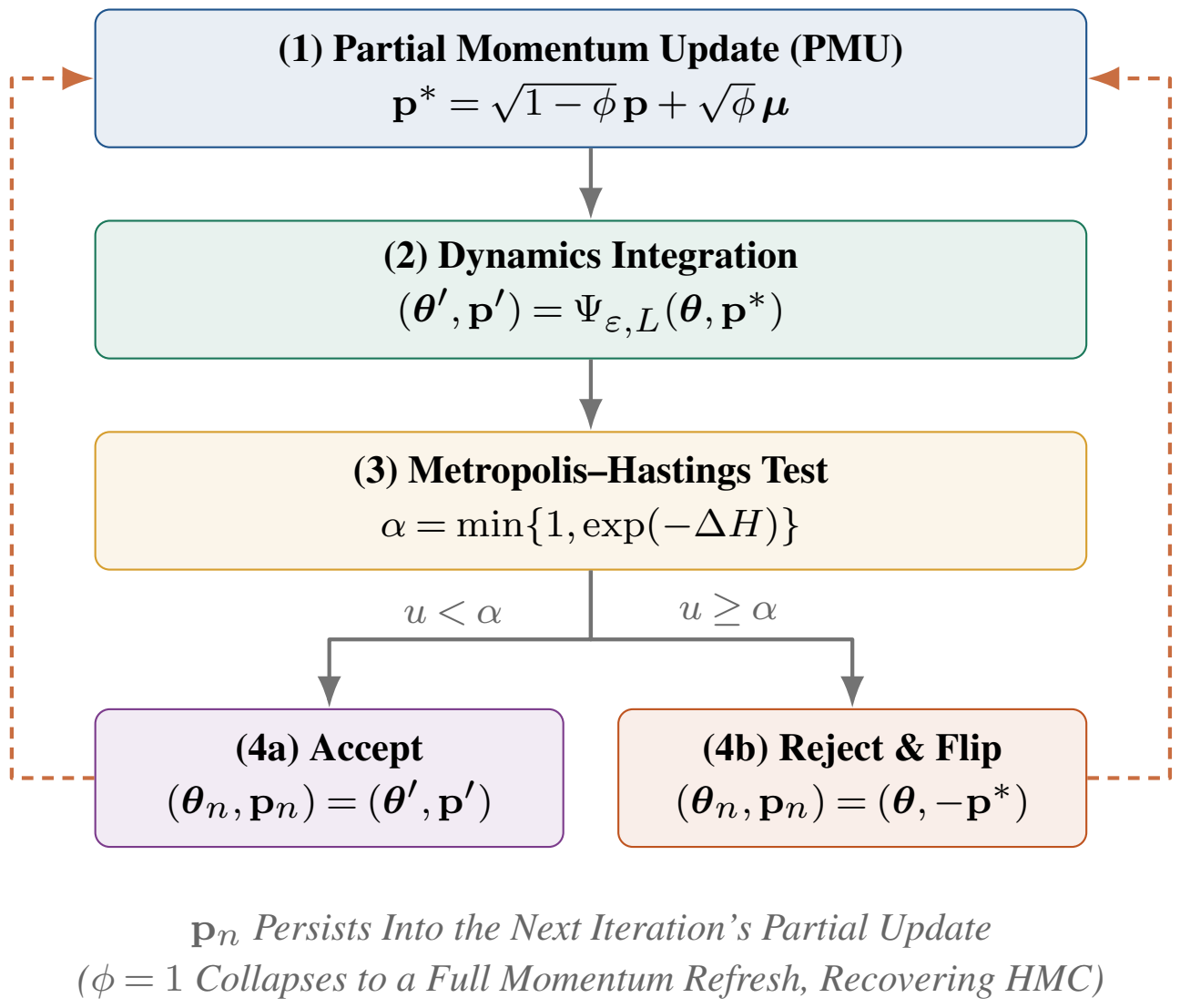

When the momentum persists, a rejected proposal cannot simply leave the joint state unchanged without violating reversibility and the invariance of $\pi$. GHMC therefore applies instead the \emph{momentum flip} $\mathcal{F}(\vth,\vp)=(\vth,-\vp)$ before performing the PMU, reversing the direction in which the chain was travelling. The chain obtained in this way is \textit{irreversible} by construction, a property that can lower the asymptotic variance relative to a reversible chain with the same invariant distribution \cite{ottobre2016markov,duncan2016variance}, but also places it outside the assumptions of standard effective sample size estimators, as discussed in Section~\ref{sec:diagnostics}. The complete procedure is stated in Algorithm~\ref{alg:ghmc} and represented schematically in Figure~\ref{fig:ghmc-anatomy}.

\begin{algorithm}[t]
\caption{Generalized Hamiltonian Monte Carlo (GHMC)}
\label{alg:ghmc}
\begin{algorithmic}[1]
\State \textbf{Input:} \begin{tabbing}
        \hspace{0.4cm} \= $N$: \hspace{0.75cm} \= Number of Monte Carlo samples \\
        \> $\varepsilon$: \> Step size \\
        \> $L$: \> Number of integration steps (trajectory length) \\
        \> $M$: \> Mass matrix \\
        \> $\phi$: \> Momentum noise parameter, $\phi \in (0, 1]$ \\
        \> $\Psi_{\varepsilon, L}$: \> Symplectic, reversible numerical integrator
    \end{tabbing}
\State \textbf{Output:} \begin{tabbing}
        \hspace{0.4cm} \= $\{\vth_n\}_{n=1}^N$: \hspace{0.3cm} \= Samples from the target distribution
    \end{tabbing}
\State Initialize $(\vth_0, \vp_0)$
\For{$n = 1$ to $N$}
    \State $(\vth, \vp) = (\vth_{n-1}, \vp_{n-1})$
    \State \textit{// Partial momentum update:}
    \State Sample $\vmu \sim \mathcal{N}(\vzero, M)$
    \State $\vp^{*} = \sqrt{1-\phi}\,\vp + \sqrt{\phi}\,\vmu$
    \State $\vmu^{*} = -\sqrt{\phi}\,\vp + \sqrt{1-\phi}\,\vmu$
    \State \textit{// Generate proposal by simulating Hamiltonian dynamics:}
    \State $(\vth', \vp') = \Psi_{\varepsilon, L}(\vth, \vp^{*})$
    \State $\Delta H = H(\vth', \vp') - H(\vth, \vp^{*})$
    \State \textit{// Metropolis--Hastings test:}
    \State $\alpha = \min\lbrace 1, \exp(-\Delta H)\rbrace$
    \State Sample $u \sim \mathrm{Uniform}(0, 1)$
    \If{$u < \alpha$}
        \State $(\vth_n, \vp_n) = (\vth', \vp')$ \textit{// Proposal accepted}
    \Else
        \State $(\vth_n, \vp_n) = (\vth, -\vp^{*})$ \textit{// Rejected: momentum flip}
    \EndIf
\EndFor
\end{algorithmic}
\end{algorithm}

\subsection{Special Cases of GHMC}

Several familiar Hamiltonian and Langevin samplers may be recovered simply by fixing the trajectory length $L$, the refreshment parameter $\phi$, and the presence or absence of a Metropolis correction step, with the most relevant choices collected in Table~\ref{tab:ghmc-cases}.

\begin{table}[t]
\centering
\caption{Special cases of Generalized HMC. Fixing the trajectory length $L$, the momentum refreshment parameter $\phi$, and the presence of a Metropolis test recovers each named algorithm. For SLDMC, $\gamma$ denotes the friction coefficient and $h$ the discretization step.}
\label{tab:ghmc-cases}
\renewcommand{\arraystretch}{1.25}
\resizebox{\columnwidth}{!}{%
\begin{tabular}{llccc}
\toprule
\textbf{Method} & \textbf{Reference} & $L$ & $\phi$ & \textbf{Metropolis test} \\
\midrule
GHMC  & \cite{horowitz1991generalized,kennedy2001cost} & arbitrary & arbitrary & yes \\
HMC   & \cite{duane1987hybrid,neal2011mcmc}            & arbitrary & $1$       & yes \\
MALA  & \cite{rossky1978brownian,roberts1998optimal} & $1$     & $1$       & yes \\
L2MC  & \cite{horowitz1991generalized}                 & $1$       & arbitrary & yes \\
MDMC  & --                                            & arbitrary & $0$       & no  \\
SLDMC & \cite{allen2017computer}                       & arbitrary & $\sqrt{2\gamma h} \ll 1$ & no \\
\bottomrule
\end{tabular}}
\end{table}

\begin{figure*}[t]
\centering
\resizebox{\textwidth}{!}{%
\begin{tikzpicture}[
  >=Latex, line join=round,
  trunk/.style   = {draw=cSampler, fill=cSampler!85, text=white, rounded corners=3pt,
                    align=center, font=\bfseries\small, inner sep=5pt, text width=5.0cm},
  leaf/.style    = {draw=cSampler!70, fill=cSampler!12, rounded corners=2pt,
                    align=center, font=\small, inner sep=4pt,
                    minimum width=2.30cm, minimum height=1.15cm},
  axisbox/.style = {draw=#1, fill=#1!12, rounded corners=3pt, align=left,
                    font=\small, inner sep=5pt,
                    text width=3.65cm, minimum width=4.00cm, minimum height=1.60cm},
  outbox/.style  = {draw=#1!60, fill=white, rounded corners=2pt, align=left,
                    font=\small, inner sep=5pt,
                    text width=6.35cm, minimum width=6.70cm, minimum height=1.60cm},
  edge/.style    = {->, draw=cSampler!70, line width=0.7pt},
  modedge/.style = {->, draw=#1, line width=0.8pt},
  bus/.style     = {dashed, draw=black!40, line width=0.7pt},
  hdr/.style     = {font=\small, text=black!75}
]

\node[hdr, anchor=south] at (-2.4, 3.60)
  {\textbf{Specializations:} Fixing $(L,\ \phi,\ \text{MH Test})$};

\node[trunk] (ghmc) at (-2.4, 2.30)
  {Generalized HMC\\[2pt]
   \normalfont\scriptsize PMU $+$ Momentum Flip Upon Rejection};

\node[leaf] (hmc)   at (-7.4,-0.60) {\textbf{HMC}\\[1pt]\scriptsize $\phi=1$};
\node[leaf] (mala)  at (-4.9,-0.60) {\textbf{MALA}\\[1pt]\scriptsize $L=1,\ \phi=1$};
\node[leaf] (l2mc)  at (-2.4,-0.60) {\textbf{L2MC}\\[1pt]\scriptsize $L=1$};
\node[leaf] (mdmc)  at ( 0.1,-0.60) {\textbf{MDMC}\\[1pt]\scriptsize $\phi=0$, No MH};
\node[leaf] (sldmc) at ( 2.6,-0.60) {\textbf{SLDMC}\\[1pt]\scriptsize $\phi\ll1$, No MH};

\foreach \n in {hmc,mala,l2mc,mdmc,sldmc}{
  \draw[edge] (ghmc.south) -- ($(\n.north)+(0,0.02)$);
}

\node[hdr, anchor=south] at (10.70, 3.60)
  {\textbf{Extensions:} Additional Modifications to GHMC};

\node[axisbox=cDiag] (ax1) at (6.70, 2.40)
  {\textbf{The Metric}\\[1pt] $M \;\rightarrow\; G(\vth)$};
\node[outbox=cDiag] (out1) at (13.35, 2.40)
  {RMHMC; SoftAbs; Lagrangian MC;\\ Magnetic and Relativistic HMC\\[1pt]
   \textit{$\Rightarrow$ Non-Separable $H$; Implicit Integrators}};

\node[axisbox=cInteg] (ax2) at (6.70, 0.60)
  {\textbf{The Conserved Density}\\[1pt] $H \;\rightarrow\; \Hmod^{[k]}$};
\node[outbox=cInteg] (out2) at (13.35, 0.60)
  {SHMC; GSHMC; MMHMC\\[1pt]
   \textit{$\Rightarrow$ Near-Unit AR; Importance Reweighting}};

\node[axisbox=cTune] (ax3) at (6.70,-1.20)
  {\textbf{The Integrator}\\[1pt] Selection of $\Psi_{\varepsilon,L}$};
\node[outbox=cTune] (out3) at (13.35,-1.20)
  {Verlet; $k$-Stage MSSIs;\\ BCSS, Minimum-Error; s-AIA$_k$\\[1pt]
   \textit{$\Rightarrow$ Energy Error vs.\ Gradients per Step}};

\draw[modedge=cDiag]  (ax1) -- (out1);
\draw[modedge=cInteg] (ax2) -- (out2);
\draw[modedge=cTune]  (ax3) -- (out3);

\draw[dashed, black!30, line width=0.8pt] (4.15,-2.15) -- (4.15, 3.95);
\draw[bus] (4.15, 2.40) -- (ax1.west);
\draw[bus] (4.15, 0.60) -- (ax2.west);
\draw[bus] (4.15,-1.20) -- (ax3.west);
\draw[bus, ->, draw=black!45, line width=0.8pt]
  (4.15,1.50) .. controls (2.8,2.10) and (1.6,2.60) .. (ghmc.east);

\end{tikzpicture}}
\caption{A taxonomy of Hamiltonian samplers organized around Generalized HMC. \textbf{Left:} fixing the trajectory length $L$, the momentum refreshment parameter $\phi$, and whether a Metropolis test is applied recovers standard HMC, MALA, L2MC, molecular dynamics and stochastic Langevin dynamics as special cases of a single algorithm (Table~\ref{tab:ghmc-cases}). Methods toward the right of the row drop the Metropolis correction and generally have a stationary bias of order $\mathcal{O}(\varepsilon^{p})$ under suitable regularity conditions for an order-$p$ integrator, whereas those to the left remain exactly asymptotically correct. \textbf{Right:} the remaining literature is obtained by replacing one of three ingredients, namely the metric, the density that the integrator conserves, or the integrator itself: RMHMC \cite{girolami2011riemann}, SoftAbs \cite{betancourt2013general}, Lagrangian Monte Carlo \cite{lan2015markov}, magnetic \cite{tripuraneni2017magnetic} and relativistic \cite{lu2017relativistic} HMC for the metric; SHMC \cite{izaguirre2004shadow}, GSHMC \cite{akhmatskaya2008gshmc} and MMHMC \cite{radivojevic2020modified} for the conserved density; multi-stage splitting integrators \cite{blanes2014numerical} and s-AIA \cite{nagar2024adaptive} for the integrator. Each axis composes freely with any sampler on the left.}
\label{fig:taxonomy}
\end{figure*}
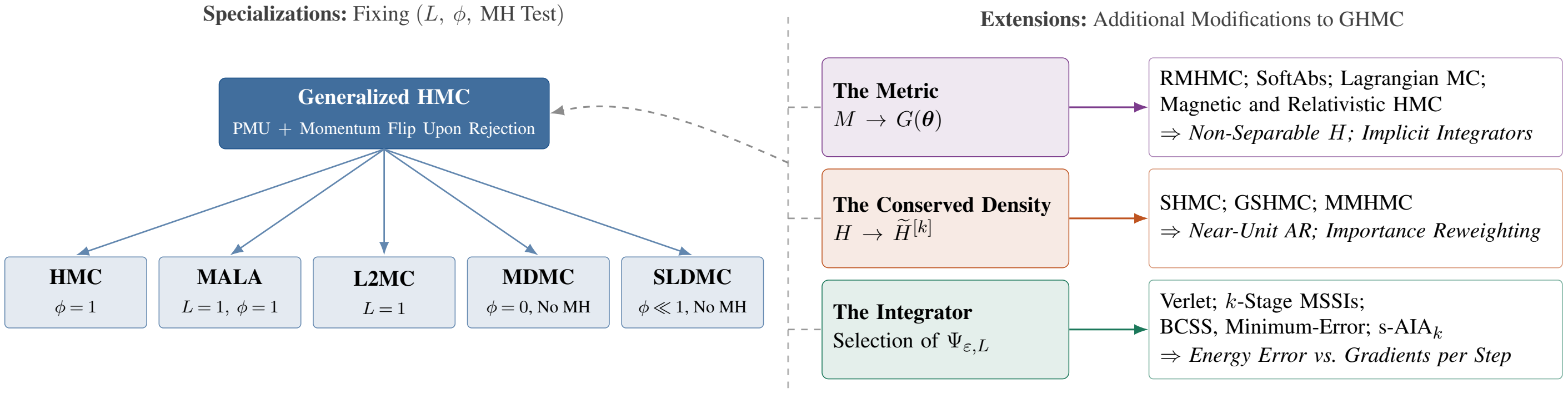

Setting $\phi=1$ makes $\vp^{*}=\vmu$ and hence recovers standard HMC through complete momentum refreshment. If one additionally takes $L=1$, the resulting method is MALA, whose $D^{-1/3}$ Euler-step scaling was reported in Table~\ref{tab:scaling}, whereas retaining $L=1$ but allowing $\phi<1$ produces second-order Langevin Monte Carlo (L2MC), in which momentum persists across iterations. If the Metropolis correction is instead removed, these choices lead to unadjusted dynamics such as deterministic molecular dynamics when $\phi=0$ and stochastic Langevin dynamics for small $\phi$, whose stationary distributions are generally biased relative to $\pi$, often by order $\varepsilon^{p}$ for an order-$p$ integrator under suitable regularity conditions. The microcanonical samplers discussed in Section~\ref{sec:geometry} belong to this latter group. Figure~\ref{fig:taxonomy} places these special cases on a single \textit{taxonomy} together with the three orthogonal directions of extension introduced in Section~\ref{sec:introduction}.

\subsection{Choosing \texorpdfstring{$\phi$}{ϕ} \& \texorpdfstring{$L$}{L}}

The choices of $\phi$ and $L$ are \textit{coupled}, and the optimal values depend upon the target distribution and the integrator. A small $\phi$ preserves the direction of travel across iterations, which is beneficial \emph{only} if the trajectories being chained together are in fact accepted, since a rejection flips the momentum and undoes the progress accumulated so far. The expected number of consecutive accepted steps is approximately $1/(1-\AR)$, and the sampler is correspondingly sensitive to the acceptance rate. GHMC therefore demands higher acceptance rates than HMC, typically $\AR \gtrsim 0.9$ rather than the $0.651$ that is optimal for the standard method, and attaining them demands either a smaller step size or a higher order integrator. This coupling is what motivates the modified Hamiltonians of Section~\ref{sec:mhmc} and the adaptive integrators of Section~\ref{sec:integrators}, since both raise the acceptance rate at fixed gradient cost, which in turn allows for a smaller $\phi$ and hence a longer effective trajectory. The optimal trajectory length $L$ is therefore a function of $\phi$, and vice versa, and the two must be tuned together.

Conversely, a small $\phi$ lets the sampler accumulate motion across iterations and thus use shorter individual trajectories. In the adaptive integration family of Section~\ref{sec:tuning} the recommended refreshment parameter scales as $1/D$ \cite{nagar2025optimising,akhmatskaya2026adaptive},
\begin{equation}\label{eq:phi-opt}
  \phi_{\mathrm{opt}}(\bar{\varepsilon}) \;=\; \min\left\{1,\;
  -\frac{\ln 0.999}{D}\cdot
  \frac{1 + 2\bar{\varepsilon}^{2}\lambda_k(\bar{\varepsilon})}
       {2\bar{\varepsilon}^{4}\lambda_k(\bar{\varepsilon})^{2}}\right\},
\end{equation}
where $\bar{\varepsilon}$ is the \textit{dimensionless} step size and $\lambda_k$ depends upon the integrator coefficients introduced in Section~\ref{sec:integrators}. Within that tuning scheme, very short trajectories are recommended, frequently $L = 1$ and otherwise $L$ drawn uniformly from $\{2, \dots, 6\}$. This contrasts with the $L = \mathcal{O}(D^{1/4})$ steps needed to maintain a fixed integration time for HMC under the product-target scaling of Table~\ref{tab:scaling}.

\begin{remark}
The flip \textit{upon rejection} is the principal structural weakness of GHMC, since a single rejection reverses the chain and, unless the acceptance rate remains close to unity, may cause the sampler to oscillate back and forth instead of continuing to explore the state space. Proposed remedies include delayed rejection schemes \cite{modi2024delayed}, which attempt a second and more conservative proposal before resorting to the flip, and non-reversible updates of the acceptance variable \cite{neal2020nonreversible}, which correlate successive acceptance decisions in order to reduce its frequency. The tuning-free scheme of Hoffman and Sountsov \cite{sountsov2022meads}, known as MEADS, instead adapts $\phi$ and $\varepsilon$ from an ensemble of chains so that the acceptance rate remains high enough for flips to be rare, and is designed for the multi-chain setting that arises naturally in modern parallel computing. Metropolis-adjusted Langevin trajectories \cite{rioudurand2022metropolis} provide a rigorous contemporary treatment of essentially the same underlying construction.
\end{remark}

\section{Modified Hamiltonian Monte Carlo (MHMC)}\label{sec:mhmc}

\subsection{Backward Error Analysis \& Shadow Hamiltonians}

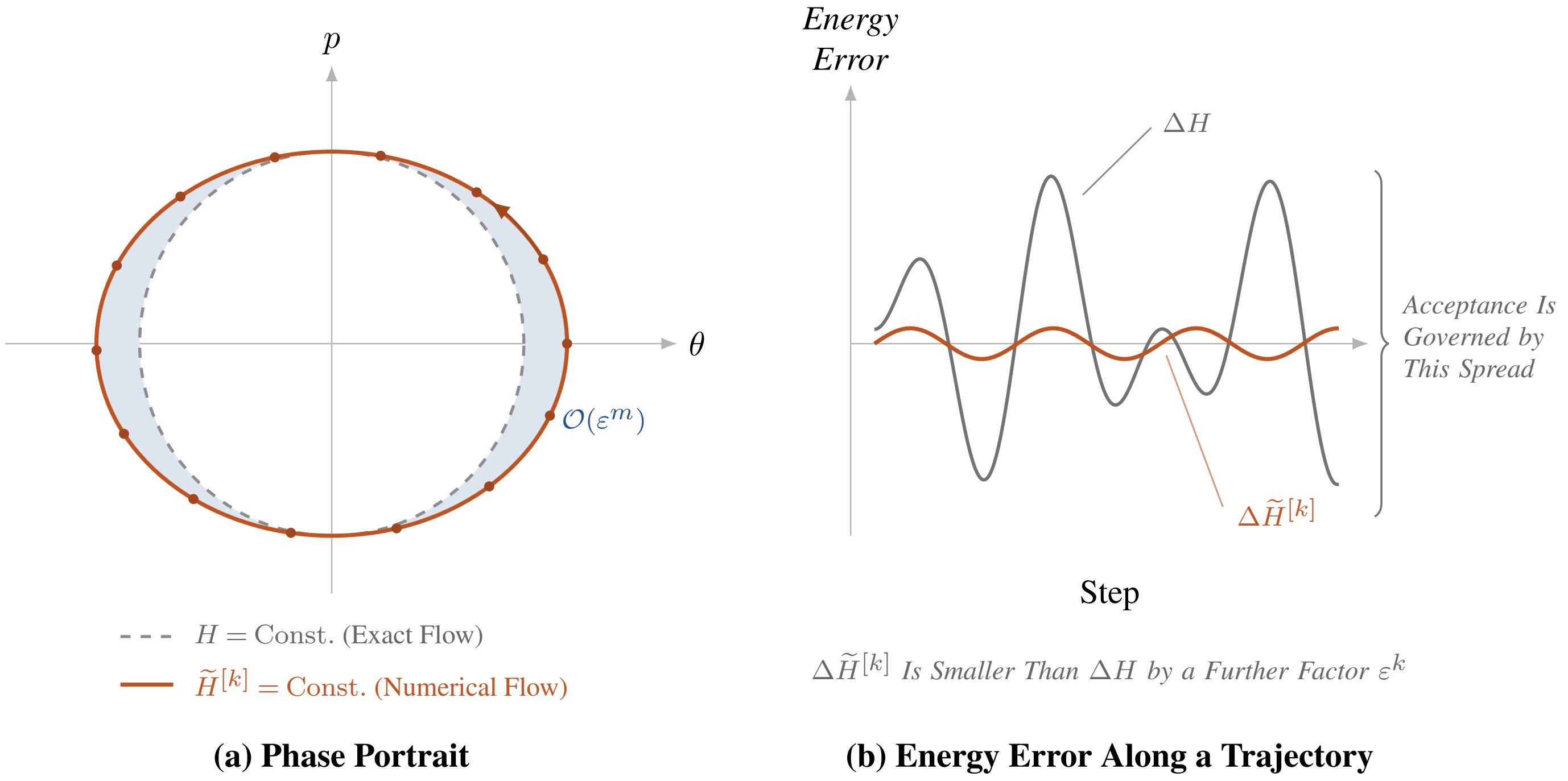
\begin{figure*}[t]
\centering
\resizebox{\textwidth}{!}{%
\begin{tikzpicture}[>=Latex, line join=round,
  sublbl/.style = {font=\small\bfseries, anchor=north},
  note/.style   = {font=\scriptsize\itshape, text=black!60, align=center,
                   anchor=north}]

\begin{scope}[x=1.0cm, y=1.0cm]
  \draw[->, gray!60] (-3.40,0) -- (3.60,0) node[right, black, font=\small] {$\theta$};
  \draw[->, gray!60] (0,-2.60) -- (0,2.90) node[above, black, font=\small] {$p$};

  \fill[cSampler!15, even odd rule]
    (0,0) ellipse [x radius=2.45, y radius=2.0]
    (0,0) circle [radius=2.0];
  \draw[black!45, dashed, line width=0.9pt] (0,0) circle [radius=2.0];
  \draw[cInteg, line width=1.2pt] (0,0) ellipse [x radius=2.45, y radius=2.0];

  \foreach \a in {0,26,...,346}{
    \fill[cInteg!85!black] ({2.45*cos(\a)},{2*sin(\a)}) circle [radius=1.5pt];
  }
  \draw[->, cInteg!85!black, line width=0.7pt]
    ({2.45*cos(26)},{2*sin(26)}) arc [start angle=26, end angle=48,
       x radius=2.45, y radius=2.0];

  \node[cSampler, font=\scriptsize, anchor=north west] at (2.28,-0.56)
    {$\mathcal{O}(\varepsilon^{m})$};

  \draw[black!45, dashed, line width=0.9pt] (-2.20,-3.05) -- (-1.65,-3.05);
  \node[black!60, font=\scriptsize, anchor=west] at (-1.55,-3.05)
    {$H = \mathrm{Const.}$ (Exact Flow)};
  \draw[cInteg, line width=1.2pt] (-2.20,-3.55) -- (-1.65,-3.55);
  \node[cInteg, font=\scriptsize, anchor=west] at (-1.55,-3.55)
    {$\Hmod^{[k]} = \mathrm{Const.}$ (Numerical Flow)};

  \node[sublbl] at (0.10,-4.05) {(a) Phase Portrait};
\end{scope}

\begin{scope}[xshift=5.65cm, x=0.62cm, y=1.0cm]
  \draw[->, gray!60] (-0.40,0) -- (8.30,0);
  \draw[->, gray!60] (-0.40,-2.00) -- (-0.40,2.70)
    node[above=1pt, black, font=\small, align=center] {\textit{Energy}\\\textit{Error}};

  \draw[black!55, line width=1.0pt, domain=0:7.8, samples=240, smooth]
    plot ({\x},{1.70*sin(\x*150)*sin(\x*37)+0.15});
  \draw[cInteg, line width=1.2pt, domain=0:7.8, samples=240, smooth]
    plot ({\x},{0.16*sin(\x*150)});

  \node[black!60, font=\scriptsize, anchor=west] at (4.65,2.30) {$\Delta H$};
  \draw[black!45, line width=0.5pt] (4.65,2.25) -- (3.50,1.55);
  \node[cInteg, font=\scriptsize, anchor=west] at (5.90,-1.75) {$\Delta \Hmod^{[k]}$};
  \draw[cInteg!60, line width=0.5pt] (5.85,-1.70) -- (4.90,-0.12);

  \draw[decorate, decoration={brace, amplitude=4pt}, black!55, line width=0.6pt]
    (8.40,1.80) -- (8.40,-1.80);
  \node[black!60, font=\scriptsize, anchor=west, align=left, text width=1.95cm]
    at (8.7,0.05) {\textit{Acceptance Is Governed by This Spread}};

  \node[black, font=\small, anchor=north] at (3.95,-2.35) {Step};
  \node[note, text width=6.4cm] at (3.95,-3.10)
    {$\Delta\Hmod^{[k]}$ Is Smaller Than $\Delta H$ by a Further Factor
     $\varepsilon^{k}$};

  \node[sublbl] at (3.95,-4.05) {(b) Energy Error Along a Trajectory};
\end{scope}

\end{tikzpicture}}
\caption{Modified Hamiltonian interpretation of the Verlet integrator for the harmonic oscillator $H = \tfrac12(p^2+\theta^2)$. \textbf{(a)} The dashed circle is a level set of the true Hamiltonian. The numerical trajectories lie on a nearby level set of the modified Hamiltonian $\Hmod^{[k]}$ (solid). For Verlet, this level set is an ellipse with eccentricity $\mathcal{O}(\varepsilon^2)$. \textbf{(b)} The error in the true Hamiltonian, $\Delta H$, oscillates along the numerical trajectory. The truncated modified Hamiltonian has a smaller error $\Delta\Hmod^{[k]}$. MHMC uses this quantity in the Metropolis test and reweights the resulting draws with \eqref{eq:mhmc-weights}.}
\label{fig:shadow}
\end{figure*}

Although Section~\ref{sec:hmc} described numerical integration through the discretization error $\Delta H\neq0$, \textit{backward} error analysis offers a more informative interpretation, according to which a symplectic integrator follows the flow of a nearby \emph{modified}, or \emph{shadow}, Hamiltonian represented by an asymptotic expansion in the step size $\varepsilon$ \cite{hairer2006geometric,sanzserna1994numerical}. Skeel and Hardy \cite{skeel2001practical} give a practical construction of these expansions for simulation use. Truncating this series gives the \emph{$k$-th order truncated modified Hamiltonian}
\begin{equation}\label{eq:modified-hamiltonian}
\begin{aligned}
  \Hmod^{[k]}(\vth, \vp) = H(\vth,\vp)
  &+ \varepsilon^{m} H_{m+1}(\vth,\vp) \\
  &+ \cdots + \varepsilon^{k-1} H_{k}(\vth,\vp),
\end{aligned}
\end{equation}
in which $m$ is the order of the integrator and the \textit{correction} terms $H_j$ are determined by the integrator coefficients, involving successively higher derivatives of the potential. For the palindromic schemes of Section~\ref{sec:integrators}, the odd-order corrections vanish, so only even powers of $\varepsilon$ survive: $\Hmod^{[4]}$ carries a single $\varepsilon^{2}$ correction and $\Hmod^{[6]}$ adds an $\varepsilon^{4}$ one. Both are written out in Appendix~\ref{app:modified}. For a suitably truncated expansion, the modified Hamiltonian is typically conserved to higher order than $H$ over an integration step, as illustrated in Figure~\ref{fig:shadow} for the harmonic oscillator, whose trajectory follows a nearby closed energy curve associated with the modified Hamiltonian.

\textit{Modified Hamiltonian Monte Carlo} (MHMC) methods exploit this observation by sampling not from $\pi$, but from the \emph{modified target}
\begin{equation}\label{eq:modified-target}
  \pimod(\vth, \vp) \;\propto\; \exp\!\left(-\Hmod^{[k]}(\vth, \vp)\right),
\end{equation}
employing $\Delta\Hmod^{[k]}$ rather than $\Delta H$ in the Metropolis test. Because the integrator conserves $\Hmod^{[k]}$ far more closely than $H$, acceptance rates can approach unity at step sizes that would be rejected outright under the true Hamiltonian. The original target must then be recovered by \emph{importance reweighting}. Because $\pi(\vth,\vp) \propto \pimod(\vth,\vp)\exp(-[H - \Hmod^{[k]}])$, each sample carries a weight
\begin{equation}\label{eq:mhmc-weights}
  \omega_n \;=\; \exp\!\left(-\left[H(\vth_n, \vp_n) - \Hmod^{[k]}(\vth_n, \vp_n)\right]\right),
\end{equation}
and expectations under $\pi$ are accordingly estimated by the \textit{self-normalized} estimator
\begin{equation}\label{eq:mhmc-estimator}
  \widehat{I}_N \;=\; \frac{\sum_{n=1}^{N}\omega_n f(\vth_n)}{\sum_{n=1}^{N}\omega_n}.
\end{equation}
MHMC thus combines Markov Chain sampling with \textit{importance sampling} through the modified Hamiltonian, which also affects its diagnostics (Section~\ref{sec:diagnostics}).

\begin{remark}
A modified Hamiltonian can improve acceptance and thereby allow larger step sizes or a smaller $\phi$ in a GHMC setting, but these gains come with the additional requirements of evaluating $\Hmod^{[k]}$, either through higher derivatives of $U$ or suitable numerical approximations, and of controlling the variance of the resulting importance weights. The method is thus beneficial only when its improvement in mixing outweighs the loss of precision caused by variable weights. The two are not automatically aligned: a step size can produce near-perfect acceptance under $\pimod$ and still yield weights dispersed enough that the self-normalized estimator in \eqref{eq:mhmc-estimator} is effectively driven by a handful of draws. Section~\ref{sec:tuning} considers this trade-off, together with the joint tuning of the integrator and modified Hamiltonian needed to maximize overall sampling efficiency.
\end{remark}

\subsection{Shadow Hamiltonian Methods}

Originating in molecular simulation and only later adapted to computational statistics, MHMC now comprises several related methods, of which the three described below differ principally in the manner in which shadow Hamiltonian corrections are combined with momentum updates and importance reweighting.

\medskip\noindent\textbf{Shadow Hybrid Monte Carlo.} SHMC \cite{izaguirre2004shadow} introduced the central idea of sampling from a shadow density and correcting the estimator by reweighting. It employed a fourth-order shadow Hamiltonian within an otherwise standard HMC setting and demonstrated substantial gains in acceptance for molecular systems, at the cost of a weight distribution that had to be controlled by a bounding parameter.

\medskip\noindent\textbf{Generalized Shadow HMC.} GSHMC \cite{akhmatskaya2008gshmc} combines the shadow Hamiltonian construction with the partial momentum update of Section~\ref{sec:ghmc}. Because $\Hmod^{[k]}$ is generally non-quadratic in $\vp$, the rotation \eqref{eq:pmu} no longer leaves the modified density invariant, and the momentum update must consequently be subjected to a second Metropolis test defined with respect to the \emph{extended Hamiltonian}
\begin{equation}\label{eq:extended-hamiltonian}
  \Hext(\vth, \vp, \vmu)
  \;=\; \Hmod^{[k]}(\vth, \vp) + \tfrac12\,\vmu^\top M^{-1}\vmu ,
\end{equation}
which restores invariance of the joint density in $(\vth, \vp, \vmu)$ under the rotation. For the standard Hamiltonian, $\Hext$ is invariant by construction, so GHMC needs no analogous test.

\medskip\noindent\textbf{Mix \& Match HMC.} MMHMC \cite{radivojevic2020modified,thesis_tijana} further specialized this framework to the field of computational statistics, where the potentials are Bayesian log-posteriors rather than molecular force fields and the derivative structure differs accordingly. It introduced practical fourth- and sixth-order modified Hamiltonians for multi-stage splitting integrators (Appendix~\ref{app:modified}), a numerical derivative formulation that avoids explicit Hessians for dense problems, and the first systematic evaluation on statistical targets. Algorithm~\ref{alg:mmhmc} differs from Algorithm~\ref{alg:ghmc} in three places: the momentum update is subjected to its own Metropolis test under $\Hext$, the trajectory test uses $\Hmod^{[k]}$, and an importance weight is recorded after each iteration.

\begin{algorithm}[t]
\caption{Mix \& Match Hamiltonian Monte Carlo (MMHMC). Three things distinguish it from Algorithm~\ref{alg:ghmc}: the partial momentum update carries its own Metropolis test under the extended Hamiltonian $\Hext$, the trajectory test uses the truncated modified Hamiltonian $\Hmod^{[k]}$, and each iteration records an importance weight that returns the estimator to the original target.}
\label{alg:mmhmc}
\begin{algorithmic}[1]
\State \textbf{Input:} \begin{tabbing}
        \hspace{0.4cm} \= $N$: \hspace{0.75cm} \= Number of Monte Carlo samples \\
        \> $\varepsilon$: \> Step size \\
        \> $L$: \> Number of integration steps (trajectory length) \\
        \> $M$: \> Mass matrix \\
        \> $\phi$: \> Momentum noise parameter, $\phi \in (0, 1]$ \\
        \> $k$: \> Order of the truncated modified Hamiltonian \\
        \> $\Psi_{\varepsilon, L}$: \> Symplectic, reversible numerical integrator
    \end{tabbing}
\State \textbf{Output:} \begin{tabbing}
        \hspace{0.4cm} \= $\{\vth_n, \omega_n\}_{n=1}^N$: \hspace{0.1cm}
        \= Weighted samples for estimating $\mathbb{E}_\pi[f(\vth)]$
    \end{tabbing}
\State Initialize $(\vth_0, \vp_0)$
\For{$n = 1$ to $N$}
    \State $(\vth, \vp) = (\vth_{n-1}, \vp_{n-1})$
    \State \textit{// Partial momentum update:}
    \State Sample $\vmu \sim \mathcal{N}(\vzero, M)$
    \State $\vp^{*} = \sqrt{1-\phi}\,\vp + \sqrt{\phi}\,\vmu$
    \State $\vmu^{*} = -\sqrt{\phi}\,\vp + \sqrt{1-\phi}\,\vmu$
    \State $\Delta \Hext = \Hext(\vth, \vp^{*}, \vmu^{*}) - \Hext(\vth, \vp, \vmu)$
    \State \textit{// Metropolis test on the momentum update:}
    \State $P = \min\lbrace 1, \exp(-\Delta \Hext)\rbrace$
    \State Sample $v \sim \mathrm{Uniform}(0, 1)$
    \If{$v < P$}
        \State $\vp^{\dagger} = \vp^{*}$ \textit{// Momentum update accepted}
    \Else
        \State $\vp^{\dagger} = \vp$ \textit{// Momentum update rejected}
    \EndIf
    \State \textit{// Generate proposal by simulating Hamiltonian dynamics:}
    \State $(\vth', \vp') = \Psi_{\varepsilon, L}(\vth, \vp^{\dagger})$
    \State $\Delta \Hmod^{[k]} = \Hmod^{[k]}(\vth', \vp') - \Hmod^{[k]}(\vth, \vp^{\dagger})$
    \State \textit{// Metropolis test on the trajectory:}
    \State $\alpha = \min\lbrace 1, \exp(-\Delta \Hmod^{[k]})\rbrace$
    \State Sample $u \sim \mathrm{Uniform}(0, 1)$
    \If{$u < \alpha$}
        \State $(\vth_n, \vp_n) = (\vth', \vp')$ \textit{// Proposal accepted}
    \Else
        \State $(\vth_n, \vp_n) = (\vth, -\vp^{\dagger})$ \textit{// Rejected: momentum flip}
    \EndIf
    \State \textit{// Importance weight recovering the original target:}
    \State $\omega_n = \exp\!\left(-\left[H(\vth_n, \vp_n) - \Hmod^{[k]}(\vth_n, \vp_n)\right]\right)$
\EndFor
\State \textbf{Estimator:} $\displaystyle \widehat{I}_N =
  \Big(\sum\nolimits_{n=1}^{N} \omega_n f(\vth_n)\Big) \Big/ \Big(\sum\nolimits_{n=1}^{N} \omega_n\Big)$
\end{algorithmic}
\end{algorithm}

\subsection{Which Modified Hamiltonian \& at What Cost}

The order $k$ at which the expansion is truncated is itself a parameter to be adapted to the problem at hand: increasing it allows the integrator to conserve $\Hmod^{[k]}$ more accurately, and may therefore raise the acceptance rate while concentrating the weights, but each additional order also introduces higher derivatives of the potential. Thus, whereas the fourth-order modified Hamiltonian already requires the Hessian $U_{\vth\vth}$ contracted against $M^{-1}\vp$, the sixth-order expression additionally involves third derivatives, a requirement that is prohibitive for most statistical models of realistic size \cite{radivojevic2018multi}.

Two practical formulations are used to circumvent this problematic obstacle. When the required derivatives are inexpensive or sparse, as happens for many molecular potentials and for models with structured likelihoods, they may be evaluated directly through the \emph{analytic} formulation. For statistical problems with dense Hessians, by contrast, the more common \emph{numerical time-derivative} formulation replaces the higher derivatives with finite differences of gradients already computed along the trajectory, at the cost of introducing a second discretization error. The special case of a quadratic potential, which describes a Gaussian target and a sufficiently regular target locally near its mode, is simpler still, because all third and higher derivatives vanish, and it is this simplification that underlies the Gaussian analysis employed by the s-AIA and s-MAIA schemes of Section~\ref{sec:tuning}.

Because the coefficients $c_{ij}$ entering these expressions depend upon the splitting parameters $a$ and $b$, the choice of modified Hamiltonian is inseparable from that of the integrator, and the MAIA, e-MAIA, and s-MAIA schemes in Section~\ref{sec:tuning} were therefore developed specifically for modified Hamiltonians instead of transferred directly from the standard HMC setting \cite{akhmatskaya2017adaptive,parga2024tackling}.

\begin{remark}
MHMC is not the only way of using integration accuracy more efficiently: extra-chance GHMC \cite{campos2015extra} continues a rejected trajectory rather than flipping the momentum at once, delayed rejection \cite{modi2024delayed} makes a second attempt at a smaller step size, and Calvo et al. \cite{calvo2021hmc} obtain higher acceptance by replacing leapfrog with schemes designed around the acceptance rate directly. All three leave the invariant distribution of the chain untouched. However, MHMC does not: it changes the \emph{invariant distribution of the chain} and repairs the change afterwards through reweighting, so its assessment requires the combined effective sample size of Section~\ref{sec:ess-mcis}.
\end{remark}

\section{Riemannian, Non-Separable \& Alternative Dynamics}\label{sec:geometry}

The methods considered in Sections~\ref{sec:ghmc} and~\ref{sec:mhmc} retain three features of standard HMC: a \textit{constant} mass matrix $M$, a separable Hamiltonian, and an explicit integrator. The methods collected in this section relax one or more of these conditions in order to adapt more directly to difficult posterior geometries such as the funnel in Figure~\ref{fig:funnel}, albeit often at a substantial increased computational cost.

\subsection{Riemannian Manifold HMC (RMHMC)}

When posterior curvature varies considerably across the parameter space, constant preconditioning is insufficient to resolve the geometry equally well throughout the domain. To address this, \textit{Riemannian manifold HMC} \cite{girolami2011riemann} replaces the constant mass matrix with a position-dependent \emph{metric tensor} $G(\vth)$ and draws momenta as $\vp \sim \mathcal{N}(\vzero, G(\vth))$. The resulting Hamiltonian includes a log-determinant term,
\begin{equation}\label{eq:rmhmc-hamiltonian}
\begin{aligned}
  H(\vth, \vp) \;=\; U(\vth)
  &\;+\; \tfrac12 \log\lvert G(\vth)\rvert \\
  &\;+\; \tfrac12\, \vp^\top G(\vth)^{-1}\vp .
\end{aligned}
\end{equation}
The canonical choice for the metric is the expected Fisher information of the likelihood, augmented by the negative Hessian of the log-prior,
\begin{equation}\label{eq:fisher}
\resizebox{0.87\columnwidth}{!}{$\displaystyle
\begin{aligned}
  G(\vth) \;=\; &\mathbb{E}_{\vx \sim p(\vx \mid \vth)}\!
  \left[\nabla_{\vth} \log p(\vx \mid \vth)\,
        \nabla_{\vth} \log p(\vx \mid \vth)^\top\right] \\
  &-\; \nabla^2_{\vth} \log p(\vth),
\end{aligned}$}
\end{equation}
which is positive definite when the prior contribution makes the sum positive definite and transforms appropriately under smooth reparameterization. This geometric invariance is an advantage of RMHMC over Euclidean HMC, particularly when parameterization affects conditioning. Moreover, a key difference from the methods of Sections~\ref{sec:ghmc} and~\ref{sec:mhmc}, however, is that the Hamiltonian in \eqref{eq:rmhmc-hamiltonian} is \textbf{non-separable}, since $\vth$ and $\vp$ now appear together in the kinetic term. Hamilton's equations thus become
\begin{align}
  \dot{\vth} &= G(\vth)^{-1}\vp, \label{eq:rmhmc-theta}\\
  \begin{split}
  \dot{\vp}  &= -\nabla_{\vth} U(\vth)
     - \tfrac12 \nabla_{\vth}\log\lvert G(\vth)\rvert \\
     &\quad\; - \tfrac12 \vp^\top \left(\nabla_{\vth} G(\vth)^{-1}\right)\vp,
  \end{split} \label{eq:rmhmc-p}
\end{align}
and the usual leapfrog scheme no longer applies as neither equation can generally be solved exactly for one variable while the other is held fixed. RMHMC instead commonly employs the \emph{generalized leapfrog} integrator \cite{leimkuhler2004simulating,hairer2006geometric}, an implicit symplectic integrator whose two half-steps require the solution of nonlinear fixed-point equations,
\begin{equation}\label{eq:glf}
\begin{aligned}
  \vp^{(l+\frac12)} &= \vp^{(l)} - \tfrac{\varepsilon}{2}\nabla_{\vth}H\!\left(\vth^{(l)}, \vp^{(l+\frac12)}\right), \\
  \vth^{(l+1)} &= \vth^{(l)} + \tfrac{\varepsilon}{2}\Big[\nabla_{\vp}H\!\left(\vth^{(l)},\vp^{(l+\frac12)}\right) \\
   &\qquad\quad + \nabla_{\vp}H\!\left(\vth^{(l+1)},\vp^{(l+\frac12)}\right)\Big],
\end{aligned}
\end{equation}
followed by an explicit half-kick. Implementing these updates requires repeated metric evaluations, linear solves, and metric derivatives, with a dense implementation incurring $\mathcal{O}(d^3)$ cost per evaluation of the Hamiltonian and its derivatives unless exploitable sparsity or other structure is present. RMHMC can therefore be substantially more expensive than \textit{euclidean} HMC, while an insufficiently converged fixed-point solve may additionally compromise reversibility and bias the resulting transition.

Several subsequent developments have nevertheless made RMHMC considerably more practical. The \emph{SoftAbs} metric \cite{betancourt2013general} smoothly maps the eigenvalues of the Hessian $\nabla^2 U$ to positive values and thereby supplies a metric when the Fisher information is unavailable or indefinite, as frequently occurs in hierarchical models with non-conjugate likelihoods. \emph{Lagrangian dynamical Monte Carlo} \cite{lan2015markov} rewrites the dynamics in velocity rather than momentum coordinates, allowing an explicit, non-symplectic integrator that can reduce the computational burden. Lastly, \emph{explicit symplectic integration by augmentation} \cite{tao2016explicit,cobb2019introducing} duplicates the phase space into two copies $(\vth_1, \vp_1)$ and $(\vth_2, \vp_2)$ and couples them. The augmented Hamiltonian is separable across the copies and so admits an explicit splitting; the strength of the binding controls how long the two copies remain close, and hence how faithfully the scheme reproduces the original non-separable flow.

\begin{remark}
On Neal's funnel, a position-dependent metric rescales the local geometry toward isotropy and thus allows the neck and the mouth to be explored with comparable efficiency. Betancourt and Girolami \cite{betancourt2015hamiltonian} demonstrate this for the SoftAbs metric where a Euclidean metric fails. Non-centering and RMHMC thus attack the same obstacle from opposite sides, one by choosing coordinates in which the geometry is benign and the other by adapting the geometry to the coordinates given. The two are not equivalent, however: non-centering is a fixed \textit{global} change of variables whose suitability depends on the prior--likelihood balance of each group, whereas the metric is re-evaluated \textit{locally} at every point of the trajectory and pays for that flexibility at every step.
\end{remark}

\subsection{Non-Canonical \& Alternative Dynamics}

Rather than modifying the metric, a second family retains the \textit{Euclidean} geometry and alters the \textit{dynamics} themselves, leading to methods that address a narrower range of difficulties but are generally much less computationally demanding than RMHMC.

\medskip\noindent\textbf{Magnetic HMC.} Adding an antisymmetric \textit{magnetic field} term to the equations of motion, so that $\dot{\vz} = (J + \mathcal{B})\nabla H$, produces a flow that remains volume preserving and still leaves $\pi$ invariant, but is no longer time reversible in the usual sense \cite{tripuraneni2017magnetic}. The rotational component this introduces can carry trajectories through regions where standard Hamiltonian paths stall, at the cost of a further matrix to specify.

\medskip\noindent\textbf{Relativistic HMC.} Replacing the quadratic kinetic energy by a relativistic one, $K(\vp) = c\sqrt{\vp^\top\vp/c^2 + m^2c^2}$, bounds the particle speed by $c$ \cite{lu2017relativistic}. This can limit the effect of large velocities and may improve numerical stability in some challenging regions, but it does not remove the need to resolve the target's geometry.

\medskip\noindent\textbf{Microcanonical Samplers.} Robnik et al.\ \cite{robnik2023microcanonical,robnik2024fluctuation} abandon the canonical ensemble altogether, evolving the state on a surface of constant energy under norm-preserving velocity dynamics and injecting noise only by rotating the velocity. The resulting Microcanonical Langevin Monte Carlo (MCLMC) carries no Metropolis test, so its stationary distribution retains a discretization bias controlled only by the step size. Against this the authors report scaling in dimension and condition number more favorable than that of NUTS, and speedups on standard benchmarks that reach orders of magnitude in the best cases. A Metropolis-adjusted variant \cite{robnik2025metropolis} removes the bias at some cost in efficiency.

\medskip\noindent\textbf{Discontinuous \& Constrained Targets.} \textit{Discontinuous} HMC \cite{nishimura2020discontinuous} extends the framework to discrete and discontinuous parameters by means of Laplace-distributed momenta and a coordinate-wise integrator that handles density jumps. \emph{Geodesic Monte Carlo} \cite{byrne2013geodesic} and \textit{constrained} HMC \cite{brubaker2012family} handle parameters living on manifolds such as simplices, spheres and orthogonal matrices, by splitting the flow into a geodesic part that is solved exactly and a potential part that is not.

\subsection{Learned \& Preconditioned Dynamics}

A third family, originating largely in machine learning, attempts to learn a transformation of the target geometry: \emph{NeuTra} \cite{hoffman2019neutra} trains a normalizing flow and subsequently runs HMC in its latent space, where the transformed target may be better conditioned. \emph{L2HMC} \cite{levy2018generalizing} parameterizes the leapfrog update with neural networks trained to increase the expected squared jump distance without sacrificing invertibility or a tractable Jacobian. Both approaches thus exchange an up-front training cost, together with the risk of learning an inadequate transformation, for the possibility of more efficient sampling once training has been completed.

Finally, \emph{stochastic gradient HMC} \cite{chen2014stochastic} replaces the exact gradient with a mini-batch estimate and introduces friction intended to counteract the noise thereby injected, extending the method to very large data sets but leaving it unadjusted and, under naive subsampling, biased because the mini-batch noise disrupts the energy conservation required for an inexpensive Metropolis correction \cite{betancourt2015fundamental}. This approach is consequently of limited relevance to the joint models considered in Section~\ref{sec:hierarchical}, where the dominant cost arises from per-subject quadrature rather than the sheer number of observations. The principal properties of the methods discussed above are summarized in Table~\ref{tab:geometry}.

\begin{table}[t]
\centering
\caption{Methods that modify the geometry or the dynamics. ``Separable'' refers to the Hamiltonian; ``explicit'' to whether an explicit symplectic integrator is available, and ``exact'' to whether the sampler targets $\pi$ without asymptotic bias.}
\label{tab:geometry}
\renewcommand{\arraystretch}{1.2}
\small
\begin{tabular}{lccc}
\toprule
\textbf{Method} & \textbf{Separable} & \textbf{Explicit} & \textbf{Exact} \\
\midrule
HMC / GHMC \cite{neal2011mcmc,kennedy2001cost}  & Yes & Yes & Yes \\
MMHMC \cite{radivojevic2020modified}            & Yes & Yes & Yes\,$^{\dagger}$ \\
RMHMC \cite{girolami2011riemann}                & No  & No  & Yes \\
SoftAbs RMHMC \cite{betancourt2013general}      & No  & No  & Yes \\
Lagrangian MC \cite{lan2015markov}              & No  & Yes & Yes \\
Tao augmentation \cite{tao2016explicit}         & No\,$^{\ddagger}$ & Yes & Yes \\
Magnetic HMC \cite{tripuraneni2017magnetic}     & Yes & Yes & Yes \\
Relativistic HMC \cite{lu2017relativistic}      & Yes & Yes & Yes \\
MCLMC \cite{robnik2023microcanonical,robnik2024fluctuation} & Yes & Yes & No \\
SGHMC \cite{chen2014stochastic}                 & Yes & Yes & No \\
NeuTra \cite{hoffman2019neutra}                 & Yes & Yes & Yes \\
\bottomrule
\end{tabular}
\\[3pt]
\begin{minipage}{0.93\columnwidth}
\scriptsize $^{\dagger}$ Exact after importance reweighting, as in \eqref{eq:mhmc-estimator}.\\
$^{\ddagger}$ Non-separable, but separable in the augmented phase space.
\end{minipage}
\end{table}

\section{Numerical Integration}\label{sec:integrators}

In practice, the Hamiltonian dynamics of Section~\ref{sec:hmc} can rarely be solved \textit{exactly} and must instead be approximated \textit{numerically}, making the choice of integrator considerably more than an implementation detail: it determines whether the resulting proposal map remains \textit{reversible} and \textit{volume preserving}, while the energy error it introduces governs the eventual acceptance rate. Since gradient evaluations ordinarily account for most of the cost of sampling, the design of an HMC integrator may consequently be viewed as the problem of extracting as much accuracy as possible from every such evaluation.

\subsection{Integrator Splitting, Drifts \& Kicks}

Standard HMC splits the Hamiltonian into kinetic and potential parts, $H(\vth, \vp) = A(\vp) + B(\vth)$, each of which generates a \textit{flow} that can be solved exactly,
\begin{equation}\label{eq:flows}
\begin{aligned}
  \varphi_t^{A}(\vth, \vp) &= \left(\vth + t M^{-1}\vp,\; \vp\right), \\
  \varphi_t^{B}(\vth, \vp) &= \left(\vth,\; \vp - t\,\nabla U(\vth)\right),
\end{aligned}
\end{equation}
customarily termed a \emph{position drift} and a \emph{momentum kick}, respectively. Since both are exact Hamiltonian flows, and thus symplectic, any composition of them inherits symplecticness. When the coefficients of that composition are also palindromic, the resulting map is also time reversible. This simple splitting of the Hamiltonian thus generates an entire family of useful integrators, of which the traditional Störmer--Verlet, or \textit{leapfrog}, scheme \cite{verlet1967computer} is the simplest representative, and is obtained from the composition
\begin{equation}\label{eq:verlet}
  \Psi_{\varepsilon}^{(\text{VV-1})}
  \;=\; \varphi_{\varepsilon/2}^{B} \circ \varphi_{\varepsilon}^{A} \circ \varphi_{\varepsilon/2}^{B},
\end{equation}
which is second order accurate at the cost of a single gradient evaluation per step, since the half-kicks of consecutive steps merge into one when the trajectory is iterated.

\subsection{Multi-Stage Splitting Integrators (MSSIs)}

Nothing, of course, compels the coefficients to be those in \eqref{eq:verlet}, and the leapfrog scheme is but one member of a broader family of \emph{multi-stage splitting integrators} (MSSIs) that generalize the $(\tfrac12, 1, \tfrac12)$ composition to more stages and free coefficients. Symmetric compositions of the flows in \eqref{eq:flows} with free coefficients are a classical construction in geometric integration \cite{mclachlan1995numerical,hairer2006geometric}, surveyed for this setting by Bou-Rabee and Sanz-Serna \cite{bourabee2018geometric}, and the multi-stage splitting integrators they generate have been developed systematically for HMC \cite{blanes2014numerical,akhmatskaya2017adaptive,radivojevic2018multi,nagar2024adaptive}. The general family of \emph{$k$-stage splitting integrators}, which leaves $(k-1)$ free parameters, is defined by
\begin{equation}\label{eq:mssi}
\resizebox{0.88\columnwidth}{!}{$\displaystyle
\Psi_{\varepsilon}^{(k)} = \begin{cases}
\varphi_{b_1 \varepsilon}^{B} \circ \varphi_{a_1 \varepsilon}^{A} \circ \cdots \circ
  \varphi_{a_{k'} \varepsilon}^{A} \circ \varphi_{b_{k'+1} \varepsilon}^{B} \circ
  \varphi_{a_{k'} \varepsilon}^{A} \circ \cdots \circ \varphi_{a_1 \varepsilon}^{A} \circ
  \varphi_{b_1 \varepsilon}^{B}, & k = 2k', \\[4pt]
\varphi_{b_1 \varepsilon}^{B} \circ \varphi_{a_1 \varepsilon}^{A} \circ \cdots \circ
  \varphi_{b_{k'} \varepsilon}^{B} \circ \varphi_{a_{k'} \varepsilon}^{A} \circ
  \varphi_{b_{k'} \varepsilon}^{B} \circ \cdots \circ \varphi_{a_1 \varepsilon}^{A} \circ
  \varphi_{b_1 \varepsilon}^{B}, & k = 2k'-1.
\end{cases}$}
\end{equation}
with coefficients $a_j, b_i \in \mathbb{R}^{+}$ subject to the \textit{consistency} conditions
\begin{equation}\label{eq:consistency}
\resizebox{0.85\columnwidth}{!}{$\displaystyle
\begin{cases}
  2\sum_{i=1}^{k'} b_i + b_{k'+1} = 2\sum_{j=1}^{k'} a_j = 1, & k = 2k', \\[2pt]
  2\sum_{i=1}^{k'} b_i = 2\sum_{j=1}^{k'-1} a_j + a_{k'} = 1, & k = 2k'-1,
\end{cases}$}
\end{equation}
which ensure that the composition approximates the exact flow $\Phi_{\varepsilon}$ to at least first order. Time reversibility again follows from the palindromic structure, and \textit{symplecticness} from the composition of Hamiltonian flows, so the requirements identified in Section~\ref{sec:hmc} are met by construction for every admissible choice of coefficients. Note also that the number of stages $k$ coincides with the number of gradient evaluations $\nabla U(\vth)$ performed per step of size $\varepsilon$.

\begin{figure}[t]
\centering
\begin{tikzpicture}[
  >=Latex, line join=round,
  x=5.3cm, y=1.0cm,
  tl/.style    = {draw=black!55, line width=0.8pt},
  drift/.style = {draw=cTune, fill=cTune!18, line width=0.6pt},
  kick/.style  = {draw=cInteg!90!black, fill=cInteg!75, line width=0.4pt},
  cl/.style    = {font=\scriptsize, text=cTune!70!black, inner sep=1pt},
  kl/.style    = {font=\scriptsize, text=cInteg!80!black, anchor=north,
                  inner sep=1pt},
  nm/.style    = {font=\scriptsize\bfseries, anchor=east},
  gc/.style    = {font=\scriptsize, text=black!55, anchor=west},
  ax/.style    = {font=\scriptsize, text=black!60}
]

\def\kscale{0.80}
\def\khalf{0.017}
\def\dbot{0.10}
\def\dtop{0.82}
\def\dmid{0.46}

\newcommand{\dblock}[3]{%
  \draw[drift] ({#1},\dbot) rectangle ({#2},\dtop);
  \node[cl] at ({(#1+#2)/2},\dmid) {#3};
}
\newcommand{\kbar}[3]{%
  \pgfmathsetmacro{\khgt}{-\kscale*#2}%
  \draw[kick] ({#1-\khalf},\dtop) rectangle ({#1+\khalf},{\khgt});
  \node[kl, anchor=south] at ({#1},{\dtop+0.15}) {#3};
}
\newcommand{\rhead}[2]{%
  \node[nm] at (-0.030,\dmid) {#1};
  \node[gc] at ( 1.030,\dmid) {#2};
  \draw[tl] (0,0) -- (1,0);
}

\begin{scope}
  \rhead{VV-1}{\textit{1 Gradient}}
  \dblock{0.0}{1.0}{$\varphi^{A}_{\varepsilon}$}
  \kbar{0.0}{0.5}{$\dfrac12$}
  \kbar{1.0}{0.5}{$\dfrac12$}
\end{scope}

\begin{scope}[yshift=-1.92cm]
  \rhead{2-Stage}{\textit{2 Gradients}}
  \dblock{0.0}{0.5}{$\varphi^{A}_{\varepsilon/2}$}
  \dblock{0.5}{1.0}{$\varphi^{A}_{\varepsilon/2}$}
  \kbar{0.0}{0.211781}{$b$}
  \kbar{0.5}{0.576438}{$1\!-\!2b$}
  \kbar{1.0}{0.211781}{$b$}
\end{scope}

\begin{scope}[yshift=-3.84cm]
  \rhead{3-Stage}{\textit{3 Gradients}}
  \dblock{0.0}{0.296195}{$a$}
  \dblock{0.296195}{0.703805}{$1\!-\!2a$}
  \dblock{0.703805}{1.0}{$a$}
  \kbar{0.0}{0.118880}{$b$}
  \kbar{0.296195}{0.381120}{$\dfrac12\!-\!b$}
  \kbar{0.703805}{0.381120}{$\dfrac12\!-\!b$}
  \kbar{1.0}{0.118880}{$b$}
\end{scope}

\begin{scope}[yshift=-4.62cm]
  \draw[->, black!45, line width=0.7pt] (0,0) -- (1.04,0);
  \draw[black!45, line width=0.7pt] (0,-0.055) -- (0,0.055);
  \draw[black!45, line width=0.7pt] (1,-0.055) -- (1,0.055);
  \node[ax, anchor=north] at (0,-0.10) {$0$};
  \node[ax, anchor=north] at (1,-0.10) {$\varepsilon$};
  \node[ax, anchor=north] at (0.5,-0.10) {\textit{Time Within One Integration Step}};
\end{scope}

\begin{scope}[yshift=-5.47cm]
  \draw[drift] (0.075,-0.060) rectangle (0.120,0.060);
  \node[cl, anchor=west, text=black!65] at (0.138,0)
    {\textit{Position Drift} $\varphi^{A}$};
  \draw[kick] (0.083,-0.450) rectangle (0.112,-0.330);
  \node[kl, anchor=west, text=black!65] at (0.138,-0.335)
    {\textit{Momentum Kick} $\varphi^{B}$};
\end{scope}

\end{tikzpicture}
\caption{Multi-stage splitting integrators (MSSIs) as sequences of kicks and drifts over one step of size $\varepsilon$. Position drifts (above the line) advance time; momentum kicks (below) are instantaneous and each costs one gradient evaluation, so the number of kicks per step is the number of stages $k$. All sequences are palindromic, which gives time-reversibility, and all are compositions of exact Hamiltonian flows, which gives symplecticity. The free coefficients, namely $b$ for $k=2$ and $(a,b)$ for $k=3$, distinguish the Velocity Verlet, BCSS and Minimum-Error schemes of Table~\ref{tab:mssi}. Values drawn here correspond to BCSS.}
\label{fig:splitting}
\end{figure}
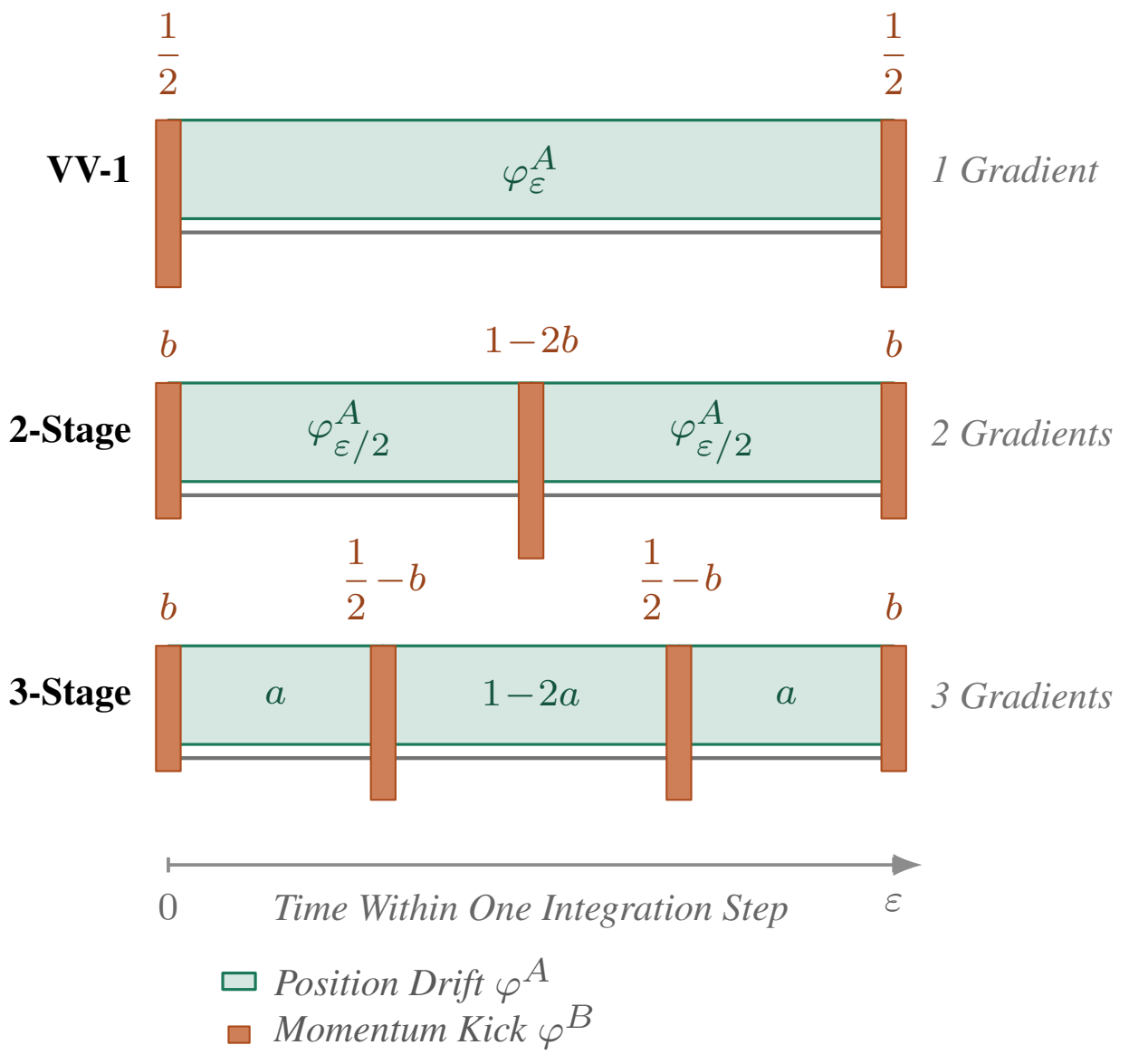

Working through \eqref{eq:consistency} for $k = 2$ leaves a single free parameter $b \in (0, \tfrac12)$ and the composition
\begin{equation}\label{eq:two-stage}
  \Psi_{\varepsilon}^{(2)} = \varphi_{b\varepsilon}^{B} \circ \varphi_{\varepsilon/2}^{A}
    \circ \varphi_{(1-2b)\varepsilon}^{B} \circ \varphi_{\varepsilon/2}^{A}
    \circ \varphi_{b\varepsilon}^{B},
\end{equation}
whereas $k = 3$ leaves two parameters $a, b \in (0, \tfrac12)$ and
\begin{equation}\label{eq:three-stage}
\resizebox{0.85\columnwidth}{!}{$\displaystyle
  \Psi_{\varepsilon}^{(3)} = \varphi_{b\varepsilon}^{B} \circ \varphi_{a\varepsilon}^{A}
    \circ \varphi_{(\frac12-b)\varepsilon}^{B} \circ \varphi_{(1-2a)\varepsilon}^{A}
    \circ \varphi_{(\frac12-b)\varepsilon}^{B} \circ \varphi_{a\varepsilon}^{A}
    \circ \varphi_{b\varepsilon}^{B},$}
\end{equation}
Campos and Sanz-Serna \cite{campos2017palindromic} show that outside the one-parameter subfamily satisfying
\begin{equation}\label{eq:three-stage-constraint-main}
  6ab - 2a - b + \tfrac12 \;=\; 0,
\end{equation}
palindromic three-stage schemes have poor stability properties, and \eqref{eq:three-stage-constraint-main} is accordingly imposed throughout the HMC literature \cite{blanes2014numerical,radivojevic2018multi,nagar2024adaptive}. The Velocity Verlet, BCSS and minimum-error values of Table~\ref{tab:mssi} all satisfy it, and under it, the error bound $\rho_3$ below is derived. Figure~\ref{fig:splitting} illustrates the drift-kick structure of both schemes, whose coefficients, together with those of other common choices, are collected in Table~\ref{tab:mssi}.

\begin{remark}[Fair comparison between integrators]\label{rem:fair}
Because a $k$-stage integrator spends $k$ gradients on every step, a comparison at equal $(\varepsilon,L)$ assigns different computational budgets to the competing schemes and is therefore misleading. The convention adopted throughout the literature \cite{radivojevic2018multi,nagar2024adaptive} fixes the \textit{gradient budget} instead: if $\widehat{L}$ steps of size $\varepsilon$ are appropriate for one-stage Verlet, a $k$-stage scheme is run for $L=\widehat{L}/k$ steps of size $k\varepsilon$, so that every scheme integrates the dynamics over the same interval $\widehat{L}\varepsilon$ and expends the same number of gradient evaluations in doing so. For precisely the same reason, the $\GRADe$ metric, introduced in Section~\ref{sec:diagnostics} as the number of gradient evaluations per \textit{effective sample}, provides a more comprehensively and fair measure of efficiency across integrators.
\end{remark}

\subsection{Choosing the Coefficients}

Within each family of integrators, the free coefficients are fixed by optimizing some proxy for the energy error, since it is this error that drives the acceptance rate. Three choices of proxy, and hence three families of coefficients, are in common use in HMC.

\medskip\noindent\textbf{Velocity Verlet Schemes.} These simply concatenate $k$ Verlet steps of length $\varepsilon/k$, which amounts to $b = 1/4$ for $k = 2$ and to $a = 1/3$, $b = 1/6$ for $k = 3$. They have the longest stability intervals in their families, $(0, 2k)$ in dimensionless units, and are the natural choice when stability, rather than accuracy, limits the step size.

\medskip\noindent\textbf{Minimum-Error Schemes.} These minimize a norm of the leading coefficients of the energy error in the limit $\varepsilon \to 0$ \cite{mclachlan1995numerical,predescu2012computationally}, namely the $c_{21}$ and $c_{22}$ of the fourth-order modified Hamiltonian of Appendix~\ref{app:modified}. The criterion is not the same in the two families. McLachlan's two-stage coefficient $b = 0.193183$ minimizes $(c_{21}^{2} + c_{22}^{2})^{1/2}$, whereas the three-stage pair of Predescu et al.\ \cite{predescu2012computationally} satisfies $c_{21} = c_{22}$, which under \eqref{eq:three-stage-constraint-main} cancels the $\mathcal{O}(\varepsilon^{4})$ term of $\rho_3$ and leaves a truncation error of size $\mathcal{O}(\varepsilon^{6})$. Both are thus optimal in the small step size limit, and by extension for high-dimensional problems in which accuracy already drives the step size down. Their stability intervals are the shortest in their families, and their advantage disappears quickly once the $\varepsilon$ approaches the stability boundary.

\medskip\noindent\textbf{BCSS Schemes.} Blanes, Casas and Sanz-Serna \cite{blanes2014numerical} argue instead that integrators are not in fact operated at small step sizes, and minimize instead the maximum of the energy error bound $\rho_k$ over the interval $0 < \bar\varepsilon < k$ on which a $k$-stage scheme is typically run. The resulting coefficients sit between the Velocity Verlet and minimum-error values, as Table~\ref{tab:mssi} shows.

\medskip
The coefficients of these schemes are collected in Table~\ref{tab:mssi}. Which of them performs best depends on the step size regime of the problem at hand, which is rarely known in advance. The adaptive procedures of Section~\ref{sec:tuning} address this caveat by estimating the regime during burn-in and then selecting $b$ to suit it.

\begin{table}[t]
\centering
\caption{Coefficients of commonly used two- and three-stage splitting integrators, in the parameterization of \eqref{eq:two-stage} and \eqref{eq:three-stage}.}
\label{tab:mssi}
\renewcommand{\arraystretch}{1.2}
\setlength{\tabcolsep}{3pt}
\resizebox{\columnwidth}{!}{%
\begin{tabular}{lcc}
\toprule
\textbf{Integrator} & \textbf{Stages $k$} & \textbf{Coefficients} \\
\midrule
1-Stage Velocity Verlet (VV-1)   & 1 & -- \\
2-Stage Velocity Verlet (VV-2)   & 2 & $b = 1/4$ \\
2-Stage BCSS (BCSS-2)            & 2 & $b = 0.211781$ \\
2-Stage Minimum-Error (ME-2)     & 2 & $b = 0.193183$ \\
3-Stage Velocity Verlet (VV-3)   & 3 & $a = 1/3$, $b = 1/6$ \\
3-Stage BCSS (BCSS-3)            & 3 & $a = 0.296195$, $b = 0.118880$ \\
3-Stage Minimum-Error (ME-3)     & 3 & $a = 0.290486$, $b = 0.108991$ \\
\bottomrule
\end{tabular}}
\end{table}

\subsection{Stability, Resonance \& Expected Energy Error}

Applied to the univariate harmonic oscillator $H = \tfrac12(p^2 + \theta^2)$, a $k$-stage palindromic integrator reduces to a linear map,
\begin{equation}\label{eq:linear-map}
  \Psi_{\varepsilon}\begin{pmatrix}\theta_i \\ p_i\end{pmatrix}
  = \begin{pmatrix} A_{\varepsilon}^{\vc} & B_{\varepsilon}^{\vc} \\
                    C_{\varepsilon}^{\vc} & D_{\varepsilon}^{\vc}\end{pmatrix}
    \begin{pmatrix}\theta_i \\ p_i\end{pmatrix},
\end{equation}
where $\vc = \{a_j, b_i\}$ denotes the set of coefficients. The scheme is stable when $\lvert A_{\varepsilon}^{\vc}\rvert < 1$, and the expected energy error admits a closed-form upper bound \cite{blanes2014numerical,nagar2024adaptive}
\begin{equation}\label{eq:rho}
  \rho_k(\varepsilon, \vc)
  \;=\; \frac{\left(B_{\varepsilon}^{\vc} + C_{\varepsilon}^{\vc}\right)^{2}}
             {2\left(1 - (A_{\varepsilon}^{\vc})^{2}\right)} ,
\end{equation}
whose explicit forms for $k = 2$ and $k = 3$ are given in Appendix~\ref{app:modified}. Although derived for a Gaussian target, the bound remains informative for non-Gaussian ones \cite{blanes2014numerical}, and near the mode of a smooth posterior. This carries two consequences of practical interest.

The first consequence concerns the \textit{scale} of the step size: for the repeated Verlet $k$-stage scheme, the dimensionless stability interval has length $2k$ and its midpoint, termed the \emph{center of the longest stability interval} (CoLSI), is $\bar{\varepsilon}_{\CoLSI}=k$. The adaptive schemes of Section~\ref{sec:tuning} use this point as an anchor when constructing their randomization intervals, although it should be understood as a convenient heuristic reference rather than a strict limit, since the stability interval is problem dependent and may be shorter or longer than the Gaussian prediction.

The second consequence concerns \textit{resonance}. Because the bound $\rho_k$ develops poles as $\lvert A_{\varepsilon}^{\vc}\rvert\to1$, at which point the expected energy error diverges and instabilities arise for step sizes related to the periods of the system. An acceptance rate target by itself may fail to identify these regions reliably, particularly before the chain has encountered them, which motivates both the explicit modelling of $\rho_k$ and the randomization of step sizes used by the procedures described below.

\subsection{Integrators for Modified Hamiltonians}

For the MHMC methods of Section~\ref{sec:mhmc}, the quantity to be conserved is $\Hmod^{[k]}$ rather than $H$, and the coefficients must be re-optimized accordingly. Radivojević et al.\ \cite{radivojevic2018multi} derive two- and three-stage families that minimize the energy error of the modified Hamiltonian, together with the coefficients $c_{ij}$ appearing in the fourth- and sixth-order expansions of Appendix~\ref{app:modified}. The resulting schemes differ from the BCSS and minimum-error coefficients of Table~\ref{tab:mssi}. Substituting the latter still gives a valid MHMC sampler, since validity rests on symplecticness and reversibility alone, but forgoes much of the advantage the modified Hamiltonian is meant to provide.

\section{Hyperparameter Tuning}\label{sec:tuning}

The methods introduced in Sections~\ref{sec:ghmc} through~\ref{sec:integrators} involve several \textit{hyperparameters}, for which convenient defaults may be unsuitable for a particular target and, even when valid, markedly inefficient. The tuning literature has consequently developed along two broad lines of work: one adapts parameters from the \textit{observed} behavior of the sampler without positing an explicit model for the target, whereas the other fits a tractable \textit{surrogate}, often multivariate Gaussian, during a \textit{burn-in} period and uses the corresponding closed-form analysis. Table~\ref{tab:tuning} compares these two approaches.

\subsection{The Step Size}

The most common approach applies stochastic approximation to the acceptance rate, adjusting $\varepsilon$ throughout warm-up until the acceptance rate approaches a nominal target. Hoffman and Gelman \cite{hoffman2014no} popularized Nesterov's dual-averaging scheme \cite{nesterov2009primal} for this purpose, within the general framework for adaptive MCMC reviewed by Andrieu and Thoms \cite{andrieu2008tutorial}. The appropriate target is itself method dependent: the asymptotically optimal rate for Verlet-based HMC on high-dimensional targets with \textit{i.i.d.}\ components is approximately $0.651$ \cite{beskos2013optimal}, \code{Stan} adopts the more conservative default of $0.8$ in order to reduce divergent transitions, and GHMC-based adaptive integration schemes generally seek values of $0.9$ or higher because every rejection also reverses the momentum \cite{nagar2024adaptive,akhmatskaya2026adaptive}.

Targeting the acceptance rate nevertheless has an important blind spot, because a scalar summary cannot distinguish a step size that is simply too large from one that is resonant in the sense of Section~\ref{sec:integrators}, nor can it reveal whether the observed acceptance is limited by ordinary integration accuracy or by proximity to the stability boundary. Procedures that estimate the stability interval directly avoid both ambiguities, although they do so at the additional cost of requiring curvature information.

\subsection{The Trajectory Length}

The trajectory length is more difficult to tune because it admits no direct analogue of the acceptance rate target: although short trajectories remain valid, they tend to produce strongly correlated draws, whereas long trajectories may spend gradients retracing regions that have already been explored. The \emph{No-U-Turn Sampler} (NUTS) \cite{hoffman2014no} addresses this problem through a local stopping criterion, recursively doubling a trajectory by integrating in randomly chosen forward or backward directions until its endpoints begin to turn back upon themselves, and subsequently selecting a state from that trajectory in a manner that preserves the target distribution. By removing the need to specify $L$ in advance, NUTS has consequently become the standard choice in \code{Stan} and a number of related frameworks.

Several later methods retain the idea of adapting the path length while changing the criterion by which this is done. \emph{ChEES-HMC} \cite{hoffman2021adaptive} replaces the U-turn test with a differentiable objective, the Change in the Estimator of the Expected Square that is optimized by stochastic gradient ascent over an ensemble of chains and, because it avoids a recursive branching tree, can be vectorized across chains more readily than NUTS (Section~\ref{sec:parallelism}). The \emph{apogee-to-apogee path sampler} \cite{sherlock2023apogee} instead segments a trajectory at successive maxima of the potential energy and reports efficiency comparable to HMC with markedly less sensitivity to its own tuning parameter, the number of apogees crossed. \emph{ATLAS} \cite{modi2024atlas} combines dynamic trajectory lengths with a step size derived from the local Hessian and a delayed rejection mechanism, aimed at targets whose curvature varies sharply from region to region.

At a more general level, \emph{GIST} \cite{bourabee2024gist} treats the tuning parameters as auxiliary variables and samples them conditionally upon the current state $(\vth,\vp)$, using a Metropolis-within-Gibbs correction to preserve the invariant distribution. Randomized HMC \cite{bourabee2017randomized}, multinomial HMC, NUTS, and the apogee-to-apogee sampler are recovered by different conditional distributions for the path length, so that algorithms ordinarily presented as separate constructions become instances of the same representation.

\subsection{The Mass Matrix \& the Riemannian Metric}

During warm-up, samplers in the \code{Stan} family estimate either a diagonal or a dense mass matrix from draws collected over a sequence of expanding windows. A diagonal $M$ rescales the individual coordinates without accounting for their correlations, whereas a dense matrix performs both tasks at the price of $\mathcal{O}(D^2)$ storage, $\mathcal{O}(D^3)$ factorization during warm-up, and $\mathcal{O}(D^2)$ work at every leapfrog step, costs that may be prohibitive for joint models containing several thousand latent parameters. Moreover, neither choice fully resolves a funnel because the appropriate preconditioner varies with position, leaving the Riemannian methods of Section~\ref{sec:geometry}, the reparameterizations of Section~\ref{sec:hierarchical}, or preconditioning through a learned normalizing flow \cite{hoffman2019neutra} as possible alternatives.

\subsection{The Adaptive Integration Approach: from AIA to ATune}

The adaptive integration line of work, first developed in molecular simulation and subsequently adapted to computational statistics, models the target rather than relying solely upon the observed behavior of the sampler. Near its mode, a smooth posterior is approximately Gaussian, in which case the integrator analysis of Section~\ref{sec:integrators} becomes exact, $\rho_k(\varepsilon,b)$ is available in closed form, and the characteristic frequencies of the approximation determine both a stability interval and a suitable integrator coefficient. The frequencies are estimated once during burn-in and the map from step size to coefficient is pre-tabulated \cite{nagar2024adaptive}, so production sampling carries no adaptation overhead at all, in contrast to the per-iteration cost of the No-U-Turn family.

\begin{figure*}[t]
\centering
\def\pw{6.40}  
\def\ph{4.50}  
\def\gx{6.95}  
\resizebox{\textwidth}{!}{%
\begin{tikzpicture}[
  >=Latex, line join=round,
  stg/.style  = {draw=#1, fill=#1!8, rounded corners=4pt, line width=0.7pt},
  hd/.style   = {font=\bfseries\small, text=white, inner sep=3.5pt,
                 rounded corners=2pt, fill=#1, minimum width=\pw cm,
                 align=center},
  goal/.style = {font=\footnotesize\itshape, text=#1!75!black, anchor=north,
                 inner sep=0pt, text width=\pw cm-0.5cm, align=center},
  body/.style = {font=\scriptsize, anchor=north west, inner sep=0pt,
                 text width=\pw cm-0.5cm, align=left},
  rule/.style = {draw=#1!35, line width=0.4pt},
  axis/.style = {draw=black!55, line width=0.6pt},
  tick/.style = {draw=black!55, line width=0.5pt},
  tl/.style   = {font=\scriptsize, text=black!65, inner sep=1.5pt},
  band/.style = {fill=#1!45, draw=#1!80!black, line width=0.4pt},
  br/.style   = {draw=#1!60, fill=white, rounded corners=2pt, align=center,
                 font=\scriptsize, inner sep=3pt, minimum height=0.95cm,
                 text width=2.55cm},
  outp/.style = {draw=black!35, fill=cLight, rounded corners=3pt, align=center,
                 font=\scriptsize, inner sep=4pt, minimum width=\pw cm,
                 minimum height=1.10cm},
  flow/.style = {->, line width=1.1pt, draw=black!50},
  drop/.style = {->, line width=0.7pt, draw=black!45}
]

\newcommand{\pitem}[1]{%
  \par\hangindent=0.9em\hangafter=1\noindent
  \makebox[0.9em][l]{\textbullet}#1\par\vspace{2pt}}
\newcommand{\stage}[4]{%
  \coordinate (p#1) at ({(#1-1)*\gx-\pw/2},0);
  \draw[stg=#2] (p#1) rectangle ++(\pw,-\ph);
  \node[hd=#2, anchor=south west] at ($(p#1)+(0,0.14)$) {#3};
  \node[goal=#2] at ($(p#1)+(\pw/2,-0.22)$) {#4};
  \draw[rule=#2] ($(p#1)+(0.25,-0.62)$) -- ++(\pw-0.5,0);
}

\stage{1}{cSampler}{(1) Tuning Stage}{Find a stable reference step size.}
\begin{scope}[shift={(p1)}]
  \node[body] at (0.25,-0.78) {%
    \pitem{Short one-stage Verlet run with \mbox{$L=1$}.}
    \pitem{Every \mbox{$N_{\text{check}}$} iterations,
      \mbox{$\varepsilon \leftarrow \varepsilon \pm \delta_\varepsilon$}.}
    \pitem{Stop once \mbox{$\lvert \AR - \alpha_{\text{target}}\rvert < \xi$}.}};  
  \def\ax{0.70}\def\al{5.20}\def\ay{-3.30}
  \node[tl, anchor=east] at (\ax-0.05,\ay) {$\AR$};
  \fill[band=cSampler] ({\ax+0.89*\al},\ay-0.11) rectangle ({\ax+0.95*\al},\ay+0.11);
  \draw[axis] (\ax,\ay) -- ++(\al,0);
  \foreach \v/\lab in {0/0, 0.5/0.5, 1/1}{
    \draw[tick] ({\ax+\v*\al},\ay-0.06) -- ++(0,0.12);
    \node[tl, anchor=north] at ({\ax+\v*\al},\ay-0.08) {$\lab$};
  }
  \draw[cSampler!80!black, line width=0.8pt, <-]
    ({\ax+0.92*\al},\ay+0.14) -- ++(0,0.28)
    node[tl, anchor=south east, xshift=6pt, text=cSampler!80!black]
      {$\alpha_{\text{target}} \approx 0.92 \pm \xi$};
\end{scope}

\stage{2}{cTune}{(2) Burn-In Stage}{Quantify the departure from harmonicity.}
\begin{scope}[shift={(p2)}]
  \node[body] at (0.25,-0.78) {%
    \pitem{Cheap fitting factor $S$ from $D$ and $\AR$,
      \eqref{eq:fitting-factor}.}};
  \coordinate (b0) at (\pw/2,-1.45);
  \node[br=cTune, anchor=north] (bh) at (\pw-1.73,-1.95)
    {{\boldmath$S > 2$}\\[1pt] \mbox{$\omega_j = \sqrt{\lambda_j(\nabla^2 U)}$}\\
     \mbox{$S_f = S_\omega$, \eqref{eq:fitting-factor-omega}}};
  \path let \p1 = ($(bh.north)-(bh.south)$) in
    node[br=cTune, anchor=north, minimum height=\y1] (bl) at (1.73,-1.95)
    {{\boldmath$S \le 2$}\\[1pt] $S_f = S$\\ \textit{(no Hessian)}};
  \draw[tick] (\pw/2,-1.22) -- (b0);
  \draw[tick, ->] (b0) -| (bl.north);
  \draw[tick, ->] (b0) -| (bh.north);
  \fill[black!55] (b0) circle (1.1pt);
  \node[body] at ([xshift=0.25cm, yshift=-0.15cm]bh.south -| p2) {%
    \pitem{Spectral summaries \mbox{$\tilde\omega = \max_j \omega_j$} and $\sigma_\omega$.}
    \pitem{Stability limit \mbox{$\mathrm{SL} = 2k/(S_f\tilde\omega)$},
      \eqref{eq:stability-limit}.}};
\end{scope}

\stage{3}{cInteg}{(3) Production Stage}{Randomize all parameters every iteration.}
\begin{scope}[shift={(p3)}]
  \def\ax{0.70}\def\al{5.20}\def\ay{-1.55}
  \node[tl, anchor=east] at (\ax-0.05,\ay) {$\bar\varepsilon$};
  \fill[band=cInteg] ({\ax+2.0772/6*\al},\ay-0.11) rectangle ({\ax+0.5*\al},\ay+0.11);
  \draw[axis] (\ax,\ay) -- ++(\al,0);
  \foreach \v/\lab/\anc in {0/0/north, 2.0772/2.0772/north east, 3/k/north west, 6/2k/north}{
    \draw[tick] ({\ax+\v/6*\al},\ay-0.06) -- ++(0,0.12);
    \node[tl, anchor=\anc] at ({\ax+\v/6*\al},\ay-0.08) {$\lab$};
  }
  \node[tl, anchor=south, text=cInteg!80!black] at ({\ax+(2.0772/6+0.5)/2*\al},\ay+0.12)
    {$\bar\varepsilon \sim \mathrm{U}(2.0772,\,k)$};
  \node[body] at (0.25,-2.30) {%
    \pitem{$b_{\text{opt}}^{k}$ minimizes the worst-case $\rho_k$
      \eqref{eq:b-opt}.}
    \pitem{\textbf{GHMC:} \mbox{$\phi \sim \mathrm{U}\bigl(\phi_{\text{opt}}(k),\,
      \phi_{\text{opt}}(2.0772)\bigr)$}, \eqref{eq:phi-opt}.}
    \pitem{Dimensional $\varepsilon = \bar\varepsilon/(S_f\tilde\omega)$.}
    \pitem{\mbox{$L = 1$} if \mbox{$S_f < 1.5$}, else
      \mbox{$L \sim \mathrm{U}\{2, \dots, 6\}$}.}};
\end{scope}

\foreach \i/\txt in {%
  1/{$\varepsilon_{\text{tuned}}$},
  2/{$S_f$ (or $S_\omega$),\ $\tilde\omega$,\ $\sigma_\omega$,\ $\mathrm{SL}$},
  3/{$\Psi^{(k)}_{\varepsilon}$ with $(\varepsilon,\, b_{\text{opt}}^{k},\, \phi,\, L)$\\
     drawn afresh at every iteration}}{
  \draw[drop] ($(p\i)+(\pw/2,-\ph)$) -- ++(0,-0.30);
  \node[outp, anchor=north] (o\i) at ($(p\i)+(\pw/2,-\ph-0.32)$) {\txt};
}

\draw[flow] ($(p1)+(\pw,-\ph/2)$) -- ($(p2)+(0,-\ph/2)$);
\draw[flow] ($(p2)+(\pw,-\ph/2)$) -- ($(p3)+(0,-\ph/2)$);

\end{tikzpicture}}
\caption{The adaptive integration pipeline introduced by s-AIA and extended by ATune \cite{nagar2024adaptive,akhmatskaya2026adaptive}, building on AIA and MAIA \cite{fernandez2016adaptive,akhmatskaya2017adaptive}. The branch on $S$ shown in the burn-in stage follows s-AIA; ATune uses $S_\omega$ whenever the full frequency spectrum is computed. In contrast to the No-U-Turn lineage, which adapts online from the sampler's own behavior, this family fits a multivariate Gaussian surrogate to the target during burn-in and then derives step size, integrator coefficients, trajectory length and momentum noise from the closed-form analysis of Section~\ref{sec:integrators}. There is no production-time overhead, since all adaptation happens before the sampling phase, and the outputs are randomization intervals rather than point values, which breaks integrator resonances. Parameters are drawn afresh from those intervals at each production iteration.}
\label{fig:tuning-pipeline}
\end{figure*}
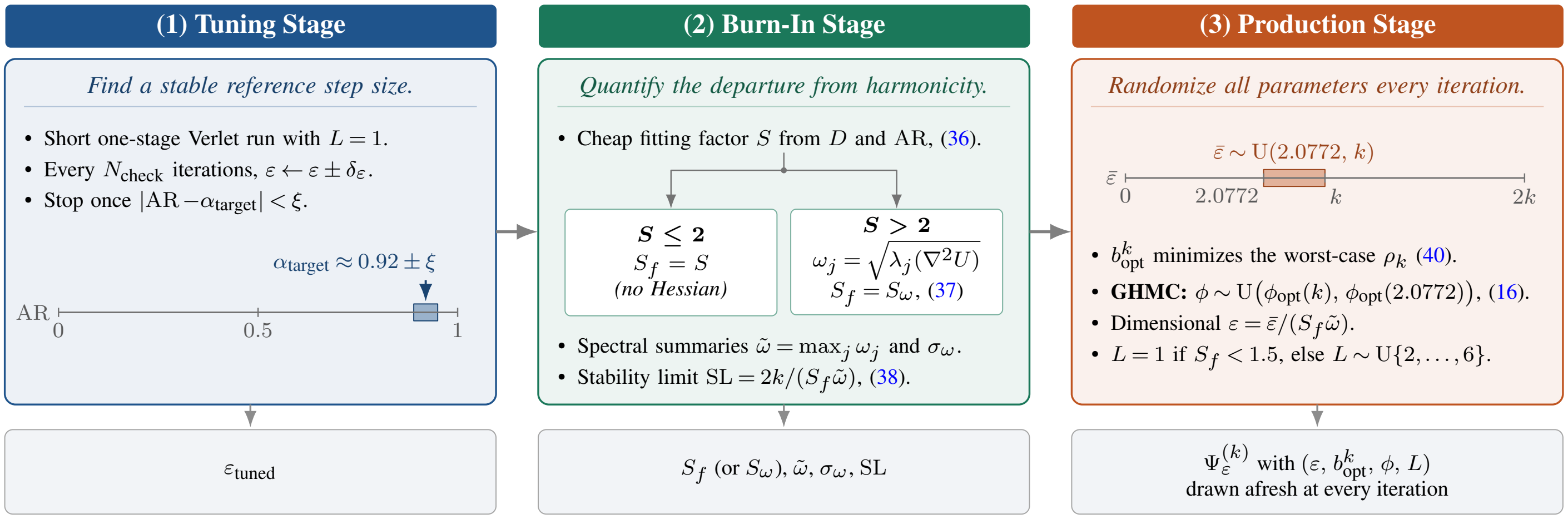

The adaptive integration family has evolved through several stages, from AIA \cite{fernandez2016adaptive}, developed for standard Hamiltonians in molecular simulation, through MAIA and its extension e-MAIA \cite{akhmatskaya2017adaptive}, which treat modified Hamiltonians and additionally tune $\phi$, to \textbf{s-AIA} \cite{nagar2024adaptive} for computational statistics, \textbf{s-MAIA} \cite{parga2024tackling} combining the two, and finally \textbf{ATune} \cite{akhmatskaya2026adaptive}, which generalizes the family so as to tune the integrator, the step size, the trajectory length and the momentum noise together, each with an accompanying randomization interval. Figure~\ref{fig:tuning-pipeline} displays the structure, which is common to all members of the family and consists of three stages.

\medskip\noindent\textbf{Tuning stage.} A short one-stage Velocity Verlet run with $L=1$ adjusts the step size in increments $\delta_\varepsilon$ until the acceptance rate lies within a tolerance $\xi$ of a target $\alpha_{\text{target}}$, derived in \cite{nagar2024adaptive} as $\alpha_{\text{target}} \approx 0.92$, the expected acceptance of this run at the center of the Verlet stability interval. The output, $\varepsilon_{\text{VV}}$, is a reference scale for the stages that follow, not a production step size.

\medskip\noindent\textbf{Burn-in stage.} The sampler runs at the tuned step size and estimates a \emph{fitting factor} $S_f \geq 1$, which measures how far the system departs from the harmonic model and by how much the harmonic \textit{stability limit} must accordingly be shortened. The cheap estimate uses only the dimension and the observed acceptance rate,
\begin{equation}\label{eq:fitting-factor}
  S \;=\; \max\!\left(1,\; \frac{2}{\tilde{\omega}\varepsilon_{\text{VV}}}
  \sqrt[6]{\frac{2\pi(1-\AR)^2}{D}}\right),
\end{equation}
where the largest frequency $\tilde\omega$ can be estimated without forming the Hessian \cite{sountsov2022meads}. In s-AIA, $S$ is adopted as $S_f$ whenever $S \leq 2$. If instead $S > 2$, indicating pronounced \textit{anharmonicity}, the algorithm computes a more informative estimate of the stability limit based on the \textit{characteristic frequencies} of the Gaussian \textit{surrogate}: at each burn-in iteration $\zeta$ it computes the \textit{eigenvalues} $\lambda_j^{(\zeta)}$ of the Hessian $\nabla^2 U(\vth^{(\zeta)})$ and the characteristic frequencies $\omega_j^{(\zeta)} = \sqrt{\lambda_j^{(\zeta)}}$, averages them over the burn-in to obtain $\omega_j$, and replaces \eqref{eq:fitting-factor} by the frequency-weighted
\begin{equation}\label{eq:fitting-factor-omega}
  S_\omega \;=\; \max\!\left(1,\; \frac{2}{\varepsilon_{\text{VV}}}
  \sqrt[6]{\frac{2\pi(1-\AR)^2}{\sum_{j=1}^{D}\omega_j^{6}}}\right).
\end{equation}
With $\tilde{\omega} = \max_j \omega_j$ and $\sigma_\omega$ the standard deviation of the frequency spectrum, the estimated stability limit for a $k$-stage scheme is
\begin{equation}\label{eq:stability-limit}
  \mathrm{SL} \;=\;
  \begin{cases}
    \dfrac{2k}{S_f\,\tilde\omega}, & \sigma_\omega \leq 1,\\[8pt]
    \dfrac{2k}{S_\omega\,(\tilde\omega - \sigma_\omega)}, & \sigma_\omega > 1,
  \end{cases}
\end{equation}
the second branch shifting the reference frequency downward when the spectrum is broad, a situation that often arises in hierarchical models.

\medskip\noindent\textbf{Production stage.} The dimensionless step size is randomized within an interval whose right endpoint is the center of the longest stability interval, $\bar{\varepsilon}_{\CoLSI} = k$, identified in Section~\ref{sec:integrators}. For the three-stage scheme that ATune adopts, the left endpoint is fixed by the local maximum of $\rho_3(\bar\varepsilon, b_{\mathrm{BCSS}_3})$ lying to the left of $\bar{\varepsilon}_{\CoLSI}$, between which and $\bar{\varepsilon}_{\CoLSI}$ the energy error stays small \cite{akhmatskaya2026adaptive}, giving
\begin{equation}\label{eq:eps-randomization}
\begin{aligned}
  \bar{\varepsilon} &\sim \mathrm{Uniform}\!\left(\bar{\varepsilon}_{\text{lower}},\;
  \bar{\varepsilon}_{\CoLSI}\right) \\
  &\approx \mathrm{Uniform}(2.0772,\; k),
\end{aligned}
\end{equation}
with the corresponding dimensional interval obtained by dividing through by $S_f\tilde\omega$, or by $S_\omega(\tilde\omega-\sigma_\omega)$ in the broad branch of \eqref{eq:stability-limit}. For each drawn $\bar\varepsilon$, the integrator coefficient is then read from the pre-tabulated map, which minimizes the \textit{worst-case} expected energy error over all dimensionless step sizes up to $\bar\varepsilon$,
\begin{equation}\label{eq:b-opt}
  b_{\text{opt}}^{k}(\bar\varepsilon)
  \;=\; \argmin_{b \in (b_{\mathrm{ME}_k},\, b_{\mathrm{VV}_k})}
  \left(\max_{0 < \bar\varepsilon' < \bar{\varepsilon}} \rho_k(\bar\varepsilon', b)\right).
\end{equation}
For GHMC, the \textit{momentum noise} is randomized over $\left(\phi_{\text{opt}}(\bar\varepsilon_{\CoLSI}),\, \phi_{\text{opt}}(\bar\varepsilon_{\text{lower}})\right)$ using \eqref{eq:phi-opt}, the ordering following from the monotonicity of $\phi_{\text{opt}}$ on that interval. The trajectory length is set to $L=1$ when $1.0 \leq S_f < 1.5$, the almost harmonic case, and otherwise drawn uniformly from $\{2, \dots, 6\}$, whose mean is the optimal length $L_{\text{opt}} = 4$ obtained by equating the expected energy errors of the harmonic and anharmonic regimes \cite{akhmatskaya2026adaptive,nagar2025optimising}.

\medskip
Unlike the No-U-Turn line of development, these methods randomize parameters and use \textit{surrogate analysis} before production sampling to avoid estimated regions of poor stability or accuracy \cite{akhmatskaya2026adaptive}. Their effectiveness, however, depends on an informative Gaussian surrogate and, whenever the full frequency spectrum is required, on the cost of Hessian eigenvalues during burn-in, which can be prohibitive in high dimension.

\begin{table*}[t]
\centering
\caption{Adaptive tuning schemes for Hamiltonian samplers. The column headed ``information used'' distinguishes methods that observe the behavior of the sampler from those that fit a surrogate model of the target.}
\label{tab:tuning}
\renewcommand{\arraystretch}{1.25}
\small
\begin{tabular}{llllc}
\toprule
\textbf{Scheme} & \textbf{Tunes} & \textbf{Information used} & \textbf{Randomizes?} & \textbf{Overhead} \\
\midrule
Dual averaging \cite{hoffman2014no}      & $\varepsilon$ & Acceptance rate & No & Warm-up only \\
NUTS \cite{hoffman2014no}                & $L$ & U-turn criterion, per iteration & Implicitly & Per iteration \\
Multinomial NUTS \cite{betancourt2017conceptual} & $L$ & Generalized U-turn & Implicitly & Per iteration \\
ChEES-HMC \cite{hoffman2021adaptive}     & $\varepsilon L$ & Ensemble second moments & No & Warm-up only \\
A2A \cite{sherlock2023apogee}            & $L$ & Potential energy apogees & Yes & Per iteration \\
GIST \cite{bourabee2024gist}             & Any & Gibbs conditional on $(\vth,\vp)$ & Yes & Per iteration \\
ATLAS \cite{modi2024atlas}               & $\varepsilon, L$ & Local Hessian, delayed rejection & Yes & Per iteration \\
Bayesian adaptive \cite{wang2013adaptive} & $\varepsilon, L$, Metric & Optimization of expected squared jump & No & Warm-up only \\
\midrule
AIA \cite{fernandez2016adaptive}         & $b$ (given $\varepsilon$) & Gaussian model, fixed safety factor & No & Setup only \\
MAIA, e-MAIA \cite{akhmatskaya2017adaptive} & $b, \phi$ (given $\varepsilon$) & Gaussian model, modified $H$ & No & Setup only \\
s-AIA \cite{nagar2024adaptive}           & $\varepsilon$, $b$ & Gaussian model, Hessian frequencies & Yes & Burn-in only \\
s-MAIA \cite{parga2024tackling}          & $\varepsilon$, $b$ (modified $H$) & As s-AIA, applied to $\Hmod^{[k]}$ & Yes & Burn-in only \\
ATune \cite{akhmatskaya2026adaptive}     & $\varepsilon, L, \phi, b$ & Gaussian model, Hessian frequencies & Yes & Burn-in only \\
\bottomrule
\end{tabular}
\end{table*}

\subsection{Which Scheme to Use}

The choice between these families depends principally upon where the computational effort is best spent. No-U-Turn methods adapt at every iteration without requiring an advance model of the target and therefore provide a useful default when gradients are inexpensive and the local geometry varies substantially. Adaptive integration methods, by contrast, concentrate their additional work in burn-in and thereafter sample from fixed randomization intervals, which may make them preferable when gradients are costly, when GHMC or MHMC is used without a direct No-U-Turn analogue, or when resonance is a particular concern. On the benchmarks reported by its authors, adaptively tuned GHMC outperforms both fixed-parameter integrators and HMC \cite{akhmatskaya2026adaptive}. Although, as the next section emphasizes, that conclusion also depends upon the effective sample size estimator on which the comparison is based.

\section{Diagnostics \& Sampling Efficiency}\label{sec:diagnostics}

The samplers, integrators, and tuning schemes of the preceding sections are only as useful as our ability to tell whether they have worked, and comparisons among them, such as the one that closed Section~\ref{sec:tuning}, are only as reliable as the efficiency measure on which they rest. The diagnostics commonly used to assess the sampling \textit{quality} and \textit{efficiency} of Hamiltonian methods focus around three important pillars: whether the chains have \textit{mixed}, how efficiently they estimate the quantities of interest, and whether their trajectories are running into unresolved geometric difficulties. All three matter and, unfortunately, no single diagnostic answers them all at once. We therefore consider each question in turn, before turning to the effective sample size, a quantity whose estimation is considerably less straightforward than its widespread and often unqualified use might suggest.

\subsection{Metrics of Convergence}

\noindent\textbf{Potential Scale Reduction Factor (PSRF).} The classical diagnostic of Gelman and Rubin \cite{gelman1992inference} runs $M$ chains from dispersed initial values and compares the \textit{within-chain} variance $W = M^{-1}\sum_m \sigma_m^2$ with the \textit{between-chain} variance $B = \frac{N}{M-1}\sum_m (\bar\theta_m - \bar\theta_G)^2$, where $\sigma_m^2$ and $\bar\theta_m$ are the variance and mean of chain $m$ and $\bar\theta_G$ the mean across all chains. Forming the \textit{pooled estimate} $\widehat{V} = (1-1/N)W + B/N + B/(MN)$ gives, in the degrees-of-freedom corrected form of Brooks and Gelman \cite{brooks1998general},
\begin{equation}\label{eq:psrf}
  \Rhat \;=\; \sqrt{\frac{d+3}{d+1}\cdot\frac{\widehat{V}}{W}},
  \qquad d \approx \frac{2\widehat{V}^2}{\Var(\widehat{V})},
\end{equation}
which tends to unity as the chains mix. Vehtari et al.\ \cite{vehtari2021rank} show that this classical form fails for heavy-tailed or infinite-variance targets, as well as for chains that share a common location but differ in scale. In its place, they propose a \emph{rank-normalized split} $\Rhat$, together with a \emph{folded} variant that is sensitive to differences in scale. Their recommended threshold of $\Rhat < 1.01$ is considerably stricter than the traditional value of $1.1$, and adopting it changes practical conclusions more often than one might expect. When many short chains are run in parallel, the natural regime for hardware accelerators examined in Section~\ref{sec:parallelism}, the \emph{nested} $\Rhat$ of Margossian et al.\ \cite{margossian2024nested} groups the chains into superchains and compares the variance within superchains with that between them. It thereby remains informative even in situations when the sampling phase is too short for the classical statistic to be meaningful.

\medskip\noindent\textbf{Simulation-based calibration.} Whereas the preceding diagnostics assess the behavior of a chain for one particular fitted model, \textit{simulation-based calibration} \cite{talts2018validating} assesses the \emph{algorithm--model pairing} over repeated experiments. Parameters are drawn from the prior, data are simulated conditionally on them, the model is fitted, and the rank of the true parameter among the resulting posterior draws is tested for uniformity. Systematic sampling bias can thus be detected even when the ordinary chain diagnostics look satisfactory, which makes the procedure particularly valuable for hierarchical models, whose narrow funnel regions may otherwise remain unexplored.

\subsection{Diagnostics Specific to Hamiltonian Methods}

Two further diagnostics, both essential for the models considered here, arise directly from the Hamiltonian construction and therefore have no exact counterpart among more general MCMC methods.

\medskip\noindent\textbf{Divergent transitions.} A divergence is recorded when the energy error accumulated along a trajectory exceeds a prescribed threshold, signalling that the integrator has encountered curvature that cannot be resolved at the current step size \cite{betancourt2016diagnosing}. Such events indicate a risk of biased exploration, and their locations often reveal the region responsible for the failure. In funnel-shaped posteriors it is therefore useful to plot the divergent draws against the population scale, since they tend to concentrate at its lower values.

\medskip\noindent\textbf{Energy Bayesian fraction of missing information.} The E-BFMI addresses the other half of the mechanism described in Section~\ref{sec:hmc}, namely whether momentum refreshment moves the chain between energy level sets at all \cite{betancourt2016diagnosing,betancourt2017conceptual}. It compares the variance of the energy changes induced by refreshment with the marginal variance of the energy. Values conventionally regarded as low, most often those below $0.3$, indicate that the chain is confined to a narrow band of energies and is therefore not exploring the tails of the target, however well it appears to move within that band. Heavy-tailed marginals and funnel geometries can both produce this behavior without necessarily causing a corresponding failure in $\Rhat$.

\subsection{Sampling Efficiency: the Many Faces of the ESS}

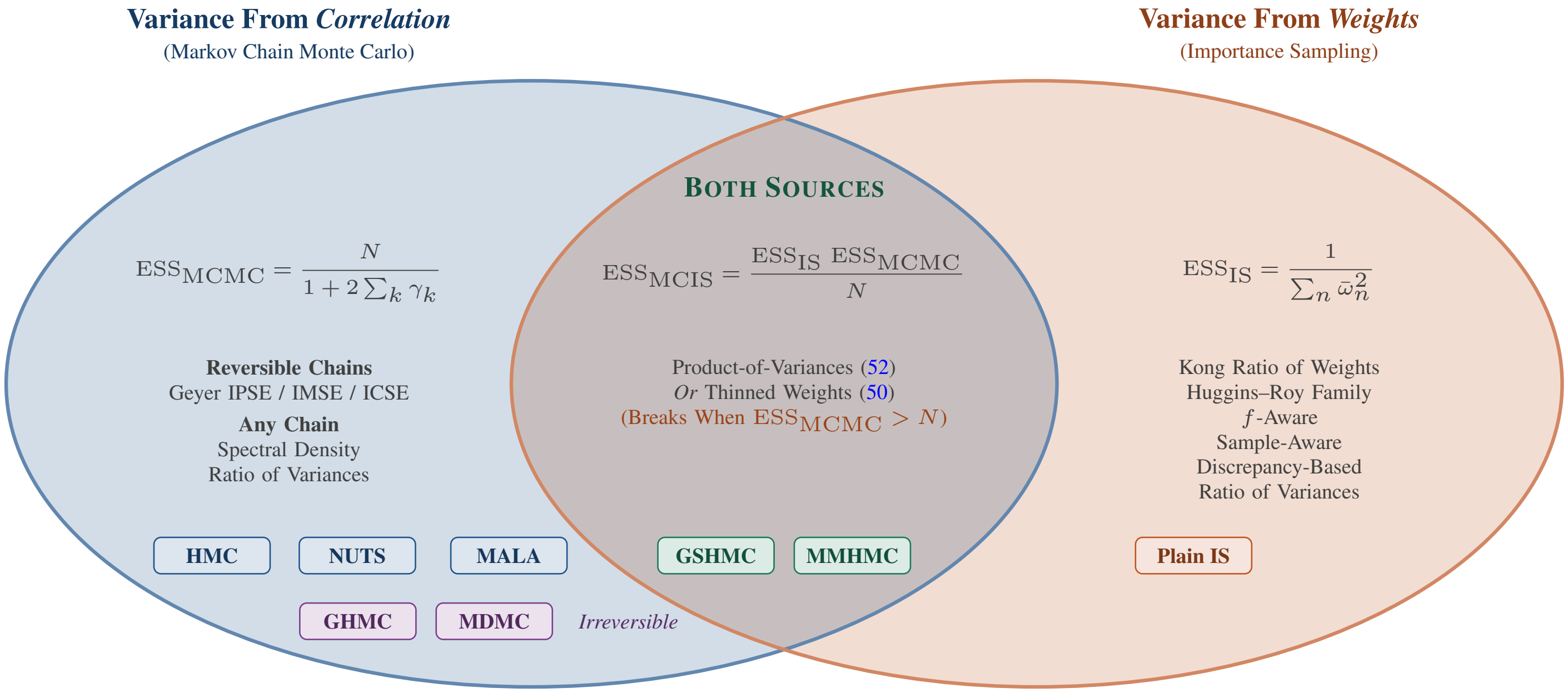
\begin{figure*}[t]
\centering
\resizebox{\textwidth}{!}{%
\begin{tikzpicture}[
  >=Latex, line join=round,
  chip/.style = {draw=#1, fill=#1!15, rounded corners=2pt, font=\tiny\bfseries,
                 inner sep=2.5pt, text=#1!65!black,
                 minimum width=1.15cm, minimum height=0.36cm},
  ttl/.style  = {font=\bfseries\scriptsize, align=center},
  est/.style  = {font=\tiny, align=center, text=black!75},
  tag/.style  = {font=\tiny\itshape, anchor=west}
]

\begin{scope}[blend group=multiply]
  \fill[cSampler!20] (-2.5,0) ellipse [x radius=5.2, y radius=3.0];
  \fill[cInteg!20]   ( 2.5,0) ellipse [x radius=5.2, y radius=3.0];
\end{scope}
\draw[cSampler!70, line width=1pt] (-2.5,0) ellipse [x radius=5.2, y radius=3.0];
\draw[cInteg!70,   line width=1pt] ( 2.5,0) ellipse [x radius=5.2, y radius=3.0];

\node[ttl, text=cSampler!70!black] at (-4.90, 3.45)
  {Variance From \emph{Correlation}\\[-1pt]\normalfont\tiny (Markov Chain Monte Carlo)};
\node[ttl, text=cInteg!75!black] at ( 4.90, 3.45)
  {Variance From \emph{Weights}\\[-1pt]\normalfont\tiny (Importance Sampling)};

\node[est] at (-4.90, 1.10)
  {$\displaystyle \ESS_{\mathrm{MCMC}}=\frac{N}{1+2\sum_k \gamma_k}$};

\node[est, text width=3.6cm, anchor=north] at (-4.90, 0.35)
  {\textbf{Reversible Chains}\\
   Geyer IPSE / IMSE / ICSE\\[2pt]
   \textbf{Any Chain}\\
   Spectral Density\\ Ratio of Variances};

\node[chip=cSampler] at (-5.6624,-1.70) {HMC};
\node[chip=cSampler] at (-4.224,-1.70) {NUTS};
\node[chip=cSampler] at (-2.724,-1.70) {MALA};
\node[chip=cDiag]    at (-4.219,-2.35) {GHMC};
\node[chip=cDiag]    at (-2.869,-2.35) {MDMC};
\node[tag, text=cDiag!70!black] at (-2.144,-2.35) {Irreversible};

\node[ttl, text=cTune!70!black] at (0, 1.95) {\textsc{Both Sources}};
\node[est] at (0, 1.10)
  {$\displaystyle \ESS_{\mathrm{MCIS}}
    = \frac{\ESS_{\mathrm{IS}}\,\ESS_{\mathrm{MCMC}}}{N}$};
\node[est, text width=3.4cm, anchor=north] at (0, 0.35)
  {Product-of-Variances \eqref{eq:ess-mcis}\\
   \emph{Or} Thinned Weights \eqref{eq:ess-thinned}\\
   \textcolor{cInteg!80!black}{(Breaks When $\ESS_{\mathrm{MCMC}} > N$)}};

\node[chip=cTune] at (-0.675,-1.70) {GSHMC};
\node[chip=cTune] at ( 0.675,-1.70) {MMHMC};

\node[est] at (4.90, 1.10)
  {$\displaystyle \ESS_{\mathrm{IS}}=\frac{1}{\sum_n \bar\omega_n^{2}}$};

\node[est, text width=3.6cm, anchor=north] at (4.90, 0.35)
  {Kong Ratio of Weights\\ Huggins--Roy Family\\
   $f$-Aware\\ Sample-Aware\\
   Discrepancy-Based\\ Ratio of Variances};

\node[chip=cInteg] at (4.0524,-1.70) {Plain IS};

\end{tikzpicture}}
\caption{The ESS landscape. A sampler's admissible ESS estimators are determined by which sources of variance it introduces, since an estimator is admissible only where its assumptions hold. Ordinary MCMC inflates variance through autocorrelation only (left). Importance sampling through weight dispersion only (right). Modified Hamiltonian methods through both (center), and require a combined estimator. Geyer's initial-sequence estimators \cite{geyer1992practical} rest upon a positivity theorem for \emph{reversible} chains, whereas GHMC and its descendants are irreversible by design. Estimators making no reversibility assumption \cite{plummer2006coda,vats2019multivariate} should be preferred there. The importance sampling estimators shown, which in turn presume independent draws, are Kong's ratio of weights \cite{kong1992note}, the Huggins--Roy family and the $f$- and sample-aware variants \cite{elvira2022rethinking}, the discrepancy-based families \cite{martino2016effective}, and the ratio-of-variances estimator \cite{agarwal2022principled}.}
\label{fig:ess-landscape}
\end{figure*}

Draws produced by MCMC are \textit{correlated}, and thus $N$ draws from a Markov Chain are not equivalent to $N$ \textit{independent} draws. The \emph{effective sample size} (ESS) formalizes this as the number of independent draws that would yield an estimator of equal variance,
\begin{equation}\label{eq:ess-definition}
  \ESS(\widetilde{I}) \;=\; N\,\frac{\Var_\pi(\bar{I})}{\Var(\widetilde{I})},
\end{equation}
in which $\bar I$ denotes the ideal estimator built from independent draws. While the definition \eqref{eq:ess-definition} is unambiguous, every \emph{estimator} of it rests on assumptions, and the samplers of Sections~\ref{sec:ghmc} and~\ref{sec:mhmc} violate those behind the estimators most commonly applied to them. Figure~\ref{fig:ess-landscape} gives an overview, organizing the estimators by the source of variance that each is designed to capture.

\subsubsection{The ESS for Markov Chains}

For a stationary chain, the variance inflation relative to independent sampling is governed by the \textit{autocorrelation sequence} $\gamma_k$, which gives
\begin{equation}\label{eq:ess-mcmc}
  \ESS_{\mathrm{MCMC}} \;=\; \frac{N}{1 + 2\sum_{k=1}^{\infty}\gamma_k}.
\end{equation}
Although the definition is straightforward, any practical estimator must \textit{truncate} the infinite sum, and the various approaches differ precisely in how they do so.

\medskip\noindent\textbf{Geyer's Initial Sequence Estimators.} Geyer \cite{geyer1992practical} begins from the result that, for an irreducible, stationary, and \emph{reversible} chain, the sums of adjacent pairs $\Gamma_k = \gamma_{2k} + \gamma_{2k+1}$ are strictly positive and decreasing, so that any departure from these properties in an observed sequence can be attributed to estimation noise. This observation gives rise to three estimators with progressively stronger smoothing. The \emph{initial positive sequence estimator} (IPSE) truncates the sum before the first non-positive $\Gamma_m$; the \emph{initial monotone sequence estimator} (IMSE), adopted by \code{Stan} and \code{ArviZ}, additionally replaces each $\Gamma_k$ by $\min\{\Gamma_0,\dots,\Gamma_k\}$; and the \emph{initial convex sequence estimator} (ICSE) further replaces the sequence by its greatest convex minorant. The positivity on which all three rely, however, holds only for \textit{reversible} chains, and so provides no justification for applying them to GHMC or its modified Hamiltonian variations, which are \textit{irreversible} by construction (Section~\ref{sec:ghmc}). In the empirical study underlying this survey, Geyer's estimators were already unreliable for GHMC and deteriorated further for MMHMC, where importance reweighting introduces a source of dependence that the ordinary autocovariance sequence does not capture.

\medskip\noindent\textbf{Spectral Density Estimators.} These exploit the fact that the power spectral density is the Fourier transform of the autocovariance function, and estimate the sum in \eqref{eq:ess-mcmc} from the spectrum evaluated at frequency zero \cite{sokal1997monte}. This is the implementation provided by \code{coda} \cite{plummer2006coda}, and makes no assumption of reversibility.

\medskip\noindent\textbf{Ratio-of-Variances Estimators.} Vats et al.\ \cite{vats2019multivariate} apply \eqref{eq:ess-definition} directly in the multivariate case, defining
\begin{equation}\label{eq:ess-vats}
  \ESS_{\mathrm{MCMC}} \;=\; N\left(\frac{\lvert\Lambda\rvert}{\lvert\Sigma\rvert}\right)^{1/p},
\end{equation}
in which $\Lambda$ is the sample covariance of the draws, $\Sigma$ a batch-means estimate \cite{flegal2010batch} of the Monte Carlo covariance, and $p$ the number of parameters. The construction does not require reversibility, handles multivariate targets natively, and comes with a principled stopping rule: for a relative tolerance $\epsilon$ and confidence level $100(1-\alpha)\%$, the required effective sample size must satisfy
\begin{equation}\label{eq:mess}
  \mathrm{mESS} \;\geq\; \frac{2^{2/p}\pi}{(p\Gamma(p/2))^{2/p}}\cdot
  \frac{\chi^2_{1-\alpha,\,p}}{\epsilon^{2}} .
\end{equation}
For \textit{irreversible} Hamiltonian samplers without importance weights, this estimator is therefore a suitable default.

\subsubsection{The ESS for Importance Sampling}

Importance sampling introduces a different inflation of variance, arising from dispersion of the weights rather than from correlation between draws. The classical estimator \cite{kong1992note,kish1965survey} is based on the \textit{ratio of weights}
\begin{equation}\label{eq:ess-is}
\begin{aligned}
  \ESS_{\mathrm{IS}} \;&=\; \frac{1}{\sum_{n=1}^{N}\bar\omega_n^{2}}
  \;=\; \frac{\left(\sum_{n=1}^{N}\omega_n\right)^{2}}{\sum_{n=1}^{N}\omega_n^{2}},
  \\[4pt]
  \bar\omega_n \;&=\; \frac{\omega_n}{\sum_j \omega_j},
\end{aligned}
\end{equation}
which is bounded within $[1,N]$ and attains $N$ exactly when all weights are equal. Elvira et al.\ \cite{elvira2022rethinking} catalogue its defects: it disregards the function $f$ being estimated and therefore claims the same quality of estimation for every integrand; it disregards the \emph{samples} entirely, depending only upon the weights, and it is consequently blind to the case in which a few well-placed samples carry most of the information. These limitations have in turn motivated three broad families of alternative estimators.

The \emph{Huggins--Roy family}, as presented by Elvira et al.\ \cite{elvira2022rethinking}, generalizes \eqref{eq:ess-is} to
\begin{equation}\label{eq:huggins-roy}
\begin{aligned}
  \ESS_{\mathrm{IS}}^{(\beta)} \;&=\;
  \Big(\textstyle\sum_{n=1}^{N}\bar\omega_n^{\beta}\Big)^{\frac{1}{1-\beta}} \\
  \;&=\; \exp\!\left(R_N^{(\beta)}(\bar\omega)\right),
  \qquad \beta \geq 0,
\end{aligned}
\end{equation}
i.e., the exponential of the Rényi entropy of order $\beta$ of the normalized weights, which recovers \eqref{eq:ess-is} at $\beta = 2$. The family is better behaved and more flexible than the classical estimator, but since it still depends on the weights alone, it shares both of the shortcomings described above.

The \emph{$f$-aware ESS} removes the first shortcoming by reweighting according to the integrand, setting $\bar\omega_n^{(f)} \propto \lvert f(\vth_n)\rvert\,\omega_n$ and substituting these weights into \eqref{eq:ess-is}. The \emph{sample-aware ESS}, however, addresses the second limitation by partitioning the sample space into $M$ regions $\mathcal{P}_m$, collecting the samples in region $m$ through the index set $\mathcal{T}_m$, and forming a summary weight $\bar\nu_m = \sum_{i \in \mathcal{T}_m}\bar\omega_i$ for each region, which gives
\begin{equation}\label{eq:ess-sample-aware}
  \ESS_{\mathrm{IS}}^{(\vx)} \;=\; \ESS_{\mathrm{IS}}(\vx)\left(
  \frac{1}{M}\sum_{m=1}^{M}\ESS_{\mathrm{IS}}(m)\right),
\end{equation}
where the global and local factors are
\begin{equation}\label{eq:ess-sample-aware-parts}
\begin{aligned}
  \ESS_{\mathrm{IS}}(\vx) &= \frac{1}{\sum_{m}\bar\nu_m^{2}},
  \\[4pt]
  \ESS_{\mathrm{IS}}(m) &= \frac{1}{\sum_{n\in\mathcal{T}_m}\bar\omega_{n,m}^{2}},
\end{aligned}
\end{equation}
with $\bar\omega_{n,m} = \bar\omega_n/\bar\nu_m$ the weights \textit{renormalized} within each region. The global factor penalizes spatial concentration of the samples, while the local factors measure how \textit{degenerate} the weights are within each region. Martino et al.\ \cite{martino2016effective} provide an axiomatic treatment of the whole question, listing the symmetry, maximum, minimum, unicity, and stability conditions that an effective sample size ought to satisfy, and constructing entire families of discrepancy-based estimators that meet them. The ratio-of-variances idea also carries over to importance sampling \cite{agarwal2022principled}, giving $\ESS_{\mathrm{IS}} = N(\lvert\Sigma\rvert/\lvert\Omega\rvert)^{1/p}$, with explicit estimators of $\Sigma$ and $\Omega$ for both the \textit{unnormalized} and the \textit{self-normalized} case and a corresponding stopping rule.

\subsubsection{The ESS for Joint MCMC \& Importance Sampling}\label{sec:ess-mcis}

Since modified HMC is simultaneously a Markov Chain and an importance sampler, neither family of estimators applies to it on its own. Two rather different attempts have therefore been made to combine their respective corrections.

\medskip\noindent\textbf{Thinned summation of weights.} The approach adopted by Radivojević \cite{thesis_tijana,radivojevic2020modified} sets $M = \ESS_{\mathrm{MCMC}}$, thins the chain by keeping every $\lfloor N/M \rfloor$-th sample, and applies \eqref{eq:ess-is} to the thinned subsample,
\begin{equation}\label{eq:ess-thinned}
  \ESS_{\mathrm{MCIS}} \;=\; \frac{\left(\sum_{n=1}^{M}\omega_n\right)^2}
                                  {\sum_{n=1}^{M}\omega_n^{2}} .
\end{equation}
In this way, the importance sampling formula is applied to the estimated number of uncorrelated draws, but the approach has three practical limitations. First, the integer thinning interval stays equal to one unless $\ESS_{\mathrm{MCMC}}<N/2$, so thinning has no effect for a well-mixing chain. Second, the expression is undefined when $M>N$, even though irreversible samplers such as MMHMC can legitimately reach an effective sample size above the nominal number of draws. Third, the procedure discards samples that have already been obtained at the cost of gradient evaluations.

\medskip\noindent\textbf{Correction as a product of variances.} The alternative avoids all three limitations by factorizing \eqref{eq:ess-definition} according to its two sources of variance, namely the Markov Chain that generates draws from $\pimod$ and the importance weights that map $\pimod$ onto $\pi$,
\begin{equation}\label{eq:ess-mcis-derivation}
  \ESS(\widetilde{I}) \;=\; N\,
  \underbrace{\frac{\Var_\pi(\bar{I})}{\Var_{\pimod}(\widehat{I})}}_{\text{MCMC}}\;
  \underbrace{\frac{\Var_{\pimod}(\widehat{I})}{\Var_{q}(\widetilde{I})}}_{\text{IS}},
\end{equation}
which yields the estimator
\begin{equation}\label{eq:ess-mcis}
  \boxed{\;\ESS_{\mathrm{MCIS}} \;=\; \frac{\ESS_{\mathrm{IS}} \cdot \ESS_{\mathrm{MCMC}}}{N}\;}
\end{equation}
Because the correction is applied only at the end, this estimator does not depend on thinning, remains defined for super-efficient chains, and uses every sample. The factorization \eqref{eq:ess-mcis-derivation} is \textit{heuristic}, however, as it treats the two sources of variance as if they acted independently. Whether \eqref{eq:ess-mcis} consequently counts the same loss twice when the weights are dispersed \emph{and} the chain mixes poorly remains an open question, to which Section~\ref{sec:challenges} returns. A second open question concerns which $\ESS_{\mathrm{IS}}$ to use in \eqref{eq:ess-mcis}. The sample-aware estimator \eqref{eq:ess-sample-aware} is the natural candidate, since weight degeneracy in a reweighted chain is typically localized in space, but it requires a partition of the sample space, and no principled rule for choosing one exists. Table~\ref{tab:ess-admissibility} summarizes the resulting recommendations, organized by the two properties of the sampler that determine which estimators are admissible.

\begin{table*}[t]
\centering
\caption{Admissibility of ESS estimators. ``Reversible'' refers to the Markov Chain and ``reweighted'' to the presence of importance weights. A dash indicates that the estimator is defined but that its supporting assumptions are violated.}
\label{tab:ess-admissibility}
\renewcommand{\arraystretch}{1.25}
\small
\begin{tabular}{lccccccc}
\toprule
\multirow{2}{*}{\textbf{Sampler}} & \multicolumn{2}{c}{\textbf{Properties}} &
\multicolumn{4}{c}{\textbf{Admissible ESS Estimator}} \\
\cmidrule(lr){2-3}\cmidrule(lr){4-7}
 & Reversible & Reweighted & Geyer & Spectral & Vats & MCIS \\
\midrule
RWMH, MALA        & Yes & No  & \checkmark & \checkmark & \checkmark & -- \\
HMC, NUTS         & Yes & No  & \checkmark & \checkmark & \checkmark & -- \\
GHMC              & No  & No  & --        & \checkmark & \checkmark & -- \\
MDMC, SLDMC       & No  & No  & --        & \checkmark & \checkmark & -- \\
Plain Importance Sampling (IS) & N/A & Yes & --        & --        & --        & \eqref{eq:ess-is} \\
GSHMC, MMHMC      & No  & Yes & --        & --        & --        & \checkmark \\
\bottomrule
\end{tabular}
\end{table*}

\subsection{Derived \& Cost-Aware Metrics}

Several quantities derived from the effective sample size are useful when the aim is not only to diagnose a chain but also to compare the statistical or computational efficiency of competing samplers. The \emph{Monte Carlo standard error} (MCSE) of an estimator $\widetilde{I}$ is the standard deviation of its sampling distribution, which may be estimated from the effective sample size as
\begin{equation}\label{eq:mcse}
  \MCSE \;=\; \sqrt{\sigma^{2}/\ESS},
\end{equation}
where $\sigma^2$ is the posterior variance of the quantity being estimated. Reported alongside a posterior summary, it expresses the simulation uncertainty that remains in that quantity. The \emph{integrated autocorrelation time} (IACT),
\begin{equation}\label{eq:iact}
  \IACT \;=\; \frac{N}{\ESS},
\end{equation}
offers a more intuitive interpretation as the average number of iterations required for an effectively independent draw \cite{sokal1997monte}. To compare computational efficiency rather than purely statistical efficiency, one may instead use a measure of sampling cost, such as the number of \emph{gradient evaluations per effective sample} ($\GRADe$),
\begin{equation}\label{eq:grad-e}
  \GRADe \;=\; \frac{N \cdot k \cdot \bar{L}}{\ESS},
\end{equation}
in which $k$ is the number of integrator stages and $\bar L$ the mean trajectory length, so that $N k \bar L$ is the total number of gradient evaluations. Its reciprocal, the effective sample size per gradient evaluation $\ESS/(N k\bar L)$, is the quantity reported in the adaptive integration literature \cite{nagar2024adaptive,akhmatskaya2026adaptive}. Either form supports the comparison \emph{across integrators} described in Remark~\ref{rem:fair}. A three-stage scheme that triples the effective sample size while also tripling the gradient cost, for example, leaves $\GRADe$ unchanged. This makes $\GRADe$ a useful metric whenever gradients dominate the runtime, although wall-clock time per effective sample is preferable when the cost of a single gradient differs between methods. Equation~\eqref{eq:grad-e} is, however, a sequential measure of efficiency, as it assumes that gradients are always evaluated one after another. Section \ref{sec:parallelism} revisits its interpretation under the paradigm of parallel execution. Table \ref{tab:metrics-summary} summarizes this measure together with the remaining diagnostics introduced above.

\begin{table}[t]
\centering
\caption{Summary of diagnostics and efficiency metrics.}
\label{tab:metrics-summary}
\renewcommand{\arraystretch}{1.25}
\small
\resizebox{\columnwidth}{!}{%
\begin{tabular}{llc}
\toprule
\textbf{Metric} & \textbf{Assesses} & \textbf{Eq./Ref.} \\
\midrule
Acceptance Rate          & Space exploration        & \eqref{eq:hmc-accept} \\
$\Rhat$ (rank-normalized)   & Convergence              & \eqref{eq:psrf}, \cite{vehtari2021rank} \\
Nested $\Rhat$              & Convergence, many chains & \cite{margossian2024nested} \\
Divergences                 & Unresolved curvature     & \cite{betancourt2016diagnosing, betancourt2017conceptual} \\
E-BFMI                      & Momentum refreshment     & \cite{betancourt2016diagnosing, betancourt2017conceptual} \\
SBC                         & Algorithm and model      & \cite{talts2018validating} \\
$\ESS$                      & Sampling efficiency      & \eqref{eq:ess-mcmc} -- \eqref{eq:ess-mcis} \\
$\MCSE$                     & Estimation precision     & \eqref{eq:mcse} \\
$\IACT$                     & Sampling efficiency      & \eqref{eq:iact} \\
$\GRADe$                    & Computational efficiency & \eqref{eq:grad-e} \\
\bottomrule
\end{tabular}}
\end{table}

\section{Parallelism \& Modern Hardware}\label{sec:parallelism}

The measures of computational cost and efficiency used thus far are intrinsically \textit{sequential}. In particular, the $\GRADe$ metric in \eqref{eq:grad-e} counts gradient evaluations as though they were necessarily performed one after another, an assumption well suited to the analyses of Sections~\ref{sec:integrators} and~\ref{sec:tuning} but no longer inevitable on modern hardware accelerators such as graphics or tensor processors, where independent evaluations may proceed simultaneously. A sampler requiring twice as many gradients can, consequently, be faster in wall-clock time if it exposes enough parallel work, so rankings established from sequential cost may not carry over to multi-process or multi-core hardware. ChEES-HMC and MEADS, discussed in Section~\ref{sec:tuning}, were designed with this setting in mind.

Although the transition from $\vth_n$ to $\vth_{n+1}$ cannot itself begin before $\vth_n$ is known, parallel work may be exposed across independent or coupled chains, within a transition through multiple proposals, within an HMC trajectory, or during evaluation of the target and its gradient. These possibilities differ not only in cost, and expected benefit, but also in whether they preserve the exact invariant distribution, and are therefore summarized in Table~\ref{tab:parallel}. The quantum methods of Section~\ref{sec:quantum} further accelerate proposal selection within this classification.

\begin{table}[t]
\centering
\caption{Axes of parallelism in Markov Chain Monte Carlo. ``Exact'' indicates whether the construction leaves the invariant distribution of the sampler unchanged.}
\label{tab:parallel}
\renewcommand{\arraystretch}{1.25}
\small
\resizebox{\columnwidth}{!}{%
\begin{tabular}{lllll}
\toprule
\textbf{Axis} & \textbf{Replicated Unit} & \textbf{Buys} & \textbf{Costs} & \textbf{Exact} \\
\midrule
Across chains        & The chain           & Variance, $1/m$        & Warm-up, $\times m$        & \checkmark \\
Coupled chains       & A pair of chains    & Removal of the bias    & Meeting time               & \checkmark \\
Multiple proposals   & The proposal        & Autocorrelation        & $O(P)$ targets per step    & \checkmark \\
Parallel-in-time$^{*}$ & The trajectory      & Depth, $L \to O(\sqrt{D})$ & Processors, iterations & \checkmark \\
Data sharding        & The gradient        & Wall-clock, exactly    & Communication              & \checkmark \\
Consensus / subposteriors & The posterior  & Wall-clock             & Recombination error        & -- \\
Quantum multiproposal & The selection step & $O(P) \to O(\sqrt{P})$ & Loading, success $(1-\epsilon)$ & -- \\
\bottomrule
\end{tabular}}
\\[3pt]
\begin{minipage}{0.97\columnwidth}
\scriptsize $^{*}$ Established for gradient-free Metropolis chains \cite{grazzi2025picard}. Its extension to Hamiltonian trajectories remains open (Section~\ref{sec:challenges}).
\end{minipage}
\end{table}

\subsection{An Exchange of Variance for Bias}

The most common form of parallelism runs $m$ chains from dispersed initial states and pools the draws they produce, thus reducing the variance of a converged estimator by a factor of $m$ and its standard error by $\sqrt{m}$. Since the chains require no communication while they run, the calculation is \textit{embarrassingly parallel}\footnote{An \textit{embarrassingly parallel computation} can be divided into independent tasks that require no communication or coordination while they run.}, but this independence also duplicates warm-up: whereas a single chain retaining $N$ draws after $b$ discarded iterations expends a fraction $b/(N+b)$ of its work on warm-up, $m$ such chains collectively spend the corresponding multiple $mb/(N+b)$.

Duplicating warm-up affects not only the computational cost but also the \textit{bias} of the estimator. Increasing the number of chains $m$ reduces only the \emph{variance} of the pooled estimate, whereas the initialization bias that remains is set by the sequential length of each chain, since its retained draws follow the law of the chain after $b$ steps rather than $\pi$. At fixed $N$, increasing $m$ can therefore make the Monte Carlo error arbitrarily small without removing initialization bias, a trade-off made especially prominent by accelerators that render many concurrent chains inexpensive while leaving the length of each chain unchanged.

Recent work addresses different aspects and consequences of this problem. The \textit{nested} $\Rhat$ \cite{margossian2024nested}, introduced in Section~\ref{sec:diagnostics}, remains informative for short chains in regimes where the classical statistic may not due to the bias of the within-chain variance estimator, and the globally centered autocovariances of Agarwal and Vats \cite{agarwal2022globally} reduce the bias of autocorrelation-based estimators when many short chains are pooled. Margossian and Gelman \cite{margossian2024forhowmany} study the allocation of a fixed budget between $N$ and $m$. Brown \cite{brown2024nonasymptotic} further derives a non-asymptotic error bound for an estimator based on many short chains under a drift condition and an importance sampling proposal. Pathfinder \cite{zhang2022pathfinder} reduces $b$ directly by replacing stochastic warm-up with variational initialization.

\medskip\noindent\textbf{Removing bias by coupling.} Jacob, O'Leary and Atchadé \cite{jacob2020unbiased} couple two chains until they meet almost surely and apply the telescoping argument of Glynn and Rhee \cite{glynn2014exact} to their difference, obtaining an exactly unbiased estimator after a random but almost surely finite meeting time, independently replicated versions of which may then be averaged across processors. Heng and Jacob \cite{heng2019unbiased} construct corresponding couplings for HMC by sharing the random numbers used for the momenta and introducing a contractive refreshment coupling. Although the approach is not yet implemented in the mainstream packages listed in Table~\ref{tab:software}, its more fundamental limitation is that the meeting time continues to depend upon the mixing of the underlying kernel, so the same geometries that prolong warm-up may also delay coupling.

\subsection{Within the Chain: Multiple Proposals}

A second, and quite different tradition, parallelizes the \textit{state transition} itself by generating multiple proposals at each step and selecting among them. Tjelmeland \cite{tjelmeland2004using} showed how to augment the current state with $P$ proposals drawn jointly from a distribution conditioned on it, forming a pool $\Theta = (\vth^{(1)}, \dots, \vth^{(P+1)})$ in which the current state occupies one position, and select the next state from this pool with probabilities
\begin{equation}\label{eq:multiproposal}
  \resizebox{\dimexpr\columnwidth-3.5em\relax}{!}{$\displaystyle
  \pi_p \;=\; \frac{\pi(\vth^{(p)})\, q(\Theta_{\setminus p} \mid \vth^{(p)})}{\sum_{j=1}^{P+1} \pi(\vth^{(j)})\, q(\Theta_{\setminus j} \mid \vth^{(j)})}, \quad p = 1, \dots, P+1,$}
\end{equation}
where $\Theta_{\setminus p}$ denotes the pool with $\vth^{(p)}$ removed and $q(\,\cdot \mid \vth^{(p)})$ is the density of the remaining $P$ points had they been proposed from $\vth^{(p)}$. This selection preserves detailed balance for every $P$. Calderhead \cite{calderhead2014general} gave a general construction of the same kind for Metropolis--Hastings, and Glatt-Holtz et al.\ \cite{glatt2024parallel} placed the accumulated variants within a single theoretic framework on an extended phase space, covering Metropolis-type and Barker-type acceptance criteria alike and clarifying which joint proposal structures admit the $O(P)$ instead of the $O(P^2)$ evaluation of the denominator in \eqref{eq:multiproposal}. Most recently, the same authors \cite{glatt2026mad} studied the kernel obtained in the limit $P \to \infty$ and use it to discriminate between schemes: if this limiting kernel is degenerate, additional proposals bring little further improvement, and the extra computational resources they occupy are largely wasted.

The benefit of multiple proposals is thus a lower autocorrelation between successive states and, for multimodal targets, a better chance of moving between modes whenever some of the proposals land beyond the low-density region that separates them. This benefit is paid for with $P$ target evaluations at every transition, on processors that could otherwise run additional independent chains. Whether the hardware is better spent on more proposals or on more chains therefore depends on the problem and on the geometry of the target.

Multiproposal MCMC has seen limited use with HMC because its proposals must be conditionally independent given the current state to be generated \textit{concurrently}. The states of an HMC trajectory instead form a recursion, with each state requiring the gradient at its predecessor. Multinomial NUTS \cite{betancourt2017conceptual, hoffman2014no} selects among trajectory states, and GIST \cite{bourabee2024gist} represents this selection as a Gibbs step, but those candidates are produced sequentially at a serial depth of $L$ gradient evaluations. A parallel proposal scheme for HMC would thus require $P$ independent trajectories and $P$ times the gradient cost. It is therefore a different use of computational budget from a single coherent Hamiltonian trajectory.

\subsection{Within the Trajectory: Parallel-in-Time}

Since the construction of $\vz_{n+1}$ requires $\nabla U(\vth_n)$, the leapfrog recursion has a serial depth of $L$, and a trajectory must nevertheless be long enough to traverse an appreciable portion of the typical set. Thus, the mechanism that gives HMC its coherent exploration is also the principal obstruction to parallelism within a trajectory. Grazzi and Zanella \cite{grazzi2025picard} circumvent an obstruction of this kind by recasting a sequence of states as the fixed point of a Picard map, which allows all of them to be updated simultaneously and, for log-concave targets, converges in $O(\sqrt{D})$ parallel iterations using $O(\sqrt{D})$ processors. Although their construction is developed for gradient-free Metropolis chains, the leapfrog update is itself a recursion of the same kind and may therefore admit a related treatment. Such a reduction cannot, however, be pushed arbitrarily far, since the lower bounds of Zhou et al.\ \cite{zhou2024adaptive} show that, for log-concave targets, depth may be reduced but not eliminated, with nearly linearly many adaptive rounds still required.

\medskip\noindent\textbf{Parallelizing the target.} For a model with $n$ conditionally independent observations, $U(\vth) = \sum_{i=1}^{n} U_i(\vth)$, and its gradient parallelizes exactly without changing the invariant distribution. Holbrook et al.\ \cite{holbrook2021massive} report speedups of more than two orders of magnitude for HMC from this strategy. In the joint models of Section~\ref{sec:hierarchical}, the dominant cost is the per-subject quadrature for the cumulative hazard, so it can be distributed over independent subjects. This approach differs from \textit{divide-and-conquer} schemes such as consensus Monte Carlo \cite{scott2016bayes}, embarrassingly parallel MCMC \cite{neiswanger2014asymptotically}, and Gaussian process subposterior merging \cite{nemeth2018merging}, which shard and approximately recombine the \emph{posterior}. For these models, parallelizing the gradient retains exactness and is a more direct route to speedup.

\subsection{Quantum Parallelization}\label{sec:quantum}

Holbrook \cite{holbrook2023quantum} identifies the selection step \eqref{eq:multiproposal} as the rate-limiting operation of a multiproposal scheme and notes that the Gumbel-max trick turns it into a discrete optimization problem: drawing $\hat p \sim \mathrm{Discrete}(\pi_1, \dots, \pi_{P+1})$ is equivalent to computing $\argmax_p \{\log \pi_p + G_p\}$ for independent standard Gumbel variates $G_p$, in which the normalizing constant making \eqref{eq:multiproposal} computationally expensive cancels. Grover search and the Dürr--H\o{}yer quantum minimization algorithm solve such a problem over $(P+1)$ items in $O(\sqrt{P})$ oracle calls rather than $O(P)$, and the current state of the chain provides a warm start that further improves the constant. The number of target evaluations per transition thus falls from $O(P)$ to $O(\sqrt{P})$, or from $O(P^2)$ to $O(\sqrt{P})$ once the pairwise terms of the Tjelmeland correction are taken into account. Lin et al.\ \cite{lin2025quantum} address the remaining $O(P)$ classical cost of \emph{generating} the proposals by adopting a Tjelmeland distribution that concentrates them near the current state, and report $O(1)$ target evaluations using $O(\log P)$ qubits. A separate line of work, in which a quantum device supplies the proposal and the acceptance test remains classical \cite{layden2023quantum}, reports polynomial speedups for Boltzmann distributions of Ising models, although Orfi and Sels \cite{orfi2024bounding} show that the spectral gap of such samplers is bounded, and that for unital proposals, including the unitary dynamics used in practice, no speedup over classical sampling is possible on worst-case problems without exploitable structure.

This approach nevertheless faces three important limitations. For a continuous target, every transition must load $O(PDb)$ classical bits, where $b$ denotes the floating-point width, and logarithmic loading depth does not eliminate that requirement. Moreover, quantum minimization succeeds only with probability $(1-\epsilon)$, so that the implemented kernel differs from the intended one and, although the resulting bias appears empirically small, no available bound relates $\epsilon$ to the distance between $\pi$ and the invariant distribution actually obtained. Most importantly for the present survey, the method accelerates the selection among proposals only after those proposals have been evaluated, whereas the gradient-based samplers considered here spend most of their time generating them, and its immediate relevance to Hamiltonian sampling is therefore limited. The Gumbel-max reduction is also useful independently of quantum hardware, since its representation of the acceptance step as a discrete optimization problem applies equally to algorithms implemented on \textit{classical} processors.

\subsection{Implications for the Surveyed Methods}

For the methods considered in this survey, the previous discussion has several practical implications.

\medskip\noindent\textbf{Control flow on accelerators.} On single-instruction multiple-data (SIMD) hardware, all chains advance in lockstep, so that the cost of an iteration is set by the longest trajectory, $\max_m L_m$, rather than by the mean $\bar L$. Because the recursive doubling of NUTS lets each chain choose its own trajectory length $2^{j_m}$, much of the hardware may sit idle while the longest trajectory completes. ChEES-HMC \cite{hoffman2021adaptive} avoids the recursive U-turn tree altogether, MEADS \cite{sountsov2022meads} adapts its parameters from an ensemble of chains, and microcanonical samplers \cite{robnik2023microcanonical,robnik2025metropolis} use a fixed number of steps per trajectory. Since gradient cost and wall-clock time on an accelerator do not necessarily rank these methods in the same order, both $\GRADe$ and the wall-clock time on clearly specified hardware should thus be reported as standard practice.

\medskip\noindent\textbf{GHMC and vectorization.} GHMC and its descendants are particularly well suited to vectorized execution: their trajectories have a fixed and often short length, they require no recursive tree construction, the partial momentum update \eqref{eq:pmu} is affine, and the multi-stage integrators of Section~\ref{sec:integrators} are fixed compositions of simple maps. Every chain therefore follows the same control flow during a GHMC or MMHMC transition and, since the reweighting of Section~\ref{sec:mhmc} acts elementwise, these methods can thus be vectorized across chains with little effort. \code{pyHaiCS} \cite{pyhaics}, a Python and JAX reimplementation of the \code{HaiCS} package in which GHMC and MMHMC have so far been available (Section~\ref{sec:software}), follows this route to run them on accelerators.

\medskip\noindent\textbf{Importance weights and many chains.} The combined effective sample size \eqref{eq:ess-mcis} is defined for a single reweighted chain, but when $m$ chains are run the importance weights may be normalized either separately within each chain or jointly across all pooled draws. Normalizing within each chain yields $m$ self-normalized estimators, each with a bias of order $1/N$ that may be appreciable when the chains are short. Normalizing across the pool reduces this bias, but couples the chains and thereby breaks the independence on which both the $\sqrt{m}$ variance argument and the nested $\Rhat$ rely. An extension of the combined estimator that accounts for this setting has yet to be developed.

\section{Hierarchical Models in Practice}\label{sec:hierarchical}

After the preceding discussions on the algorithmic samplers, integrators, and sampling diagnostics of Sections~\ref{sec:ghmc} through~\ref{sec:diagnostics}, we now return to the motivating example of hierarchical models. The hierarchical structure of these models often leads to challenging posterior geometries, such as funnels, that can impede efficient sampling. In this section, we explore practical strategies for addressing these challenges, including reparameterization techniques, marginalization approaches, and considerations specific to joint models for longitudinal and time-to-event data.

\subsection{Reparameterization \& Why It Is Not Sufficient}

The usual first response to a \textit{funnel} is to replace the centered parameterization of \eqref{eq:hier} with the non-centered version in \eqref{eq:ncp}. Papaspiliopoulos, Roberts and Sköld \cite{papaspiliopoulos2007general} showed, however, that these two are not the only options but rather endpoints of a continuum of \emph{partially non-centered} parameterizations. In the explicit per-variable form used by Gorinova, Moore and Hoffman \cite{gorinova2020automatic}, the Gaussian hierarchy of \eqref{eq:hier} becomes
\begin{equation}\label{eq:pncp}
  \theta_j^{(w)} \;=\; w_j\,\mu \;+\; \tau^{\,w_j-1}\left(\theta_j - \mu\right),
  \qquad w_j \in [0,1],
\end{equation}
so that $\theta_j^{(w)} \mid \varphi \sim \mathcal{N}(w_j\mu,\, \tau^{2w_j})$, with $w_j = 1$ recovering the centered coordinate $\theta_j$ and $w_j = 0$ the non-centered coordinate $\tilde\theta_j$ of \eqref{eq:ncp}. The efficiency of a Gibbs-type sampler depends considerably on where each $w_j$ is placed between these two extremes \cite{papaspiliopoulos2007general}.

Moreover, Betancourt and Girolami \cite{betancourt2015hamiltonian} provide a geometric interpretation of this result in terms of the \emph{relative contribution of the prior and the likelihood} to the posterior of each group $j$. When the likelihood dominates, as it may for a group contributing many informative observations, the centered parameterization is often preferable, whereas for a data poor group dominated by the prior, non-centering frequently improves the conditioning by removing the prior dependence on $\tau$. In their analysis, non-centering is effective precisely when it removes a dependence between layers that the data have not already broken, and ineffective when the data have already broken it. They also add an important caveat: models for which the centered parameterization is clearly preferable often contain enough data per group that the partial pooling of a hierarchical model is hardly needed in the first place.

Real hierarchical models, however, seldom exhibit a uniform balance between prior and likelihood. Some patients in a clinical study may contribute dozens of measurements and others only a few, just as site sizes may vary widely in a multi-center trial. The likelihood may therefore dominate for some groups and the prior for others, so that no single parameterization is ideal for all of them. This motivates either a hand-crafted mixture of centered and non-centered coordinates, the partially non-centered form in \eqref{eq:pncp} with tuned weights $w_j$, or a learned interpolation such as the variationally inferred parameterization of Gorinova, Moore and Hoffman \cite{gorinova2020automatic}.

\subsection{Marginalization \& Hybrid Schemes}

Marginalization provides a different way of removing the source of the difficult geometry. Whenever the \textit{latent} parameters enter the model as conditionally Gaussian variables, they may be integrated out analytically, leaving HMC to target only the hyperparameters. INLA \cite{rue2009approximate} carries out this integration deterministically for latent Gaussian models and has been extended specifically to joint models \cite{vanniekerk2023joint}. Margossian et al. \cite{margossian2020hamiltonian} instead embed an adjoint-differentiated Laplace approximation within HMC, so that the approximate marginal density and its gradient remain inexpensive to evaluate. HMC then samples the hyperparameters with respect to this approximate marginal, while the latent field itself is only approximated.

For non-Gaussian latent structures, a class that unfortunately includes most survival submodels, exact marginalization is not available, and the resulting approximation error must be assessed by other means, typically through simulation-based calibration. For joint models specifically, we are aware of no such assessment, which makes this an obvious target for future work.

\subsection{Joint Models for Longitudinal \& Time-to-Event Data}

A joint model \cite{rizopoulos2012joint,alsefri2020bayesian} links a longitudinal submodel for a repeatedly measured biomarker to a survival submodel for an event time through shared random effects. In the shared-parameter formulation, for subject $i = 1, \dots, n$,
\begin{align}
  y_i(t) &= m_i(t) + \epsilon_i(t), \nonumber\\
  m_i(t) &= \vx_i(t)^\top\beta + \vz_i(t)^\top \mathbf{b}_i, \nonumber\\
  \mathbf{b}_i &\sim \mathcal{N}(\vzero, \Sigma_b), \label{eq:jm-long}\\
  h_i(t) &= h_0(t)\exp\!\left\{\gamma^\top \mathbf{w}_i + \alpha\, m_i(t)\right\}\!,
  \label{eq:jm-surv}
\end{align}
in which $\mathbf{b}_i$ are subject-specific random effects, $h_0$ is a baseline hazard, and $\alpha$ couples the two submodels. The posterior over $(\beta, \Sigma_b, \gamma, \alpha, \{\mathbf{b}_i\})$ and the baseline-hazard parameters, is hierarchical in the sense of Section~\ref{sec:hier-pathology}, and three of its features are particularly important from a computational point of view.

\medskip\noindent\textbf{Multiple funnels.} Each variance component in $\Sigma_b$ forms a funnel with the corresponding random-effect coordinates, so that a model with $q$ random effects per subject contains $q$ such structures, which are further coupled through the off-diagonal entries when $\Sigma_b$ is unstructured. Non-centering is standard practice here, but by the argument above it suits only subjects with few measurements, whereas longitudinal studies may contain subjects whose numbers of measurements vary widely.

\medskip\noindent\textbf{Expensive gradients.} The survival likelihood depends on the cumulative hazard $\int_0^{T_i} h_i(s)\,ds$, which has no closed form when $m_i(t)$ varies with time and must therefore be approximated by numerical methods, typically Gauss--Kronrod or Gauss--Legendre quadratures, ordinarily with seven to fifteen nodes for every subject at every likelihood evaluation. Since each node requires a fresh evaluation of the longitudinal trajectory and of its derivative with respect to every parameter involved, the resulting gradients may be one or two orders of magnitude more expensive than those of a comparable generalized linear mixed model. The cost per gradient evaluation is consequently far more informative here than the ESS per iteration alone.

\medskip\noindent\textbf{Dimension growing with the cohort.} The latent dimension is $nq$, so that a cohort of one thousand subjects with a random intercept and slope already implies two thousand latent parameters before any fixed effects are counted. At this scale, dense mass matrices often become impractical, and the $\mathcal{O}(D^3)$ metric operations of naive RMHMC implementations can become prohibitive. The resulting need for diagonal or low-rank approximations to the mass matrix, however, may further exacerbate the funneling problem.

\medskip
In practice, NUTS with a non-centered parameterization, as implemented in \code{stan\_jm} \cite{brilleman2020joint}, remains a sensible default, although its cost varies from one iteration to the next and may become substantial when long trajectories are built in flatter regions of the posterior. A GHMC sampler with short randomized trajectories and an ATune-family tuning scheme \cite{akhmatskaya2026adaptive} offers a more predictable alternative, in which persistent momentum may preserve coherent movement across iterations and modified Hamiltonians may further improve acceptance, provided that the importance weights remain well behaved. Whether this combination actually outperforms NUTS on realistic joint models has yet to be established. A credible comparison will require an ESS estimator suited to a sampler that is both irreversible and reweighted, a setting in which the standard Geyer-based estimator reported by \code{Stan} is not justified.

\subsection{A Practical Checklist}

Previous sections have discussed the challenges of Hamiltonian sampling in the context of hierarchical models and the methods available to address them. We conclude with four recommendations for applying the methods discussed here to hierarchical models, which we summarize below as a checklist.

\medskip\noindent\textit{Parameterization.} Begin with a non-centered parameterization for data-poor groups, but examine the balance between prior and likelihood across the hierarchy. Where this balance differs appreciably between groups, consider a mixed or learned parameterization \cite{gorinova2020automatic}.

\medskip\noindent\textit{Diagnosis.} Inspect divergences, trace plots, E-BFMI, $\Rhat$ and ESS together, since each detects a different mode of failure in the sampling process. Report the rank-normalized $\Rhat$ against the $1.01$ threshold \cite{vehtari2021rank}, and accompany every posterior summary with its Monte Carlo standard error.

\medskip\noindent\textit{Comparison.} Compare samplers by gradient evaluations or wall-clock time per effective draw rather than by ESS per iteration, and state which ESS estimator was used and why its assumptions hold for the transitions in question.

\medskip\noindent\textit{Validation.} Check every new model--algorithm pairing by simulation-based calibration \cite{talts2018validating} before trusting its output. Trace plots alone are insufficient and a single successful run does not guarantee correctness and robustness.

\section{Software}\label{sec:software}

The methods surveyed here differ considerably in their practical availability. The distance between an algorithm described in a paper and one supported by a maintained, well-tested implementation determines which methods reach applied work at all. Table~\ref{tab:software} summarizes where that line currently falls. The entries reflect the state of these packages in September 2026 and will date quickly.

The \code{Stan} ecosystem \cite{carpenter2017stan} is the reference implementation for most users: NUTS with dual averaging, adaptation of either a diagonal or dense mass matrix, divergence and E-BFMI diagnostics, and rank-normalized $\Rhat$ together with its associated ESS estimators through the \code{posterior} and \code{ArviZ} packages \cite{kumar2019arviz,vehtari2021rank}. Those ESS estimators are Geyer's, with the consequences discussed in Section~\ref{sec:diagnostics}. For joint models, this sampling machinery is exposed directly by \code{stan\_jm} in \code{rstanarm} \cite{brilleman2020joint}, whereas \code{JMbayes2} \cite{rizopoulos2024jmbayes2} relies upon a bespoke C++ sampler with problem-specific blocking that may be faster for the models it was designed to fit, but is correspondingly less transferable to other hierarchical structures.

JAX-based libraries have become a particularly common setting for methodological development because they combine automatic differentiation and execution on hardware accelerators with readily vectorized transition kernels. \code{NumPyro} \cite{phan2019composable} and \code{BlackJAX} \cite{cabezas2024blackjax} both provide vectorized NUTS, but their scope beyond it differs. \code{NumPyro} concentrates on variants of HMC suited to particular model classes, among them energy-conserving subsampling and Gibbs updates for discrete latent variables. \code{BlackJAX} is the broader of the two for the methods surveyed here, offering GHMC, MEADS adaptation, MCLMC and its Metropolis-adjusted variant, RMHMC, ChEES adaptation, and the two- and three-stage splitting integrators of Section~\ref{sec:integrators}. Lao et al.\ \cite{lao2020tfp} set out the design considerations behind this class of library, whose statistical implications were examined in Section~\ref{sec:parallelism}. \code{TensorFlow Probability} \cite{dillon2017tensorflow} includes ChEES-HMC and several of the ESS estimators of Section~\ref{sec:diagnostics}, while \code{PyMC} \cite{abril2023pymc} sits between these ecosystems in interface and scope, providing HMC and NUTS but none of the generalized or modified variants.

The modified Hamiltonian and adaptive integration branch of the literature has been, however, available almost exclusively through the in-house \code{HaiCS} package \cite{thesis_tijana,akhmatskaya2026adaptive}, which implements GHMC, MMHMC, multi-stage splitting integrators, and the s-AIA and ATune tuning procedures. This narrow implementation base has probably constrained the use of these methods outside the groups that developed them, and provides the principal motivation for our Python and JAX port \code{pyHaiCS} \cite{pyhaics}.

\begin{table}[t]
\centering
\caption{Availability of the surveyed methods. A check mark denotes a first-class implementation, an open circle denotes availability through our pyHaiCS extension, and a blank entry denotes unavailability.}
\label{tab:software}
\renewcommand{\arraystretch}{1.2}
\small
\setlength{\tabcolsep}{4pt}
\begin{tabular}{lccccc}
\toprule
& \rotatebox{60}{\code{Stan}}
& \rotatebox{60}{\code{PyMC}}
& \rotatebox{60}{\code{NumPyro}}
& \rotatebox{60}{\code{BlackJAX}}
& \rotatebox{60}{\code{HaiCS}} \\
\midrule
HMC / NUTS            & \checkmark & \checkmark & \checkmark & \checkmark & \checkmark \\
GHMC                  &            &            &            & \checkmark & \checkmark \\
MMHMC                 &            &            &            &            & \checkmark \\
RMHMC                 &            &            &            & \checkmark & $\circ$           \\
MCLMC                 &            &            &            & \checkmark &            \\
MEADS                 &            &            &            & \checkmark &            \\
ChEES-HMC             &            &            &            & \checkmark &            \\
Multi-stage integrators &          &            &            & \checkmark & \checkmark \\
s-AIA / ATune         &            &            &            &            & \checkmark \\
Rank-normalized $\Rhat$ & \checkmark & \checkmark & \checkmark & \checkmark & \checkmark \\
Divergences / E-BFMI  & \checkmark & \checkmark & \checkmark & \checkmark &            \\
$\GRADe$              &            &            &            &            & \checkmark \\
Combined MCIS ESS     &            &            &            &            & $\circ$    \\
\bottomrule
\end{tabular}
\end{table}

Table \ref{tab:software} also makes two practical points about the current state of the field. RMHMC still lacks a general production quality implementation because each model requires its own metric and metric derivatives, objects that mainstream probabilistic programming systems do not presently supply automatically. At the same time, those systems report neither $\GRADe$ nor a combined effective sample size accounting for both Markov Chain and importance sampling variance together. The former restricts routine use of position dependent geometries, while the latter prevents standardized comparisons among integrators and complicates the evaluation of the reweighted importance samplers discussed in Section \ref{sec:mhmc}.

\section{Challenges \& Open Directions}\label{sec:challenges}

We finally conclude by distinguishing limitations that remain intrinsic to the available methods from those gaps that arise in their implementation, diagnosis, and empirical evaluation in current practice.

\subsection{Limitations of the Methods}

\noindent\textbf{Multimodality.} HMC struggles to cross regions of negligible density, since energy conservation confines each trajectory to a level set and leaves any mode behind a barrier higher than the available kinetic energy out of reach from the current state. Momentum resampling helps only when the energy distribution has sufficient spread, as measured by the E-BFMI (Section~\ref{sec:diagnostics}). The remaining options, namely parallel or continuous tempering \cite{graham2017continuously}, Wormhole HMC \cite{lan2014wormhole}, transport by normalizing flows \cite{hoffman2019neutra}, and hybrids with sequential Monte Carlo (SMC), all require either prior knowledge of the modal structure or a separate training stage. Because label switching and unidentified variance components are common sources of multimodality in hierarchical models, the lack of general gradient-based solutions remains a significant practical limitation.

\medskip\noindent\textbf{Mixed-regime hierarchical models.} When the balance between prior and likelihood varies across groups, neither the centered nor the non-centered parameterization is appropriate throughout. RMHMC can in principle accommodate this variation, but may be prohibitively expensive at the dimensions involved, whereas automatic reparameterization \cite{gorinova2020automatic} introduces an approximation of its own. A method that is simultaneously inexpensive, exact, and sufficiently general for routine use has yet to be developed.

\medskip\noindent\textbf{Discrete parameters.} Although discontinuous HMC \cite{nishimura2020discontinuous} extends the framework to discrete or discontinuous targets, there remains no automatic and general purpose solution comparable to that provided by ordinary HMC in continuous spaces, leaving latent-class joint models, mixture cure models, and change-point survival models beyond the reach of standard Hamiltonian tooling.

\medskip\noindent\textbf{Non-separable dynamics.} RMHMC relies upon an expensive implicit generalized leapfrog whose fixed-point iteration may fail to converge and thereby compromise both reversibility and correctness. Explicit alternatives based upon phase-space augmentation \cite{tao2016explicit,cobb2019introducing} or Lagrangian coordinates \cite{lan2015markov} avoid this particular difficulty, but their behavior on statistical targets is not yet characterized as thoroughly as that of the standard method.

\medskip\noindent\textbf{Modified Hamiltonians beyond Euclidean metrics.} MHMC changes the Hamiltonian used in the Metropolis correction, whereas RMHMC changes the underlying metric. Combining the two would require a modified Hamiltonian for a position-dependent metric, which does not appear to have been developed so far. Such a construction would nevertheless be attractive, as it could address integration error and geometric anisotropy within a single sampler.

\medskip\noindent\textbf{The cost of gradients.} When evaluating a likelihood requires additional quadratures, the solution of differential equations, or large scale linear algebra, its gradient may dominate the total runtime, and naive data subsampling does not remove this limitation for HMC \cite{betancourt2015fundamental}. For joint models in particular, the relevant expense lies in per-subject quadrature rather than simply in the number of observations. Improvements to the integrator, together with surrogate and multi-fidelity models, may therefore prove more valuable in this setting than changes to the transition kernel alone.

\subsection{Gaps in Practice}

\medskip\noindent\textbf{Effective sample size.} Section~\ref{sec:diagnostics} identified a methodological gap between the default estimators of widely used software, which generally assume \textit{reversibility}, and an important class of samplers that are irreversible by construction. No consensus estimator exists for draws that are both autocorrelated and importance reweighted, and the combined estimator \eqref{eq:ess-mcis} has yet to receive a complete theoretical analysis. Several questions therefore remain open: whether its product-of-variances correction counts the same loss twice when the weights are dispersed and mixing is poor, how the partition for the sample-aware $\ESS_{\mathrm{IS}}$ should be chosen, and whether Geyer's positivity argument admits an irreversible analog or must give way to batch-means estimators such as that of Vats et al.\ \cite{vats2019multivariate}.

\medskip\noindent\textbf{Autocorrelation estimators for irreversible chains.} Fang, Cao and Skeel \cite{fang2022quasi} show that generalized hybrid Monte Carlo belongs to a class of ``modified reversible'' propagators, each composed of a reversible substep and a measure preserving involution, whose transfer operators are self-adjoint. Since self-adjointness underlies Geyer's positivity and monotonicity results in the reversible case, the natural question is whether those results carry over to this extended class. If they do, the default estimators reported by mainstream software would be justified for GHMC after all, and much of the difficulty described above would disappear. The same authors also observe that the momentum flip defeats the very purpose of introducing momenta, which raises the complementary question of whether a GHMC variant can avoid the flip without losing the variance benefits of irreversibility.

\medskip\noindent\textbf{Initialization and warm-up.} The choice of initial state $(\vth_0,\vp_0)$ receives comparatively little attention, even though running $m$ chains reduces the Monte Carlo error only by a factor of $\sqrt{m}$, multiplies the warm-up overhead from $b/(N+b)$ to $mb/(N+b)$, and leaves the initialization bias untouched. Pathfinder \cite{zhang2022pathfinder} attempts to lighten this burden through variational initialization, and many short chain methods \cite{margossian2024nested,margossian2024forhowmany} revisit the allocation between chain length and the number of them. Initialization nevertheless remains consequential whenever accelerators make concurrent chains inexpensive without shortening the warm-up that each of them requires.

\medskip\noindent\textbf{Multi-chain reweighted samplers.} With $m$ reweighted chains, the importance weights may be normalized either within each chain or across the pooled draws. Normalizing within each chain yields $m$ self-normalized estimators, each with a bias of order $1/N$ that may be appreciable when the chains are short. Normalizing across the pool reduces this bias, but couples the chains and thereby breaks the independence on which both the $\sqrt{m}$ variance argument and the nested $\Rhat$ rely. How \eqref{eq:ess-mcis} should be extended to this setting has not yet been studied.

\medskip\noindent\textbf{Parallel-in-time dynamics.} The serial depth of a trajectory limits parallelization within HMC, although the Picard-map construction of Grazzi and Zanella \cite{grazzi2025picard} shows that, for log-concave targets, this depth can be reduced by a factor of $\sqrt{D}$. Their method was developed for gradient-free Metropolis chains, but the leapfrog update is also a recursion and may admit a comparable fixed-point representation. This raises the question of whether the multi-stage integrators of Section~\ref{sec:integrators} would retain their symplecticness, and with it the validity of the acceptance test, under such an iteration. Whatever the answer, the lower bounds of Zhou et al.\ \cite{zhou2024adaptive} limit the reduction that can ultimately be achieved.

\medskip\noindent\textbf{Two approaches to tuning.} GIST \cite{bourabee2024gist} samples the tuning variables conditionally on the current state, whereas the adaptive integration family \cite{akhmatskaya2026adaptive} derives them from an explicit surrogate model, and the two approaches have so far developed largely in isolation. However, a common framework relating GIST-style conditional distributions to the Gaussian analysis underlying $\rho_k(\varepsilon,b)$ and its associated stability interval would clarify whether their respective advantages can be combined.

\medskip\noindent\textbf{Benchmarking.} Published comparisons differ in their effective sample size estimators, their cost measures, their acceptance rate targets, and their tested models, and most are reported by the authors of the method being favoured. Since Section \ref{sec:diagnostics} shows that the choice of ESS estimator alone can change which sampler appears better, these comparisons cannot be meaningfully set against one another. A shared benchmark for hierarchical models should therefore report both $\GRADe$ and wall-clock cost, together with an ESS estimator whose assumptions hold for each of the samplers being compared.

\section{Conclusion}\label{sec:conclusion}

Hamiltonian Monte Carlo is most usefully viewed not as a single algorithm but as a family of related design choices, for which generalized HMC provides one convenient organizing framework: varying the trajectory length, the degree of momentum refreshment, and the presence of a Metropolis correction recovers several familiar samplers, with further extensions modifying the metric, the Hamiltonian used by that correction, or the numerical integrator itself. The associated tuning problem has a similarly broad structure, ranging from NUTS-style adaptation driven by the observed behavior of the chain to adaptive integration methods whose parameters are derived from a surrogate fitted during burn-in.

Hierarchical models bring several of these choices into play simultaneously, since funnel-like geometry can render a single global step size or metric inadequate while the latent dimension itself continues to grow with the data. For the joint models considered here, gradients may also be sufficiently expensive that evaluations per effective sample and wall-clock time convey more than ESS per iteration, a distinction that exposes a further weakness in current benchmarking: effective sample size has no universally valid estimator, and the defaults used in practice may be inappropriate for irreversible or reweighted chains. Any comparison involving such methods should consequently state both the ESS estimator and the computational cost measure explicitly, so that an apparent advantage can be interpreted in light of the assumptions on which it rests.

\begin{appendix}
\section{Modified Hamiltonians \& Integrator Coefficients}\label{app:modified}

For ease of reference, this appendix collects the expressions used in Sections~\ref{sec:mhmc} and~\ref{sec:integrators}, following the derivations of Radivojević et al.\ \cite{radivojevic2018multi,radivojevic2020modified} and Nagar et al.\ \cite{nagar2024adaptive}.

\subsection{Fourth- \& Sixth-Order Modified Hamiltonians}

For the symmetric splitting integrators of \eqref{eq:mssi}, the truncated modified Hamiltonian in the \emph{analytic} formulation is, to fourth order,
\begin{equation}\label{eq:mod4}
\begin{aligned}
  \Hmod^{[4]}(\vth, \vp) \;=\;& H(\vth, \vp) \\
  &+\; \varepsilon^{2}\, c_{21}\, \vp^\top M^{-1} U_{\vth\vth} M^{-1}\vp \\
  &+\; \varepsilon^{2}\, c_{22}\, U_{\vth}^\top M^{-1} U_{\vth},
\end{aligned}
\end{equation}
and, to sixth order,
\small
\begin{equation}\label{eq:mod6}
\begin{aligned}
  \Hmod^{[6]}(\vth, \vp) \;=\;& H(\vth, \vp) \\
  &+ \varepsilon^{2} c_{21}\, \vp^\top M^{-1} U_{\vth\vth} M^{-1}\vp \\
  &+ \varepsilon^{2} c_{22}\, U_{\vth}^\top M^{-1} U_{\vth} \\
  &+ \varepsilon^{4} c_{41}\, U_{\vth\vth\vth\vth}\!
       \left[M^{-1}\vp,\, M^{-1}\vp,\, M^{-1}\vp,\, M^{-1}\vp\right] \\
  &+ \varepsilon^{4} c_{42}\, U_{\vth}^\top M^{-1} U_{\vth\vth\vth}\!
       \left[M^{-1}\vp,\, M^{-1}\vp\right] \\
  &+ \varepsilon^{4} c_{43}\, U_{\vth}^\top M^{-1} U_{\vth\vth} M^{-1} U_{\vth} \\
  &+ \varepsilon^{4} c_{44}\, \vp^\top M^{-1} U_{\vth\vth} M^{-1} U_{\vth\vth} M^{-1}\vp ,
\end{aligned}
\end{equation}
\normalsize
in which $U_{\vth}$, $U_{\vth\vth}$, $U_{\vth\vth\vth}$ and $U_{\vth\vth\vth\vth}$ denote the first to fourth derivatives of the potential, the last two acting as multilinear forms on their bracketed arguments. For a quadratic potential such as a Gaussian target, and approximately for a target near its mode, the third and fourth derivatives vanish and \eqref{eq:mod6} reduces to
\begin{equation}\label{eq:mod6-gaussian}
\begin{aligned}
  \Hmod^{[6]} \;&=\; \Hmod^{[4]}
   \;+\; \varepsilon^{4} c_{43}\, U_{\vth}^\top M^{-1} U_{\vth\vth} M^{-1} U_{\vth} \\
  &\quad\; +\; \varepsilon^{4} c_{44}\, \vp^\top M^{-1} U_{\vth\vth} M^{-1} U_{\vth\vth} M^{-1}\vp .
\end{aligned}
\end{equation}
It is this simplification that underlies the Gaussian analysis employed by the s-AIA and s-MAIA schemes of Section~\ref{sec:tuning}. When the Hessian is instead dense and expensive to evaluate directly, an equivalent \emph{numerical time-derivative} formulation replaces its higher derivatives by finite differences of quantities already computed along the trajectory \cite{radivojevic2018multi}.

\subsection{The Coefficients \texorpdfstring{$c_{ij}$}{cij}}

\medskip\noindent\textbf{Two-stage integrators.} For the one-parameter family of \eqref{eq:two-stage} with $b \in (0, \tfrac12)$, the coefficients are
\begin{align*}
  c_{21} &= \tfrac{1}{24}(6b - 1), \\
  c_{22} &= \tfrac{1}{12}(6b^{2} - 6b + 1), \\
  c_{41} &= \tfrac{1}{5760}(7 - 30b), \\
  c_{42} &= \tfrac{1}{240}(-10b^{2} + 15b - 3), \\
  c_{43} &= \tfrac{1}{120}(-30b^{3} + 35b^{2} - 15b + 2), \\
  c_{44} &= \tfrac{1}{240}(20b^{2} - 1).
\end{align*}

\medskip\noindent\textbf{Three-stage integrators.} For the two-parameter family of \eqref{eq:three-stage} with $a, b \in (0, \tfrac12)$, they become
\begin{align*}
  c_{21} &= \tfrac{1}{12}\left(1 - 6a(1-a)(1-2b)\right), \\
  c_{22} &= \tfrac{1}{24}\left(6a(1-2b)^{2} - 1\right), \\
  c_{41} &= \tfrac{1}{720}\left(30a^{2}(1-a)^{2}(1-2b) - 1\right), \\
  c_{42} &= \tfrac{1}{120}\big(30a^{3}(1-2b)^{2} - 5a^{2}(24b^{2} - 22b + 5)\\
         &\qquad\quad - 10ab(1-2b) + 1\big), \\
  c_{43} &= \tfrac{1}{240}\left(20a(1-2b)\left(b + a(1 - 6b + 6b^{2})\right) - 1\right), \\
  c_{44} &= \tfrac{1}{120}\big(-30a^{3}(1-2b)^{2} + 5a^{2}(12b^{2} - 20b + 7)\\
         &\qquad\quad - 5a(1-2b)(3+2b) + 2\big).
\end{align*}
For $a = \tfrac13$, $b = \tfrac16$ the scheme consists of three Verlet steps of length $\varepsilon/3$, and the coefficients reduce to the one-stage values $c_{21} = \tfrac{1}{12}$, $c_{22} = -\tfrac{1}{24}$, $c_{41} = -\tfrac{1}{720}$, $c_{42} = \tfrac{1}{120}$, $c_{43} = -\tfrac{1}{240}$ and $c_{44} = \tfrac{1}{60}$ divided by $3^{2}$ for the first two and by $3^{4}$ for the rest. The three-stage expressions for $c_{41}, \dots, c_{44}$ printed in \cite{radivojevic2018multi} do not satisfy this check, and those given above were rederived from the Baker--Campbell--Hausdorff expansion.

\subsection{The Expected Energy Error Bound \texorpdfstring{$\rho_k(\varepsilon,b)$}{rho}}

Applied to the harmonic oscillator $H = \tfrac12(p^{2}+\theta^{2})$, whose equations of motion are $\dot\theta = p$ and $\dot p = -\theta$, a $k$-stage palindromic integrator acts as the linear map of \eqref{eq:linear-map}, and the bound \eqref{eq:rho} specializes as follows. For $k = 2$,
\begin{equation}\label{eq:rho2}
\resizebox{0.85\columnwidth}{!}{$\displaystyle
  \rho_2(\varepsilon, b) \;=\;
  \frac{\varepsilon^{4}\left(2b^{2}\left(\tfrac12 - b\right)\varepsilon^{2}
        + 4b^{2} - 6b + 1\right)^{2}}
       {8\left(2 - b\varepsilon^{2}\right)
        \left(2 - \left(\tfrac12 - b\right)\varepsilon^{2}\right)
        \left(1 - b\left(\tfrac12 - b\right)\varepsilon^{2}\right)} ,$}
\end{equation}
whereas for $k = 3$ it is convenient to name the numerator and denominator separately, writing
\begin{equation}\label{eq:rho3}
  \rho_3(\varepsilon, b) \;=\;
  \frac{\varepsilon^{4}\,\big(R^{\mathrm{num}}_3(\varepsilon,b)\big)^{2}}
       {2\,R^{\mathrm{den}}_3(\varepsilon,b)},
\end{equation}
with
\begin{align*}
  R^{\mathrm{num}}_3 &= -3b^{4} + 8b^{3} - \tfrac{19}{4}b^{2} + b - \tfrac{1}{16} \\
    &\quad + b^{2}\varepsilon^{2}\!\left(b^{3} - \tfrac54 b^{2} + \tfrac{b}{2} - \tfrac{1}{16}\right), \\[3pt]
  R^{\mathrm{den}}_3 &= \left(3b - b\varepsilon^{2}\!\left(b - \tfrac14\right) - 1\right) \\
    &\quad \times \left(1 - 3b - b\varepsilon^{2}\!\left(b - \tfrac12\right)^{2}\right) \\
    &\quad \times \Big(-9b^{2} + 6b - 1 \\
    &\qquad\quad - \varepsilon^{2}\!\left(b^{3} - \tfrac54 b^{2} + \tfrac{b}{2} - \tfrac{1}{16}\right)\Big),
\end{align*}
where the three-stage coefficients $(a, b)$ are restricted to the one-parameter family \eqref{eq:three-stage-constraint-main}, outside which the scheme has poor stability properties \cite{campos2017palindromic}; $\rho_3$ is written as a function of $b$ alone on that account. The optimal coefficient $b_{\text{opt}}^{k}$ of \eqref{eq:b-opt} is then obtained by minimizing the maximum of $\rho_k$ over the interval of step sizes determined during burn-in, searching over values of $b$ lying between the minimum-error and Velocity Verlet values recorded in Table~\ref{tab:mssi}.

\subsection{The Extended Hamiltonian for the Partial Momentum Update}

Because $\Hmod^{[k]}$ is generally non-quadratic in the momentum, the rotation \eqref{eq:pmu} no longer preserves the modified density and the momentum update must therefore be corrected by a Metropolis test with respect to
\begin{equation}\label{eq:extended-appendix}
  \Hext(\vth, \vp, \vmu) \;=\; \Hmod^{[k]}(\vth, \vp)
  \;+\; \tfrac12\,\vmu^\top M^{-1}\vmu ,
\end{equation}
which restores invariance of the joint density in $(\vth,\vp,\vmu)$ and corresponds to the momentum update test of Algorithm~\ref{alg:mmhmc}. Under the standard Hamiltonian, $K(\vp)+K(\vmu)$ is already invariant under the rotation, so GHMC needs no analogous test.

\end{appendix}

\begin{acks}[Acknowledgments]
The authors would like to thank the Basque Center for Applied Mathematics (BCAM), particularly the \textit{Modeling \& Simulation in Life and Materials Sciences} group, for its stimulating research environment and support. We are especially grateful to Elena Akhmatskaya, Tijana Radivojevi\'c, Mario Fern\'andez-Pend\'as, Mart\'in Parga Pazos, Lorenzo Nagar, and Leonardo Gavira Balmacz for their valuable discussions and insights that helped shape the perspective of this survey.
\end{acks}

\begin{funding}
This work was supported by the Basque Government under grant DEUSTEK6 -- Humanized Computing for Smart Sustainable and Healthier Communities and Environments (IT1901-26), and by the European Union's Horizon Europe Research and Innovation Programme under the LATE-AYA project (Grant Agreement No. 101214326).

M. Fernandez-de-Retana acknowledges additional support from the ELKARTEK project CICe21 -- ``MODELIZACI\'ON ELECTROQU\'IMICA Y T\'ERMICA PARA EL DESARROLLO EFICIENTE DE BATER\'IAS DE NUEVA GENERACI\'ON -- ION-MODEL'' (KK-2021/ 00064), funded by the Basque Government (ELKARTEK, Basque Country).
\end{funding}



\bibliographystyle{imsart-number} 
\bibliography{References}

\end{document}